\documentclass[12pt,english,floatfix,nofootinbib,superscriptaddress,aps,prd,preprint]{revtex4}
\usepackage[utf8]{inputenc}
\usepackage{float}
\usepackage{array}
\usepackage{bbold}
\usepackage{lipsum}
\usepackage{dsfont}
\usepackage{graphicx}
\usepackage{amsmath,amsthm,amsfonts,amssymb}
\usepackage{makecell}
\usepackage{graphicx}
\usepackage[english]{babel} 
\usepackage{color}
\usepackage{tensor}
\usepackage{esint}
\usepackage[dvips]{epsfig}
\usepackage[dvips]{graphicx}
\usepackage{float}
\usepackage{units}
\usepackage{textcomp}
\usepackage{mathrsfs}
\usepackage{amsmath}
\usepackage[makeroom]{cancel}
\usepackage{amssymb}
\usepackage{amsbsy}
\usepackage{amsfonts}
\usepackage{amssymb,mathrsfs,xcolor}
\usepackage{esint}
\usepackage{braket}
\usepackage{array}
\usepackage{graphicx}

\usepackage{wasysym}
\usepackage{multirow}
\usepackage{wrapfig}
\usepackage{subfig}

\usepackage{stmaryrd}
\usepackage{upgreek}

\makeatletter

\makeatletter\usepackage{babel}

\usepackage{hyperref}
\hypersetup{
    colorlinks,
    citecolor=blue,
    filecolor=green,
    linkcolor=purple,
    urlcolor=red,
}

\usepackage{slashed}

\newcommand{\ie}{\begin{equation}}
\newcommand{\fe}{\end{equation}}
\newcommand{\se}{\begin{eqnarray}}
\newcommand{\ff}{\end{eqnarray}}

\begin{document}

\title{Analysis of a nonlinear electromagnetic generalization of the Reissner--Nordström black hole}

\author{A. A. Ara\'{u}jo Filho}
\email{dilto@fisica.ufc.br}

\affiliation{Departamento de Física, Universidade Federal da Paraíba, Caixa Postal 5008, 58051-970, João Pessoa, Paraíba,  Brazil.}


\date{\today}

\begin{abstract}

In this work, we investigate the gravitational signatures of a nonlinear electromagnetic extension of the Reissner--Nordström solution. We conduct an analysis of light propagation, focusing on the photon sphere, shadow formation, and geodesic trajectories in this spacetime. The constraints on the parameter $\xi$, which characterizes the nonlinear extension of the Reissner--Nordström black hole, are derived from observational data provided by the Event Horizon Telescope (EHT). The time delay effects are also considered. In the thermodynamic analysis, we examine the Hawking temperature, entropy, heat capacity, and the emission of Hawking radiation via the tunneling process. The remnant mass and evaporation time of the black hole at its final stage are estimated. In addition, we compute the quasinormal modes using the WKB approximation, taking into account the characteristic oscillations of the system under scalar, vector, and tensor perturbations. Additionally, the time--domain solution is analyzed for all these perturbations to examine their evolution over time.

\end{abstract}

\maketitle


\section{Introduction}

Strongly magnetized compact objects, such as magnetars and neutron stars, exhibit nonlinear electromagnetic (NLE) effects that require modifications to classical Maxwell theory \cite{mosquera2004non,bronnikov2001regular,sorokin2022introductory,lobo2007gravastars,denisov2017vacuum,javed2019deflection,soares2023thermodynamics,lyutikov2011electromagnetic,denisova2019compact,denisov2016pulsar,torres2011unified}. Solutions to the Einstein--NLE equations offer improtant features about the physics of highly magnetized black holes, and they are useful tools for testing numerical models. Moreover, stationary solutions involving NLE fields can reveal new aspects of rotating astrophysical objects, particularly in addressing the issue of singularities. Many static black holes sourced by nonlinear electrodynamics avoid these singularities \cite{NEL03}, and recently a significant development was the introduction of a Kerr--Newman black hole extended by Euler--Heisenberg nonlinearities \cite{NEL04}.

Various forms of NLE Lagrangians, expressed as nonlinear functions of electromagnetic invariants, offer innovative approaches for developing solutions that generalize the Kerr--Newman model \cite{dymnikova2015electromagnetic,galindo2024nonlinear,hassaine2008higher,dymnikova2015regular,ayon2024unveiling,lammerzahl2018slowly,burinskii1999super,burinskii2014kerr,burinskii2008dirac,zilhao2014testing,hintz2018non,kubizvnak2022slowly}. Recent studies \cite{NEL05,NEL06} have introduced an exact solution to the Einstein--NLE equations, describing a rotating black hole characterized by mass, angular momentum, a cosmological constant, electric charge, and a nonlinear electromagnetic parameter. In this context, theories which preserve Lorentz and gauge invariance, have been systematically studied by Plebański \cite{NEL07} and further refined by Boillat \cite{NEL08}. These theories have significant implications for light propagation, predicting that rays follow null geodesics of two distinct optical metrics. Novello and collaborators \cite{NEL10} revisited this result, and Obukhov and Rubilar \cite{NEL11} demonstrated that in specific NLE models, the Fresnel equation for wave covectors factorizes, leading to birefringence.

In Ref. \cite{galindo2024nonlinear}, the authors introduced a nonlinear electromagnetic extension of the Reissner--Nordström solution with a cosmological constant and briefly discussed its {\\color{features}}. However, a detailed analysis of this case was lacking up to date. To fulfill this gap in the literature, we explore the gravitational features of it. Our study focuses on light propagation by examining the photon sphere, shadow formation, and the geodesic trajectories within this modified spacetime. Additionally, we analyze the time delay effects. In the thermodynamic aspect, we investigate key quantities such as the Hawking temperature, entropy, heat capacity, and the emission of Hawking radiation via the tunneling mechanism. We also estimate the remnant mass and evaporation time as the black hole approaches its final state. Furthermore, we determine the quasinormal modes using the WKB approximation, considering the system's characteristic oscillations under scalar, vector, and tensor perturbations. Moreover, we analyze the time--domain solution for these perturbations to investigate their temporal evolution.


\section{The static black hole solution}

In the scenario where rotational effects are absent, the solution reduces to a static configuration that represents a nonlinear electrodynamics extension of the Reissner--Nordström metric \cite{galindo2024nonlinear} 
\ie 
\begin{split}
\mathrm{d}s^{2} = & - \left[  1 - \frac{2M}{r} + \frac{{Q^{2}}}{r^{2}}(1 + \xi r^{3})    \right] \mathrm{d} t^{2} + \frac{1}{\left[  1 - \frac{2M}{r} + \frac{{Q^{2}}}{r^{2}}(1 + \xi r^{3})     \right]} \mathrm{d}r^{2} + r^{2} \mathrm{d}\theta^{2} + r^{2} \sin^{2}\theta \mathrm{d}\phi^{2} ,
\end{split}
\fe
where \( Q \equiv Q^{2}_{e} + Q^{2}_{m} \) denotes the effective charge, with \( Q_{e} \) and \( Q_{m} \) corresponding to the electric and magnetic charges, respectively. Here, let us define $f(r) \equiv 1 -\frac{2 M}{r}+\frac{Q^2 \left(\xi  r^3+1\right)}{r^2}$. This black hole solution exhibits three distinct horizons, depending on the very particular choices of parameters $Q$, $M$, $\xi$. However, we shall focus on the event horizon for addressing the next calculations, which is expected as 
\ie
\begin{split}
\label{evennt}
& r_{h} = \, M + \sqrt{M^2-Q^2}  + \frac{1}{2}  Q^2 \left(\frac{M \left[3 Q^2-4 M \left(\sqrt{M^2-Q^2}+M\right)\right]}{\sqrt{(M-Q) (M+Q)}}+Q^2\right) \xi,
\end{split}
\fe
where we have assumed $\xi$ to be small. The first term after the equality corresponds to the pure Reissner--Nordström black hole, while the second one reflects its modified extension governed by the parameter $\xi$. Additionally, it is crucial to note that the event horizon remains real and positive if and only if the following conditions are satisfied simultaneously:
\ie
\begin{split}
& M^{2} > Q^{2}, \quad \xi<0, \quad M>0, \quad Q>0.
\end{split}
\fe  
It is worth noting that, in the limit of the Reissner--Nordström solution ($\xi=0$), these constraints reduce to the well--known classical condition $M^{2} \geq Q^{2}$, ensuring the existence of real and positive values in this case.

\begin{figure}
    \centering
     \includegraphics[scale=0.51]{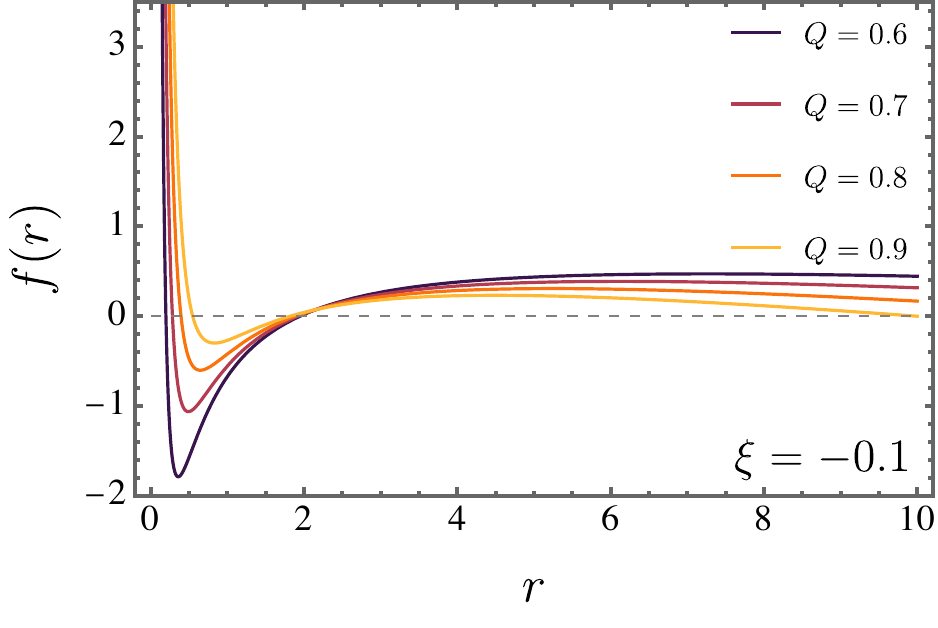}
     \includegraphics[scale=0.51]{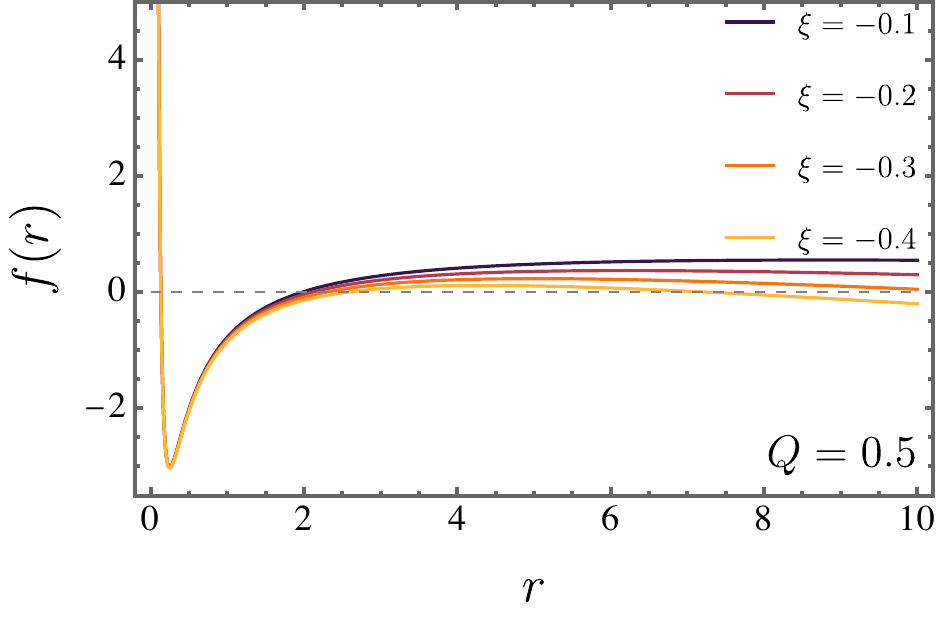}
    \caption{The representation of $f(r)$ is analyzed for different values of the effective $Q$ and the coupling parameter $\xi$.}
    \label{frs}
\end{figure}

\begin{figure}
    \centering
     \includegraphics[scale=0.51]{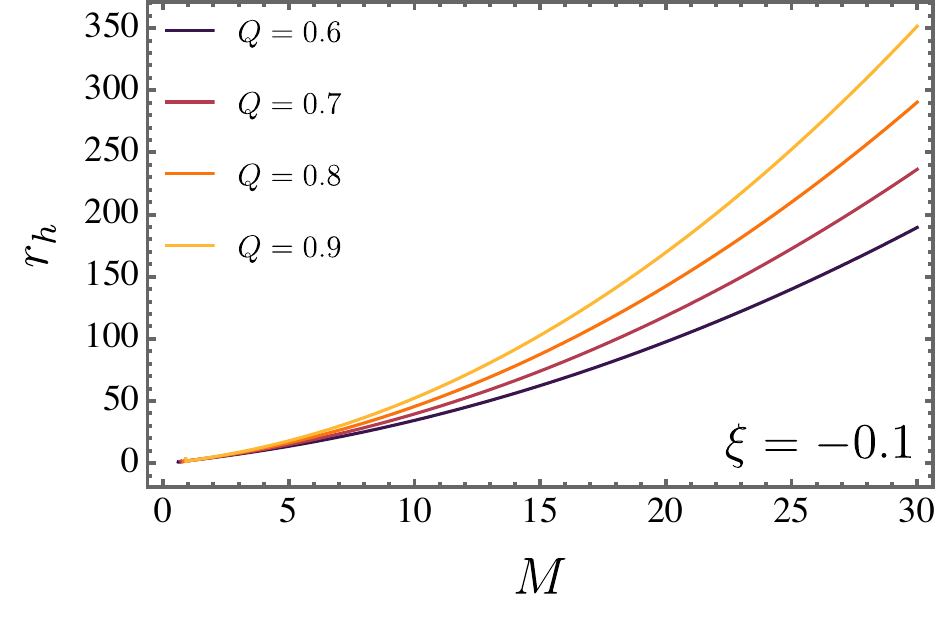}
     \includegraphics[scale=0.51]{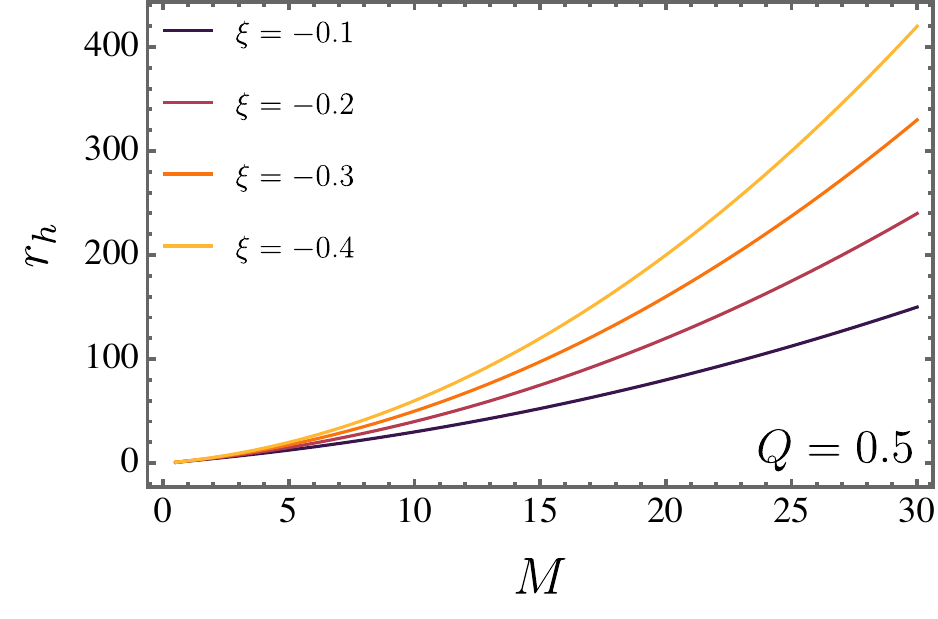}
    \caption{The event horizon $r_{h}$, is examined for various values of the effective $Q$ and the coupling parameter $\xi$.}
    \label{horizons}
\end{figure}

To validate the results obtained so far, we present a graphical interpretation in Figs. \ref{frs} and \ref{horizons}. Before proceeding, it is important to highlight that, consistent with Ref. \cite{galindo2024nonlinear}, the parameter $xi$ is considered to take negative values. Specifically, in Fig. \ref{frs}, we investigate the influence of the effective charge $Q$ and the nonlinear electromagnetic parameter $\xi$ on the metric function $f(r)$.

In contrast, Fig. \ref{horizons} illustrates the impact of varying the charge and nonlinear parameter on the event horizon radius $r_{h}$. An increase in the effective charge  $Q$ leads to a decrease of the event horizon \( r_{h} \) (for $\xi = -0.1$). Analogously, as the parameter $\xi$ decreases, the magnitude of $r_{h}$ increases (for $Q = 0.5$). Additionally, Tab. \ref{eventhorizonreal} illustrates the quantitative behavior of the event horizon. In general, for a given $M=1$ and $Q = 0.99$, a decrease in the parameter $\xi$ results in a corresponding increase in the horizon radius $r_{h}$. On the other hand, as the charge $Q$ increases, the magnitude of $r_{h}$ correspondingly decreases (for $\xi = -0.1$).

\begin{table}[!h]
\begin{center}
\begin{tabular}{c c c || c c c} 
 \hline\hline\hline
 $\xi$ & $Q$ &  $r_{h}$ & $\xi$ & $Q$ &  $r_{h}$ \\ [0.2ex] 
 \hline
 {-0.1}  & {0.99} & 1.65718 & {-0.1}  & {0.60} & {1.93122}  \\ 

 {-0.2}  & {0.99} & {2.17330} & {-0.1}  & {0.70} & {1.88693}  \\
 
 {-0.3}  & {0.99} & {2.68942} & {-0.1}  & {0.80} & {1.81845}  \\
 
 {-0.4}  & {0.99} & {3.20553} & {-0.1}  & {0.90} & {1.71096}  \\
 
 {-0.5}  & {0.99} & {3.72165} & {-0.1}  & {0.99} & {1.65718}  \\ 
 [0.2ex] 
 \hline \hline \hline
\end{tabular}
\caption{\label{eventhorizonreal} Quantitative analysis of the event horizon $r_{h}$ for different values of  $\xi$ and  $Q$.}
\end{center}
\end{table}

\begin{figure}
    \centering
     \includegraphics[scale=0.51]{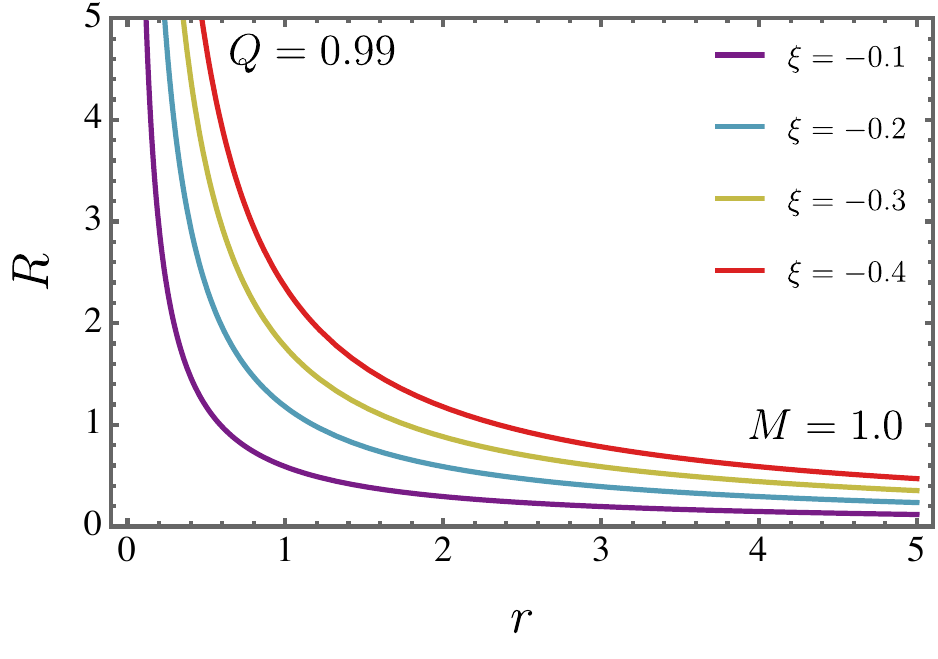}
     \includegraphics[scale=0.51]{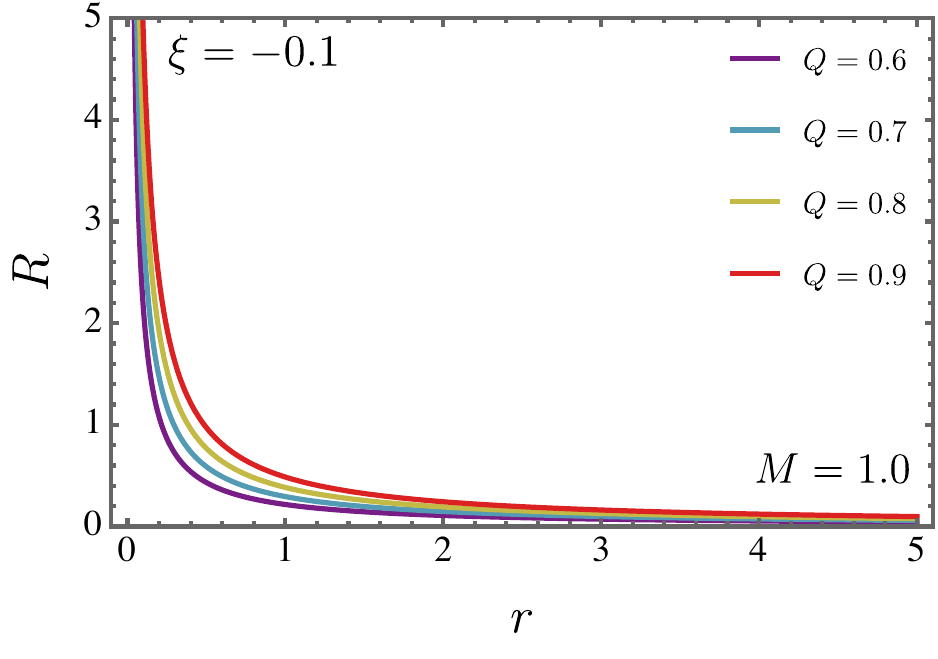}
    \caption{The Ricci scalar $R$ is represented for different values of $\xi$ and $Q$.}
    \label{Ricci}
\end{figure}

\begin{figure}
    \centering
     \includegraphics[scale=0.49]{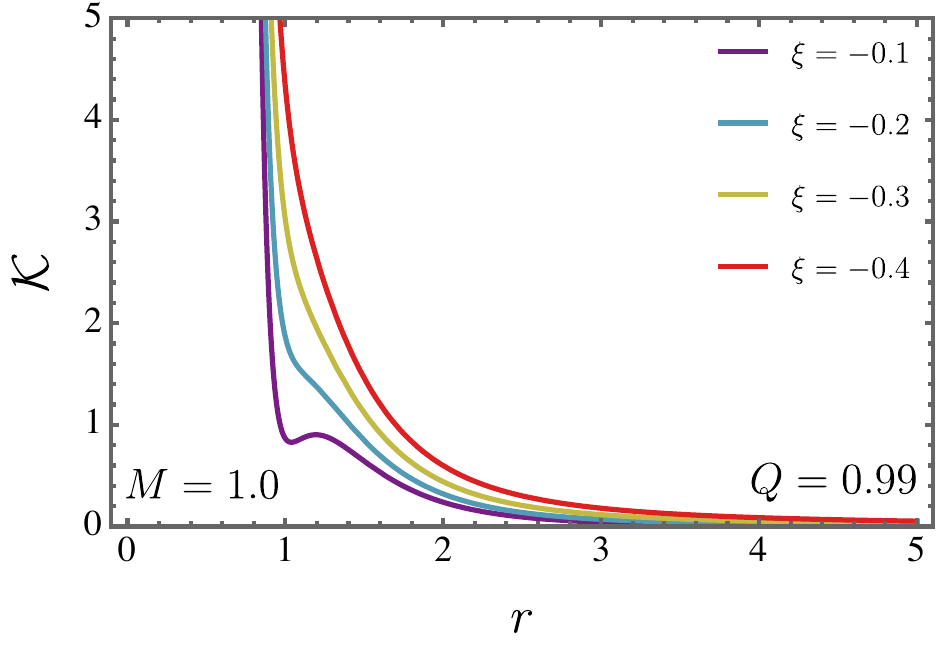}
     \includegraphics[scale=0.51]{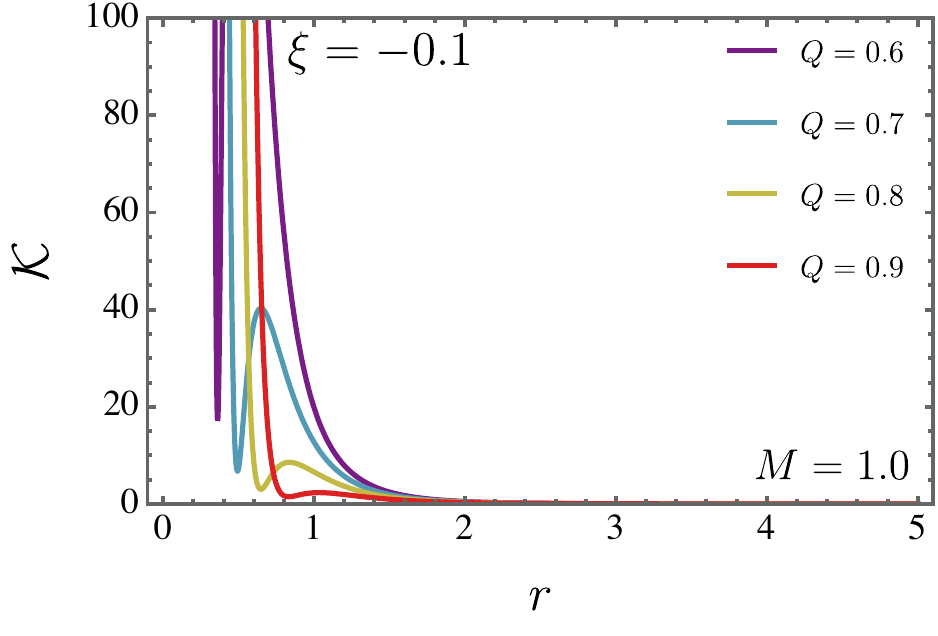}
    \caption{The Kretschmann scalar $\mathcal{K}$ is represented for different values of $\xi$ and $Q$.}
    \label{Kretschmann}
\end{figure}

Now, let us compute the Ricci scalar, i.e., $R \equiv g_{\mu\nu}R^{\mu\nu}$,
\ie
R = -\frac{6 \xi  Q^2}{r},
\fe
which possesses the following behavior addressed in Fig. \ref{Ricci}. Here, it is evident that there exists a singularity in $r \to 0$. In addition, the Kretschmann scalar is analyzed in order to verify the possible physical divergences in the black hole under consideration, which reads
\ie
\mathcal{K} = \frac{48 M^2}{r^6}-\frac{96 M Q^2}{r^7}+\frac{48 Q^4}{r^8}-\frac{8 \xi  Q^4}{r^5}+\frac{8 \xi ^2 Q^4}{r^2}.
\fe
The above expression clearly indicates the presence of a physical singularity as \( r \to 0 \). To better visualize the behavior of \( \mathcal{K} \) for various values of \( \xi \) and \( Q \), we refer to Fig. \ref{Kretschmann}. At this stage, it is important to highlight the non--vanishing Christoffel symbols, \( \Gamma\indices{^\mu_{\alpha\beta}} \), as they play a crucial role in calculating the geodesics within the framework of the theory. Then, they are
\ie
\begin{split}
\nonumber
& \Gamma\indices{^1_{00}} = \frac{\left(2 M r+Q^2 \left(\xi  r^3-2\right)\right) \left(r (r-2 M)+Q^2 \left(\xi  r^3+1\right)\right)}{2 r^5},\\
& \Gamma\indices{^1_{11}} = \frac{Q^2 \left(2-\xi  r^3\right)-2 M r}{2 r \left(r (r-2 M)+Q^2 \left(\xi  r^3+1\right)\right)},\\
& \Gamma\indices{^1_{22}} = 2 M+Q^2 (-r) \left(\frac{1}{r^2}+\xi  r\right)-r,\\
& \Gamma\indices{^1_{33}} = -r \sin ^2(\theta ) \left(-\frac{2 M}{r}+Q^2 \left(\frac{1}{r^2}+\xi  r\right)+1\right),  \\
& \Gamma\indices{^2_{21}} = 1/r,\\
& \Gamma\indices{^2_{33}} = \sin (\theta ) (-\cos (\theta )),\\
& \Gamma\indices{^3_{13}} = 1/r,\\
& \Gamma\indices{^3_{23}} = \cot (\theta ), \\
& \Gamma\indices{^3_{31}} = 1/r\\ 
& \Gamma\indices{^3_{32}} =  \cot (\theta ),\\
& \Gamma\indices{^0_{10}} = \frac{2 M r+Q^2 \left(\xi  r^3-2\right)}{2 r \left(r (r-2 M)+Q^2 \left(\xi  r^3+1\right)\right)}.
\end{split}
\fe
In this way, using the above expressions, the light trajectories can be accurately analyzed, as demonstrated in the following section.

\section{The light path}

The study of light trajectories in a nonlinear electromagnetic generalization of the Reissner--Nordström black hole serves as a complementary analysis to the rotating case \cite{galindo2024nonlinear}, providing a clearer picture of how modified field dynamics influence the spacetime structure and its observational signatures. Specifically, photon spheres determine the stability of light orbits and play a key role in characterizing gravitational lensing properties, while shadows link directly to the observable image of the black hole, offering a way to test theoretical models against astronomical data. Moreover, examining geodesic motion reveals important details about light propagation in such a spacetime. All these features will be presented below in the forthcoming subsections.

An important point should be emphasized regarding the behavior of light in nonlinear electrodynamics: unlike in linear theories, light rays do not generally trace the null geodesics of the background spacetime. Instead, they propagate along null curves defined by an effective, or optical, metric—see, for example, Refs. \cite{NEL08, NEL10}. Consequently, the analysis carried out in this work pertains to massless test particles rather than actual photons. This distinction arises because the nonlinear character of the theory introduces self--interactions in the electromagnetic field that affect photon propagation. To accurately describe light trajectories, one must first obtain the corresponding effective metric, as outlined in \cite{NEL10} by Novello et al. However, deriving this metric is often a nontrivial task, especially since it involves calculating derivatives of the Lagrangian with respect to the electromagnetic invariants—a procedure that becomes particularly challenging in the context of the specific spacetime considered here.


\subsection{Photon sphere and shadows}

To proceed, we introduce a generic form for the metric
\ie
g_{\mu \nu}\mathrm{d}x^\mu \mathrm{d}x^\nu  =  - A(r)\mathrm{d}t^2 + B(r)\mathrm{d}{r^2} + C(r)\mathrm{d}\theta ^2 + D(r)\,{{\mathop{\rm \sin}\nolimits} ^2}\theta \mathrm{d}\varphi ^2,
\fe
where, naturally, $A(r)$, $B(r)$, $C(r)$ and $D(r)$ are the respective components of such a tensor. Next, we regard the Lagrangian method as follows 
\ie
\mathcal{L} = \frac{1}{2}{g_{\mu \nu }}{{\dot x}^\mu }{{\dot x}^\nu },
\fe
so that
\begin{equation}
\mathcal{L} = \frac{1}{2}[ - A(r){{\dot t}^2} + B(r){{\dot r}^2} + C(r){{\dot \theta }^2} + D(r){{\mathop{\rm \sin}\nolimits} ^2}\, \theta {{\dot \varphi }^2}].
\end{equation}
Employing the Euler–Lagrange equation and restricting the motion to the equatorial plane (\(\theta = \frac{\pi}{2}\)), we obtain two constants of motion, namely the energy \(E\) and angular momentum \(L\), which can be expressed as:
\begin{equation}\label{constant}
E = A(r)\dot t \quad\mathrm{and}\quad L = D(r)\dot \varphi,
\end{equation}
and considering the light, then, it reads
\begin{equation}\label{light}
- A(r){{\dot t}^2} + B(r){{\dot r}^2} + D(r){{\dot \varphi }^2} = 0.
\end{equation}
Notice that, after performing some algebraic manipulations to substitute Eq. (\ref{constant}) into Eq. (\ref{light}), we have:
\begin{equation}\label{rdot}
\frac{{{{\dot r}^2}}}{{{{\dot \varphi }^2}}} = {\left(\frac{{\mathrm{d}r}}{{\mathrm{d}\varphi }}\right)^2} = \frac{{D(r)}}{{B(r)}}\left(\frac{{D(r)}}{{A(r)}}\frac{{{E^2}}}{{{L^2}}} - 1\right).
\end{equation}
Also, it is crucial to note that
\ie
\frac{\mathrm{d}r}{\mathrm{d}\lambda} = \frac{\mathrm{d}r}{\mathrm{d}\varphi} \frac{\mathrm{d}\varphi}{\mathrm{d}\lambda}  = \frac{\mathrm{d}r}{\mathrm{d}\varphi}\frac{L}{D(r)}, 
\fe
in which
\ie
\Dot{r}^{2} = \left( \frac{\mathrm{d}r}{\mathrm{d}\lambda} \right)^{2} =\left( \frac{\mathrm{d}r}{\mathrm{d}\varphi} \right)^{2} \frac{L^{2}}{D(r)^{2}}.
\fe
Thus far, we have outlined a general approach to determine the critical orbits (photon sphere) for a generic spherically symmetric spacetime. Now, let us specialize this framework to our specific case, yielding: $A(r) = -\frac{2 M}{r}+\frac{Q^2 \left(\xi  r^3+1\right)}{r^2}+1$, $B(r) =  \left( -\frac{2 M}{r}+\frac{Q^2 \left(\xi  r^3+1\right)}{r^2}+1\right)^{-1}$, $C(r) = r^{2}$ and $D(r) = r^{2}\sin^{2}\theta$. Thereby,
\ie
\Dot{r}^{2} = E^{2} + \mathcal{V}(r,\xi,Q),
\fe
where $\mathcal{V}(r,\xi,Q)$ reads
\ie
\mathcal{V}(r,\xi,Q) = \frac{L^2 \left(-\frac{2 M}{r}+\frac{Q^2 \left(\xi  r^3+1\right)}{r^2}+1\right)}{r^2}.
\fe
To determine the location of the light sphere, we solve \(\mathrm{d}\mathcal{V}/\mathrm{d}r = 0\). Interestingly, this equation yields three distinct roots; however, only two of them, \(r_{1c}\) and \(r_{2c}\), correspond to physical solutions, as shown below:
\ie
\begin{split}
& r_{1c} =  \frac{1}{2} \left(\sqrt{9 M^2-8 Q^2}+3 M\right) + \frac{1}{4} \xi  Q^2 \left(-\frac{9 M \left(3 M^2-2 Q^2\right)}{\sqrt{9 M^2-8 Q^2}}-9 M^2+2 Q^2\right),
\end{split}
\fe
and
\ie
\begin{split}
& r_{2c} = \frac{1}{2} \left(3 M-\sqrt{9 M^2-8 Q^2}\right) + \frac{1}{4} \xi  Q^2 \left(\frac{9 M \left(3 M^2-2 Q^2\right)}{\sqrt{9 M^2-8 Q^2}}-9 M^2+2 Q^2\right).
\end{split}
\fe

An important point to highlight is the existence of a critical photon sphere when $M = Q$, which still yields real and positive values. However, for $Q > M$ with $\xi < 0$, the photon spheres become complex. Another clarification is that, although the calculations yield two critical orbits, only $r_{1c}$ is considered physically relevant. This is because $r_{2c}$ (see Tab. \ref{photonrings2}) lies inside the event horizon, as can be verified by comparing Tab. \ref{eventhorizonreal} (for the event horizon) and Tab. \ref{photonrings2} (for $r_{2c}$).

\begin{table}[!h]
\begin{center}
\begin{tabular}{c c c || c c c} 
 \hline\hline\hline
 $\xi$ & $Q$ &  $r_{1c}$ & $\xi$ & $Q$ &  $r_{1c}$ \\ [0.2ex] 
 \hline
 {-0.05}  & 1.0 & {2.2}  & {-0.1}  & {0.5} & {2.92915}   \\ 

 {-0.10}  & 1.0 & {2.4}  & {-0.1}  & {0.6} & {2.88610}  \\
 
{ -0.20}  & 1.0 & {2.8}  & {-0.1}  & {0.7} & {2.82400}  \\
 
{ -0.30}  & 1.0 & {3.2} & {-0.1}  & {0.8} & {2.73415}  \\
 
 {-0.40}  & 1.0 & {3.6}  & {-0.1}  & {0.9} & {2.60160}  \\ 
 
  {-0.50}  & 1.0 & {4.0} & {-0.1}  & {1.0} & {2.42380}  \\ 
 [0.2ex] 
 \hline \hline \hline
\end{tabular}
\caption{\label{photonrings} Quantitative analysis of the photon rings $r_{1c}$ for different values of $\xi$ and $Q$.}
\end{center}
\end{table}
\begin{table}[!h]
\begin{center}
\begin{tabular}{c c c || c c c} 
 \hline\hline\hline
 $\xi$ & $Q$ &  $r_{2c}$ & $\xi$ & $Q$ &  $r_{2c}$ \\ [0.2ex] 
  \hline
 {-0.05}  & 1.0 & {0.975}  & {-0.1}  & {0.5} & {0.177098}   \\ 

 {-0.10}  & 1.0 & {0.950}  & {-0.1}  & {0.6} & {0.262936}  \\
 
{ -0.20}  & 1.0 & {0.900}  & {-0.1}  & {0.7} & {0.372493}  \\
 
{-0.30}  & 1.0 & {0.850} & {-0.1}  & {0.8} & {0.512894}  \\
 
 {-0.40}  & 1.0 & {0.800}  & {-0.1}  & {0.9} & {0.697286}  \\ 
 
  {-0.50}  & 1.0 & {0.750} & {-0.1}  & {1.0} & {0.950000}  \\ 
 [0.2ex] 
 \hline \hline \hline
\end{tabular}
\caption{\label{photonrings2} Quantitative analysis of the photon rings $r_{2c}$ for different values of $\xi$ and $Q$.}
\end{center}
\end{table}

The stability of photon spheres in black holes is closely tied to the geometric and topological properties of optical spacetime, where conjugate points play a fundamental role. When perturbations affect photon trajectories, their response depends on whether the photon sphere is stable or not. In unstable cases, small deviations cause photons to either be absorbed by the black hole or escape to infinity. Meanwhile, if the photon sphere is stable, photons can remain trapped in bounded orbits nearby \cite{qiao2022curvatures,qiao2022geometric}.

The behavior of photon spheres is directly influenced by the presence or absence of conjugate points in the spacetime manifold. Stability is associated with the existence of these points, while their absence characterizes instability. The Cartan–Hadamard theorem links the Gaussian curvature, $\Tilde{\mathcal{K}}(r,\xi,Q)$, to this structure, offering a way to determine the stability of critical orbits \cite{qiao2024existence}. Taking into account that
\ie
\mathrm{d}s^2 = g_{tt}(r) \mathrm{d} t^2 + g_{rr}(r) \mathrm{d}r^2 + g_{\theta\theta}(r) \mathrm{d}\theta^2 + g_{\phi\phi}(r,\theta) \mathrm{d}\phi^2,
\fe
and noting that null geodesics, characterized by the condition $\mathrm{d}s^2=0$, may be formulated as \cite{araujo2024effects,araujo2024impact,heidari2024absorption}:
\ie
\mathrm{d}t^2=\gamma_{ij}\mathrm{d}x^i \mathrm{d}x^j = -\frac{g_{rr}(r)}{g_{tt}(r)}\mathrm{d}r^2  -\frac{\Bar{g}_{\varphi\varphi}(r)}{g_{tt}(r)}\mathrm{d}\varphi^2   ,
\fe
where the optical metric is represented by $\gamma_{ij}$, with indices $i$ and $j$ spanning from $1$ to $3$. The function $\Bar{g}_{\varphi\varphi} (r)$ is defined as $\Bar{g}_{\varphi\varphi} (r) \equiv g_{\varphi\varphi}(r,\theta = \pi/2)$. In addition, the Gaussian curvature can be expressed as \cite{qiao2024existence}
\ie
\label{dffdsf}
\Tilde{\mathcal{K}}(r,\xi,Q) = \frac{R}{2} =  \frac{g_{tt}(r)}{\sqrt{g_{rr}(r) \,  \Bar{g}_{\varphi\varphi}(r)}}  \frac{\partial}{\partial r} \left[  \frac{g_{tt}(r)}{2 \sqrt{g^{(\Theta,l)}_{rr}(r) \, \Bar{g}_{\varphi\varphi}(r) }}   \frac{\partial}{\partial r} \left(   \frac{\Bar{g}_{\varphi\varphi}(r)}{g_{tt}(r)}    \right)    \right] ,
\fe
where the two-dimensional Ricci scalar is denoted by $R$. When $l$ and $\Theta$ are sufficiently small, its explicit formulation takes the form
\ie
\begin{split}
\label{gaussiancurvature}
\Tilde{\mathcal{K}}(r,\xi,Q) & =  \frac{3 M^2}{r^4}-\frac{6 M Q^2}{r^5}+\frac{2 Q^4}{r^6} -\frac{3 M \xi  Q^2}{r^2}-\frac{2 M}{r^3}-\frac{\xi ^2 Q^4}{4}+\frac{4 \xi  Q^4}{r^3}+\frac{3 Q^2}{r^4}.
\end{split}
\fe

\begin{figure}
    \centering
     \includegraphics[scale=0.65]{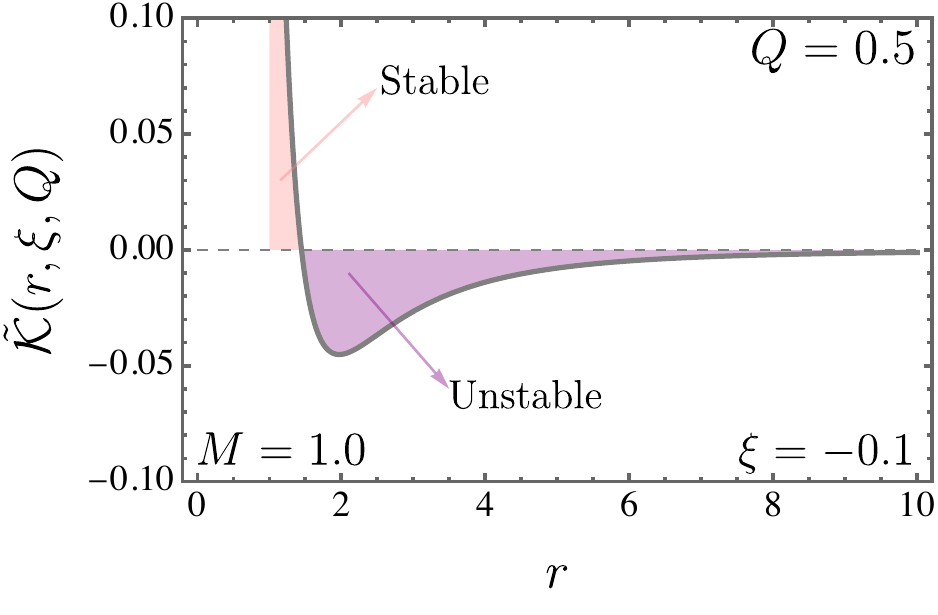}
    \caption{{The representation of the Gaussian curvature $\Tilde{\mathcal{K}}(r,\xi,Q)$ as a function of $r$ for $Q =0.5$, $M=1$, $\xi = -0.1$, highlighting the stable and unstable regions for the photon spheres.}}
    \label{Kanalysis}
\end{figure}

As discussed in \cite{qiao2022curvatures, qiao2022geometric, qiao2024existence}, the stability of photon spheres depends on the sign of $\Tilde{\mathcal{K}}(r,\xi,Q)$. When the curvature is negative, $\Tilde{\mathcal{K}}(r,\xi,Q) < 0$, the photon sphere is unstable, whereas a positive curvature, $\Tilde{\mathcal{K}}(r,\xi,Q) > 0$, indicates stability. To illustrate this distinction, Fig. \ref{Kanalysis} presents the Gaussian curvature $\Tilde{\mathcal{K}}(r,\xi,Q)$ as a function of $r$, marking the regions associated with stable and unstable photon spheres. The analysis is performed for $M = 1$, $\xi = -0.1$, and $Q = 0.5$.  

A further clarification is necessary. Since the results reveal the existence of two photon spheres, it is important to determine which one is stable. In the chosen configuration, the event horizon is located at $r = 1.95981$, while the transition between stability and instability occurs near $r \approx 1.45$. Given that $r_{1c} = 2.92915$ and $r_{2c} = 0.177098$, it follows that $r_{1c}$ corresponds to an unstable orbit, while $r_{2c}$ remains stable.

Since the spacetime under consideration is non–asymptotically flat, both the observer's position, $r_{o}$, and the photon sphere radius, $r_{ph}$, play a crucial role in determining the shadow \cite{cel4,araujo2024charged}
\ie
\mathcal{R} = r_{ph} \sqrt{\frac{f(r_{o})}{f(r_{ph})}}.
\fe

Now, let us derive the expression to the shadow radii as well
\ie
\begin{split}
 &\mathcal{R} = r_{1c} \sqrt{\frac{1 -\frac{2 M}{r_{o}}+\frac{Q^{2} \left(\xi  r_{o}^3+1\right)}{r_{o}^2}}{1-\frac{2 M}{r_{1c}}+\frac{Q^{2} \left(\xi  r_{1c}^3+1\right)}{r_{1c}^2}}} \\
& = \frac{1}{4} \left(\xi  Q^2 \left(-\frac{9 M \left(3 M^2-2 Q^2\right)}{\sqrt{9 M^2-8 Q^2}}-9 M^2+2 Q^2\right)+2 \left(\sqrt{9 M^2-8 Q^2}+3 M\right)\right) \\
& \times \sqrt{\frac{-\frac{2 M}{r_{o}}+Q^2 \left(\frac{1}{r_{o}^2}+\xi  r_{o}\right)+1}{-\frac{8 M}{\eta }+\frac{16 Q^2 \left(\frac{\eta ^3 \xi }{64}+1\right)}{\eta ^2}+1}},
\end{split}
\fe
where $\eta = \xi  Q^2 \left(-\frac{9 M \left(3 M^2-2 Q^2\right)}{\sqrt{9 M^2-8 Q^2}}-9 M^2+2 Q^2\right)+2 \left(\sqrt{9 M^2-8 Q^2}+3 M\right)$.

\begin{figure}
    \centering
     \includegraphics[scale=0.51]{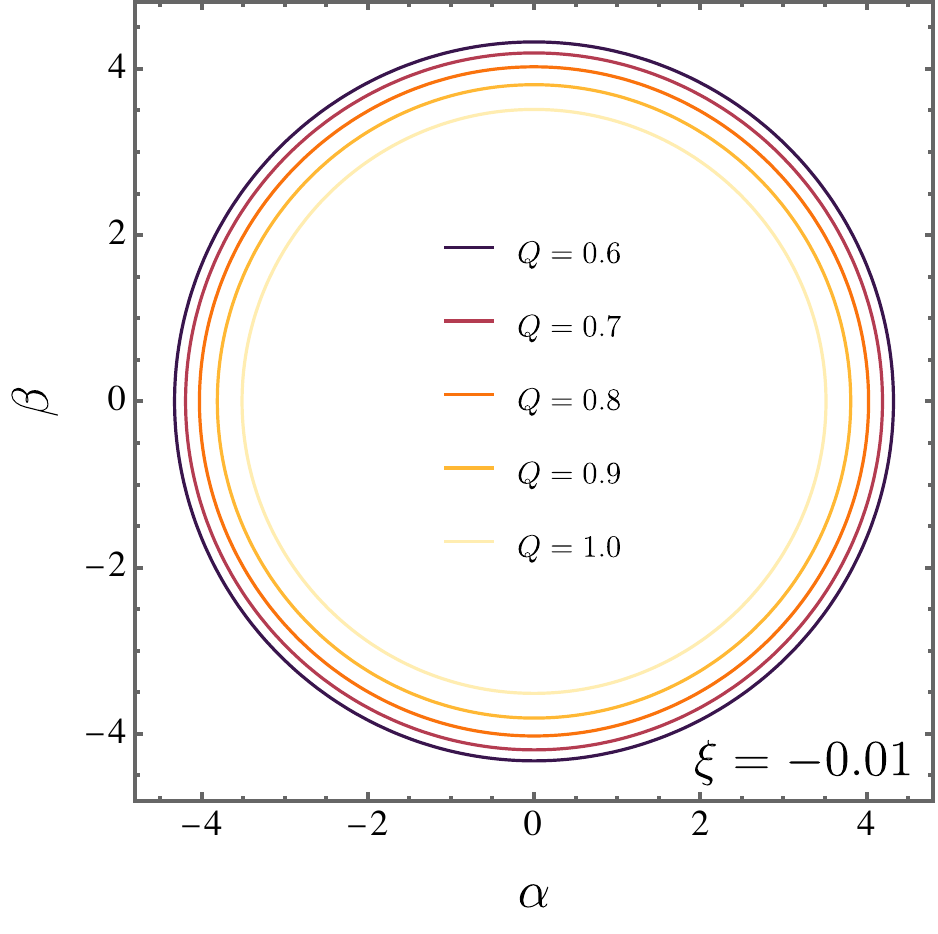}
     \includegraphics[scale=0.51]{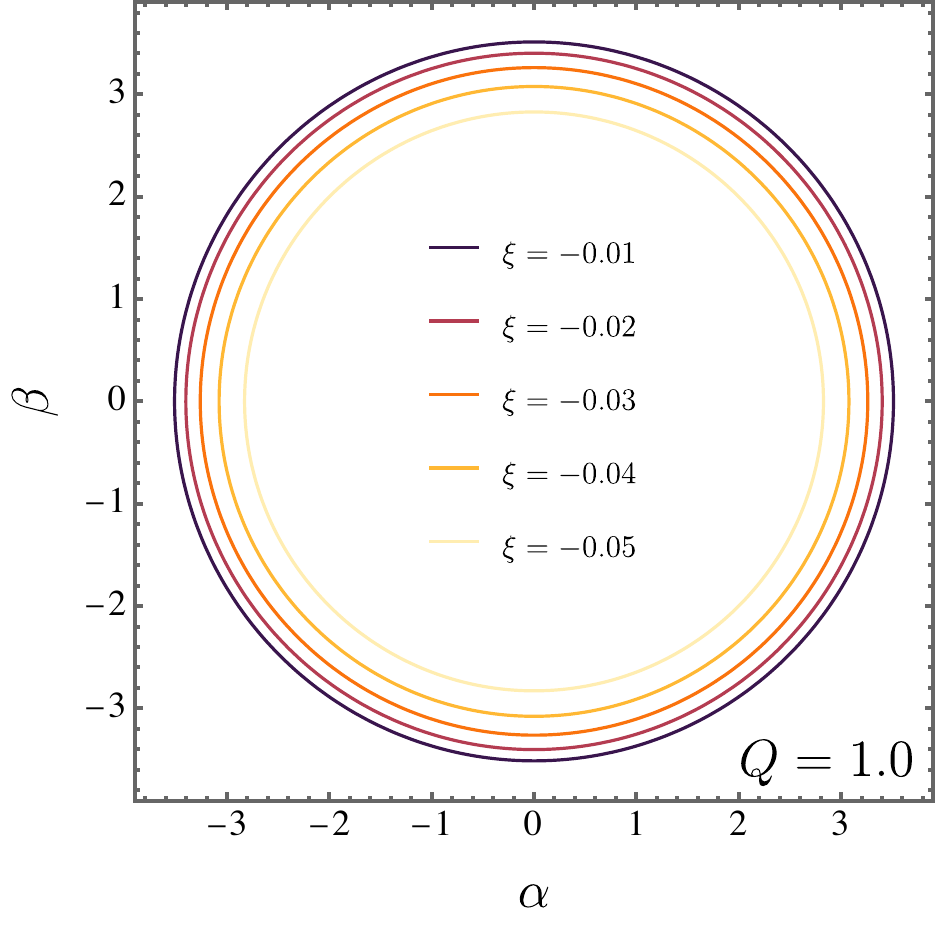}
    \caption{The shadow contours are depicted for various values of \(Q\) and \(\xi\).}
    \label{radiisha}
\end{figure}

It is worth mentioning that although there are two physical solutions (for the specific parameter values under consideration), when computing the shadow radii, only the outer critical orbit (photon sphere) is considered. Following the standard approach in the literature, we present the results as parametric plots of the celestial coordinates \(\alpha\) and \(\beta\) \cite{afrin2024testing,cel1,cel2,cel3,cel4}. Fig. \ref{radiisha} illustrates the shadow boundaries for varying values of \(Q\) and \(\xi\). In the left panel, the contours are shown for different values of \(Q\) with a fixed \(\xi = -0.01\). As \(Q\) increases, the size of the silhouette gradually diminishes. Conversely, in the right panel, the profiles correspond to different choices of $\xi$. Here, a decrease in $\xi$ results in a smaller outline of the shadow.

\begin{figure}
    \centering
     \includegraphics[scale=0.51]{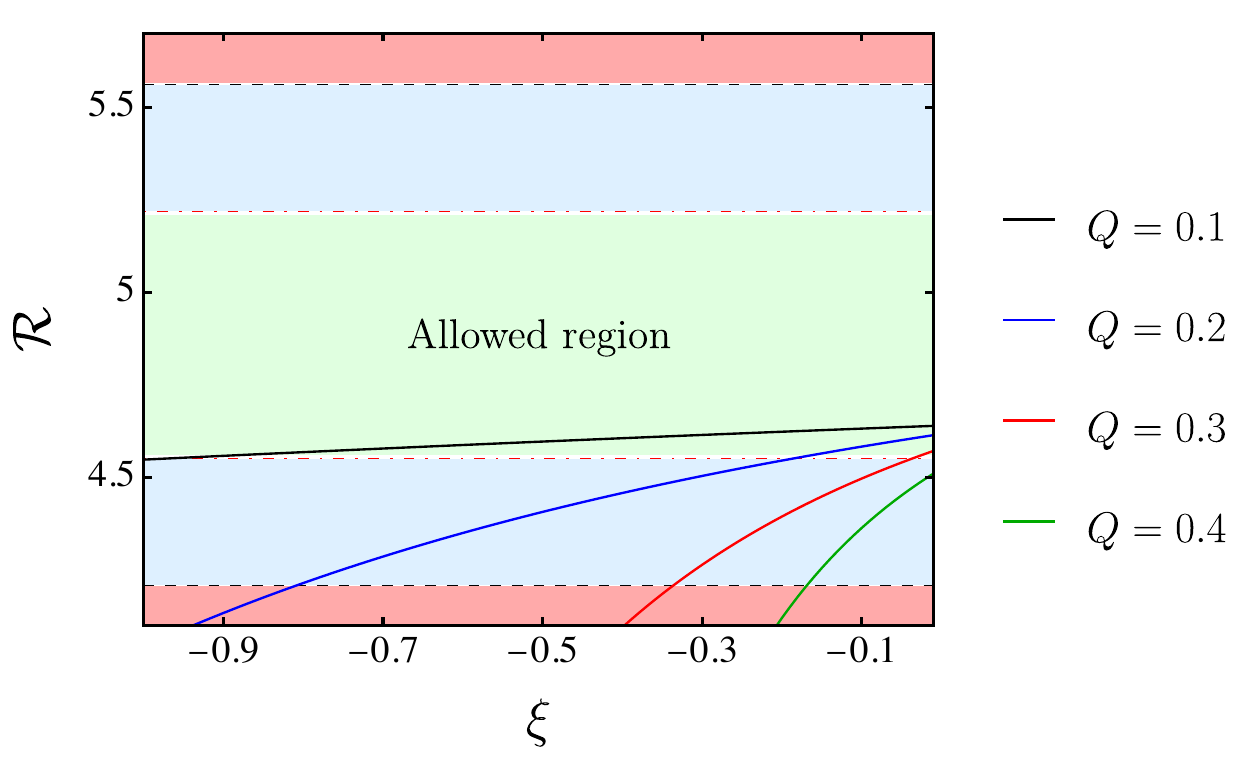}
    \caption{{Shadow radius in is plotted versus $\xi$ based on the  experimental constraints of $SgrA^{*}$ \cite{vagnozzi2022horizon,akiyama2022first}.}}
    \label{constraints}
\end{figure}

Based on the EHT horizon-scale observations of $Sgr A^*$, the mass-to-distance ratio priors from Keck and VLTI have been averaged. By incorporating a two-standard-deviation range, two constraints on the shadow radius have been derived \cite{vagnozzi2022horizon,akiyama2022first}
\ie
\label{const1asas}
4.55 < \frac{\mathcal{R}}{M} < 5.22,
\fe
and
\ie
\label{const1asas2}
4.21 < \frac{\mathcal{R}}{M} < 5.56.
\fe
The constraints on the parameter $\xi$ have been analyzed in light of observational limits from the Event Horizon Telescope. Figure \ref{constraints} illustrates the variation of the shadow radius as a function of $\xi$, expressed in units of $M$. The regions shaded in blue and green align with the experimental bounds given in Eqs. (\ref{const1asas}, \ref{const1asas2}). The intersection points between the plotted curve and the experimental constraint lines define an upper bound for $\xi$. All relevant values are summarized in Table \ref{tab:constr}.

\begin{table}[h!]
\centering
\caption{{Bounds for $\xi$ based on the observational data of EHT concering $SgrA^{*}$ \cite{vagnozzi2022horizon,akiyama2022first}.}}
\label{tab:constr}
\begin{tabular}{lc}
\hline\hline
\textbf{Parameter} & Bounds  \\
\hline
\quad  $Q =0.1$  & \makecell{$\xi < -0.89$} \\
\quad  $Q = 0.2$     & \makecell{$-0.84<\xi<-1.70$}  \\
\quad  $Q = 0.3$   & \makecell{$-0.56<\xi<-0.32$}  \\
\quad $Q = 0.4$  & \makecell{$-0.17<\xi$}  \\
\hline\hline
\end{tabular}
\end{table}


\subsection{Geodesics}

\begin{figure}
    \centering
     \includegraphics[scale=0.75]{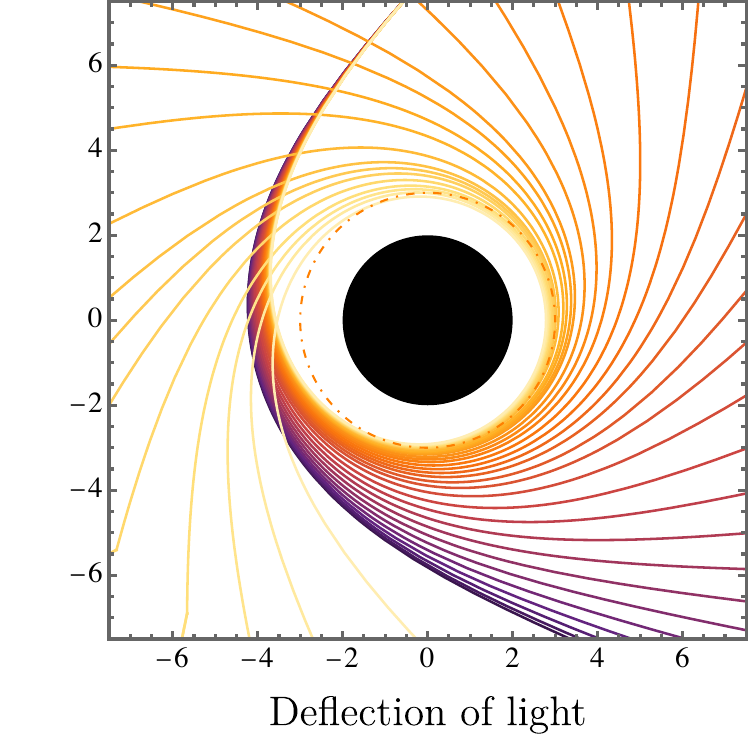}
    \caption{The light deflection is illustrated for a specific system configuration with parameters set to $\xi = -0.1$ and $M = 1$, while considering a range of values for $Q$: 0.05, 0.1, 0.15, 0.2, 0.25, 0.3, 0.35, 0.4, 0.45, 0.5, 0.55, 0.6, 0.64, 0.67, 0.7, 0.725, 0.75, 0.775, 0.8, 0.82, 0.84, 0.86, 0.869, 0.88, 0.89, 0.905, 0.915, 0.92, 0.925, 0.932, 0.939, 0.944, 0.949, and 0.956.}
    \label{deflec}
\end{figure}

Our primary objective is to thoroughly analyze the dynamics governed by the geodesic equations. To achieve this, we express these equations in the form
\ie
\frac{\mathrm{d}^{2}x^{\mu}}{\mathrm{d}\tau^{2}} + \Gamma\indices{^\mu_\alpha_\beta}\frac{\mathrm{d}x^{\alpha}}{\mathrm{d}\tau}\frac{\mathrm{d}x^{\beta}}{\mathrm{d}\tau} = 0, \label{geogeo}
\fe
where \( \tau \) represents an arbitrary affine parameter. This approach yields a set of four coupled differential equations, each describing the motion along a specific coordinate, as outlined below:
\ie
\frac{\mathrm{d} t^{\prime}}{\mathrm{d} \tau} = -\frac{r' t' \left(2 M r+Q^2 \left(\xi  r^3-2\right)\right)}{r \left(r (r-2 M)+Q^2 \left(\xi  r^3+1\right)\right)},
\fe
\ie
\begin{split}
& \frac{\mathrm{d} r^{\prime}}{\mathrm{d} \tau} = \frac{\left(r'\right)^2 \left(2 M r+Q^2 \left(\xi  r^3-2\right)\right)}{2 r \left(r (r-2 M)+Q^2 \left(\xi  r^3+1\right)\right)}\\
& -\frac{\left(r (r-2 M)+Q^2 \left(\xi  r^3+1\right)\right) \left(\left(t'\right)^2 \left(2 M r+Q^2 \left(\xi  r^3-2\right)\right)  -2 r^4 \left(\left(\theta '\right)^2+\sin ^2(\theta ) \left(\varphi '\right)^2\right)\right)}{2 r^5}, 
\end{split}
\fe
\ie
\frac{\mathrm{d} \theta^{\prime}}{\mathrm{d} \tau} = \sin (\theta ) \cos (\theta ) \left(\varphi '\right)^2-\frac{2 \theta ' r'}{r},
\fe
and, finally, 
\ie
\frac{\mathrm{d} \varphi^{\prime}}{\mathrm{d} \tau} = -\frac{2 \varphi ' \left(r'+r \theta ' \cot (\theta )\right)}{r}.
\fe
Using a numerical approach, we present Fig. \ref{deflec}, which illustrates the light trajectory for the black hole under consideration. In this depiction, the light--like geodesic is shown as a yellow curved line, the black disk represents the event horizon, while the orange dot--dashed lines correspond to $r_{1c}$. The illustration of light deflection is based on numerical computations for a specific system configuration, where the parameters are set to $\xi = -0.1$ and $M = 1$. The analysis considers a range of values for $Q$, including 0.05, 0.1, 0.15, 0.2, 0.25, 0.3, 0.35, 0.4, 0.45, 0.5, 0.55, 0.6, 0.64, 0.67, 0.7, 0.725, 0.75, 0.775, 0.8, 0.82, 0.84, 0.86, 0.869, 0.88, 0.89, 0.905, 0.915, 0.92, 0.925, 0.932, 0.939, 0.944, 0.949, and 0.956. For a given set of initial conditions of numerical calculations considered here, an increase in $Q$ results in a more curved trajectory for the light around the black hole.

\section{Time delay}

To determine the time delay experienced by light in a gravitational field, we solve the null geodesic equations within the context of a spherically symmetric spacetime. By tackling these differential equations, we obtain the behavior of light as it traverses the curved spacetime, allowing us to quantify the delay caused by the gravitational influence
The metric of the spacetime is given by:
\[
\mathrm{d}\tau^2 = f(r) \mathrm{d}t^2 - \frac{1}{f(r)} \mathrm{d}r^2 - r^2 (\mathrm{d}\theta^2 + \sin^2 \theta \, \mathrm{d}\phi^2).
\]
From this, we identify the conserved quantities arising from the spacetime symmetries. Specifically, the angular momentum \(L\) and energy \(E\) are defined as:
\ie
L \equiv r^2 \sin^2 \theta \frac{\mathrm{d}\phi}{\mathrm{d}\lambda}, \quad E \equiv f(r) \frac{\mathrm{d}t}{\mathrm{d}\lambda}.
\fe
The equation of motion for a particle is expressed through the Lagrangian:
\ie
\mathcal{L} = g_{\mu\nu} \frac{\mathrm{d}x^\mu}{\mathrm{d}\lambda} \frac{\mathrm{d}x^\nu}{\mathrm{d}\lambda}
= f(r) \left( \frac{\mathrm{d}t}{\mathrm{d}\lambda} \right)^2
- \frac{1}{f(r)} \left( \frac{\mathrm{d}r}{\mathrm{d}\lambda} \right)^2
- r^2 \left( \frac{\mathrm{d}\theta}{\mathrm{d}\lambda} \right)^2 
- r^2 \sin^2 \theta \left( \frac{\mathrm{d}\phi}{\mathrm{d}\lambda} \right)^2
\fe
where \(\lambda\) is the affine parameter, \(L\) represents the conserved angular momentum, and \(E\) corresponds to the conserved energy for a test particle. For particles restricted to motion in the equatorial plane (\(\theta = \pi/2\)), these quantities simplify the equations of motion to the following differential equations:
\ie
		\frac{1}{2} \bigg( \frac{\mathrm{d}r}{\mathrm{d}\lambda} \bigg)^{2} + \frac{1}{2} f(r) \bigg[ \frac{L^{2}}{r^{2}} + \mathcal{L} \bigg]
		= \frac{1}{2} \bigg( \frac{\mathrm{d}r}{\mathrm{d}\lambda} \bigg)^{2} + V(r)
		= \frac{1}{2}E^{2}.
\fe
In this scenario, the effective potential for a particle in a spherically symmetric gravitational field is expressed as
\ie
V(r) = \frac{f(r)}{2} \left( \frac{L^2}{r^2} + \mathcal{L} \right),
\fe
where \(L\) is the angular momentum, and \(\mathcal{L}\) represents the additional terms in the equation of motion. The impact parameter, defined as \(b = |L/E|\), plays a crucial role in determining the particle’s trajectory. For massless particles, such as photons, traveling along null geodesics, \(\mathcal{L} = 0\), simplifying the analysis. Concentrating on photon paths, we arrive at the following equation for their trajectories
\ie
	\frac{\mathrm{d}r}{\mathrm{d}t} = \frac{\mathrm{d}r}{\mathrm{d}\lambda}  \frac{\mathrm{d}\lambda}{\mathrm{d}t}
	= \pm f(r) \sqrt{1 - b^{2}\frac{f(r)}{r^{2}}}  .
\fe
We start with the expression \( E = f(r) \frac{\mathrm{d}t}{\mathrm{d}\lambda} \) and, for a photon (where \(\mathcal{L} = 0\)), the interpretation of the signs \( \pm \) becomes clear. As the photon travels along its path, it begins at a source location \( r_{\text{S}} \), with the radial coordinate \( r \) decreasing as it approaches the closest point, \( r = r_0 \), near the black hole. Once it reaches this point, the radial coordinate reverses its behavior, increasing as the photon moves away from the black hole. This setup leads to the following equations:
\ie
\frac{\mathrm{d}r}{\mathrm{d}t} = - f(r) \sqrt{1 - b^{2} \frac{f(r)}{r^{2}}} < 0.
\fe
As a photon moves from its starting point at \( r = r_{\text{S}} \) toward the closest distance, \( r = r_0 \), the radial coordinate decreases steadily. Also,
\ie
\frac{\mathrm{d}r}{\mathrm{d}t} = f(r) \sqrt{1 - b^{2} \frac{f(r)}{r^{2}}} > 0. 
\fe
Here, we consider the photon traveling from the turning point at \(r = r_0\) to the observer's location at \(r = r_{\text{O}}\). In the framework, when the light source is positioned at \(r = r_{\text{S}}\) and the observer is at \(r = r_{\text{O}}\), the time delay encountered by the light due to the gravitational field can be described below \cite{qiao2024time}
\ie
\begin{split}
\label{timeee}
 \Delta T & = T - T_{0} \\
 & = -\int_{r_{\text{S}}}^{r_{0}} \frac{\mathrm{d}r}{f(r)\sqrt{1-\frac{b^{2} f(r)}{r^{2}}}}
	      + \int_{r_{0}}^{r_{\text{O}}} \frac{\mathrm{d}r}{f(r)\sqrt{1-\frac{b^{2} f(r)}{r^{2}}}}
	      - T_{0}
	      \\
	& = \int_{r_{0}}^{r_{\text{S}}} \frac{\mathrm{d}r}{f(r)\sqrt{1-\frac{b^{2} f(r)}{r^{2}}}}
          + \int_{r_{0}}^{r_{\text{O}}} \frac{\mathrm{d}r}{f(r)\sqrt{1-\frac{b^{2} f(r)}{r^{2}}}}
          - \sqrt{r_{\text{S}}^{2}-r_{0}^{2}} - \sqrt{r_{\text{O}}^{2}-r_{0}^{2}}   
\end{split}
\fe

In this expression, \( T_{0} = \sqrt{r_{\text{S}}^{2} - r_{0}^{2}} + \sqrt{r_{\text{O}}^{2} - r_{0}^{2}} \) defines the light travel time between the source and the observer in the absence of any gravitational influence. The time delay \( \Delta T \) is found to increase continuously as both the source \( r_{\text{S}} \) and observer \( r_{\text{O}} \) move farther away from the turning point \( r_0 \), indicating a direct correlation between their positions and the delay caused by the gravitational field. As anticipated, Eq. (\ref{timeee}) does not admit an analytical solution. However, upon solving it numerically, we observe that as \(\xi\) decreases, the time delay \(\Delta T\) also diminishes.


\section{Thermal analysis}

In the early 1970s, a pivotal framework was established by Bardeen, Carter, and Hawking, who outlined a set of principles now recognized as the four laws of black hole mechanics. These principles were deliberately constructed to parallel the foundational laws of thermodynamics \cite{bardeen1973four}. The zeroth law indicates that surface gravity is uniform across the event horizon of a black hole, much like the constancy of temperature in a thermal equilibrium state \cite{page2005hawking}. The first law provides a relationship between the variation in a black hole’s mass (interpreted as its energy) and the corresponding shifts in its surface area, angular momentum, and electric charge, drawing a clear parallel to the thermodynamic rule connecting internal energy changes to heat and mechanical work \cite{carlip2014black}. According to the second law, the event horizon’s total area cannot decrease, mirroring the thermodynamic principle that entropy never decreases in an isolated system \cite{davies1978thermodynamics}. The third law establishes that it is fundamentally impossible to reduce a black hole’s surface gravity to zero by any physical means, echoing the impossibility of reaching absolute zero temperature \cite{hawking1976black}.

In addition, the Christodoulou’s contributions significantly broadened the understanding of these laws by addressing the irreversible processes that govern black hole dynamics \cite{christodoulou1970reversible}. Around the same time, Bekenstein introduced the concept of black hole entropy, marking a turning point in how black holes were viewed from a thermodynamic standpoint. His work established a direct relationship between a black hole’s entropy and the surface area of its event horizon \cite{1bekenstein2020black,2bekenstein1974generalized}. This feature led to the Bekenstein--Hawking entropy formula, which is a natural link between black hole mechanics and thermodynamic principles.


\subsection{Hawking temperature}

\begin{figure}
    \centering
     \includegraphics[scale=0.515]{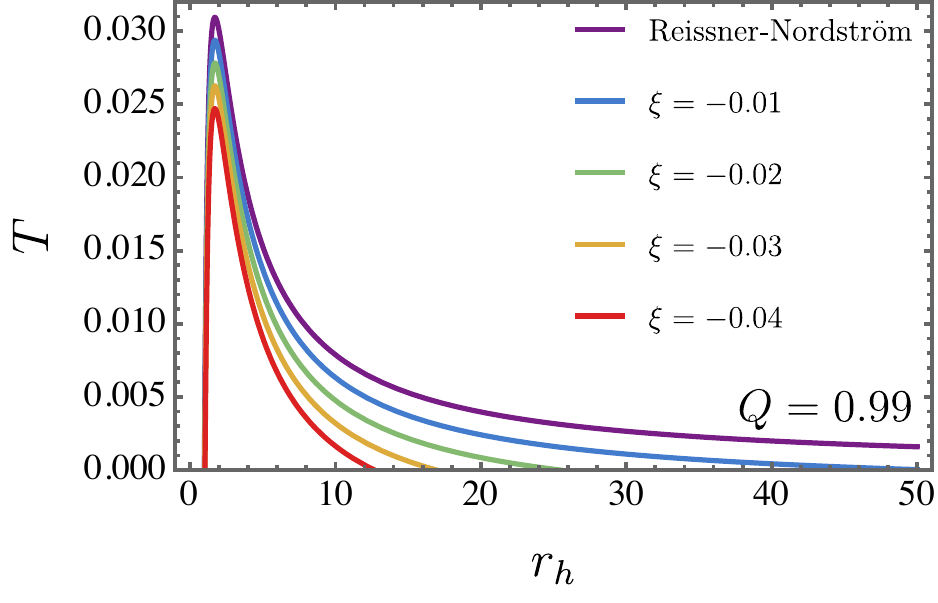}
     \includegraphics[scale=0.51]{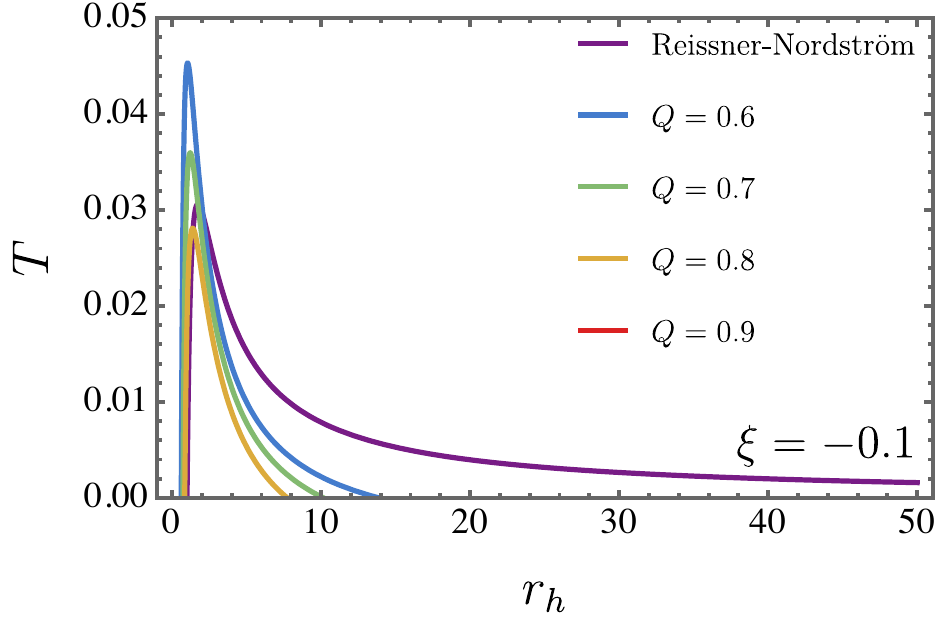}
    \caption{The Hawking temperature as a function of $r_{h}$ is shown for several values of $\xi$ and $Q$}
    \label{hawtemp}
\end{figure}

In this subsection, we analyze the key aspects of the Hawking temperature. As will be demonstrated in the following sections, this thermal property is crucial for determining the evaporation process as the black hole approaches its final stage. To achieve this, we derive the following expression based on the procedure for computing the surface gravity \cite{heidari2023gravitational}
\ie
\label{hahaha}
T = \frac{1}{{4\pi \sqrt {{g_{tt}}{g_{rr}}} }}{\left. {\frac{{\mathrm{d}{g_{rr}}}}{{\mathrm{d}r}}} \right|_{r = {r_{h}}}} = \frac{Q^2 \left(2 \xi  r_{h}^3-1\right)+r_{h}^2}{4 \pi  r_{h}^3}.
\fe
To facilitate a clearer understanding of our results, we present Fig. \ref{hawtemp}. As $\xi$ decreases, the intensity of the Hawking temperature diminishes correspondingly. Conversely, increasing $Q$ results in a further reduction in the magnitude of $T$. In addition, all these configurations are compared to the Reisser--Nordström black hole. Unless specified otherwise, comparisons with the Reissner–Nordström case assume $Q = 1$ exclusively for the pure Reissner–Nordström scenario to simplify the analysis. Another important point to examine is the existence of a remnant mass, which can be determined by expressing $T$ as a function of mass. This requires substituting the event horizon expression (Eq. (\ref{evennt})) into the Hawking temperature (Eq. (\ref{hahaha})), which leads to
\ie
\label{approt}
T \approx \frac{1}{8 \pi  M}+\frac{3 \xi  Q^2}{4 \pi } - \frac{Q^4 \left(1 + 16 M^3 \xi\right)}{128 \pi  M^5} + ... \, .
\fe

\begin{figure}
    \centering
     \includegraphics[scale=0.515]{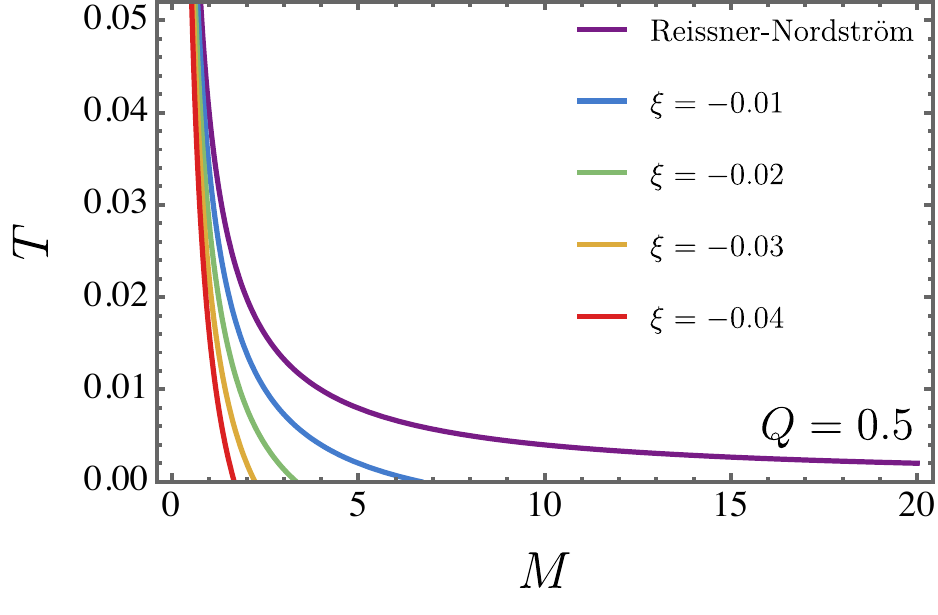}
     \includegraphics[scale=0.51]{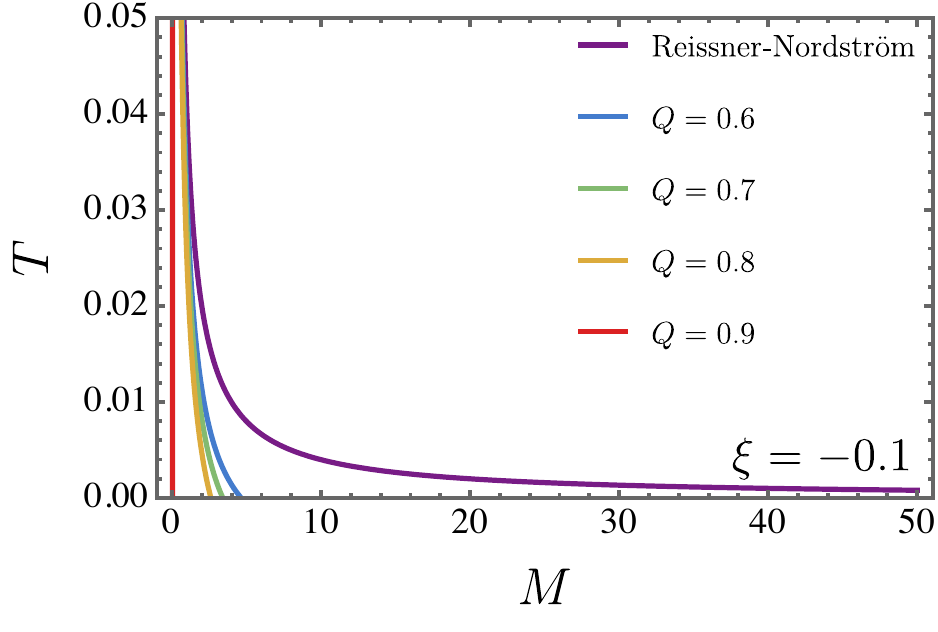}
    \caption{The Hawking temperature as a function of $M$ is shown for several values of $\xi$ and $Q$}
    \label{hawtempmass}
\end{figure}

Observing the limiting cases, setting $Q \to 0$ retrieves the well--known Hawking temperature for the Schwarzschild spacetime, given by $T = 1/8\pi M$. Additionally, in the limit $\xi \to 0$, the expression reduces to the Reissner--Nordström case: $T = 1/8\pi M - Q^{4}/128\pi M^{5}$. To advance the analysis, we consider only the first two terms of Eq. (\ref{approt}), as they suffice to account for the corrections introduced by $Q$ and $\xi$ in the Hawking temperature. For better visualization, Fig. \ref{hawtempmass} illustrates Eq. (\ref{approt}) by plotting the Hawking temperature as a function of mass. The temperature approaches zero at a specific mass value, indicating the presence of a remnant mass. Further discussion on this aspect is provided in the evaporation section of this paper.

\subsection{Heat capacity}

Complementing the results derived so far, we address here the behavior of the heat capacity $C_{V}$. In this manner, we write
\ie
C_{V} = T \frac{\partial S}{\partial T}= \frac{2 \pi  r_{h}^2 \left(Q^2 \left(2 \xi  r_{h}^3-1\right)+r_{h}^2\right)}{3 Q^2-r_{h}^2},
\fe
where $S$ represents the entropy which is give by $S = \pi r_{h}^{2}$.

\begin{figure}
    \centering
     \includegraphics[scale=0.51]{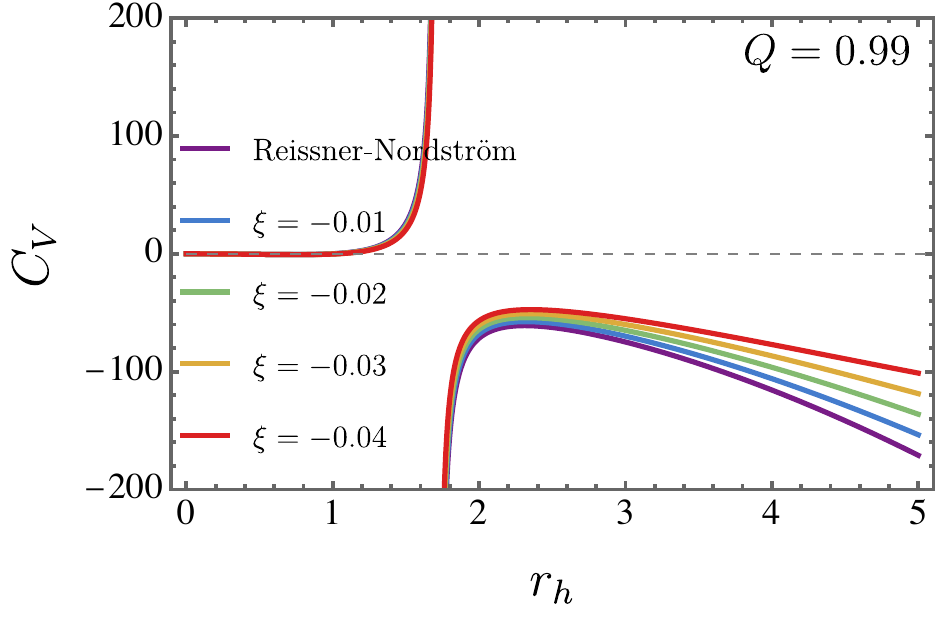}
     \includegraphics[scale=0.51]{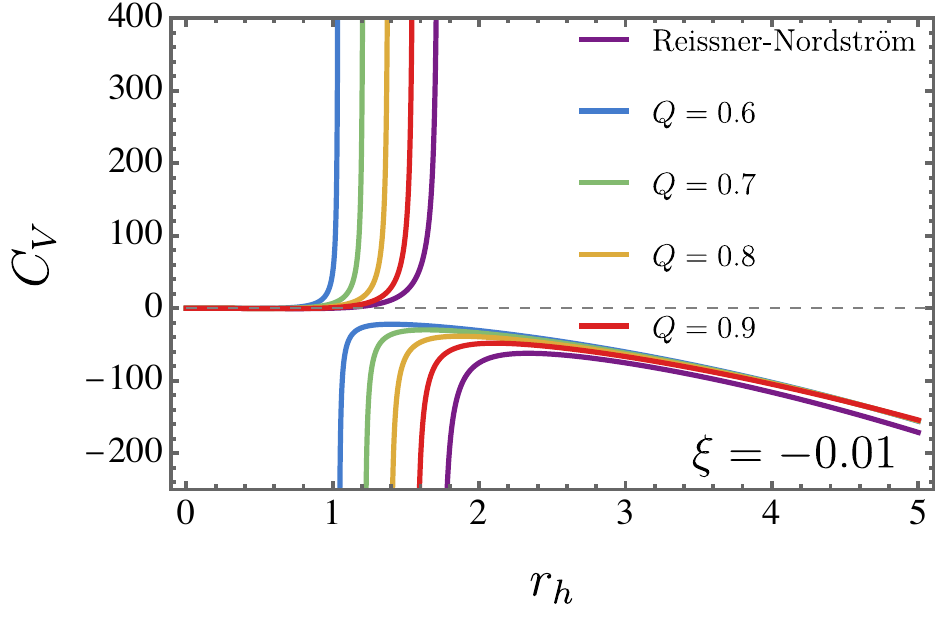}
    \caption{The heat capacity is exhibited for several values of $\xi$ and $Q$ }
    \label{heatcap}
\end{figure}

In Fig. \ref{heatcap}, we illustrate the behavior of the heat capacity for varying values of $\xi$ and $Q$. The plot reveals the presence of both stable (positive) and unstable (negative) configurations. Generally, as \(\xi\) decreases, the magnitude of \(C_{V}\) also decreases for stable configurations. Conversely, for unstable configurations, the magnitude of \(C_{V}\) increases as \(\xi\) decreases. Additionally, for different values of \(Q\), the heat capacity profile shifts to the right. It is important to note that all these configurations are compared against the Reisser--Nordström case. Similarly to the approach taken for the Hawking temperature, this thermodynamic property could also be examined as a function of mass. However, since it is not essential for the subsequent analysis, particularly in the context of black hole evaporation, it will not be included in this manuscript.

\section{Hawking radiation as a tunneling process}

\begin{figure}
    \centering
     \includegraphics[scale=0.51]{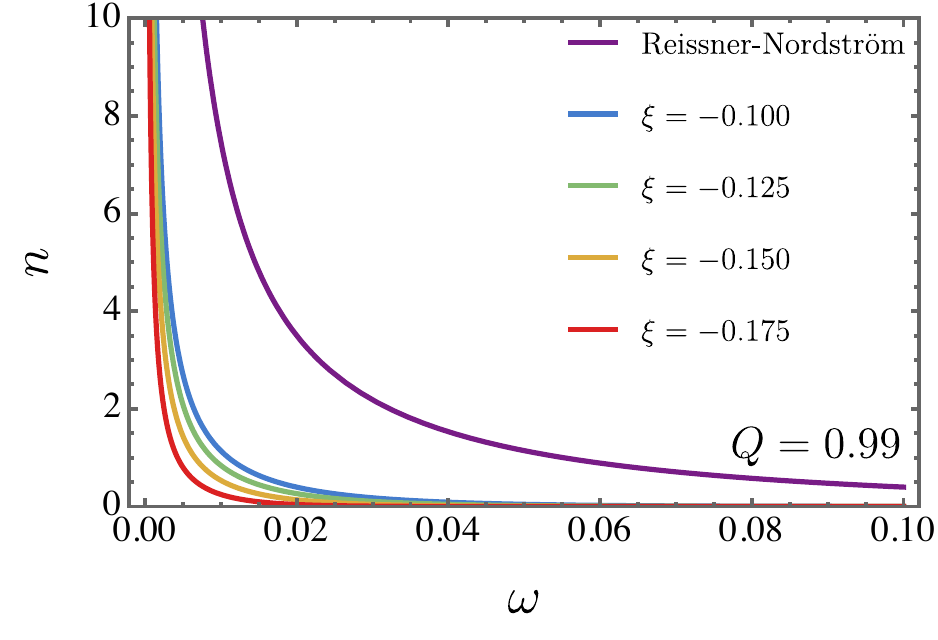}
     \includegraphics[scale=0.51]{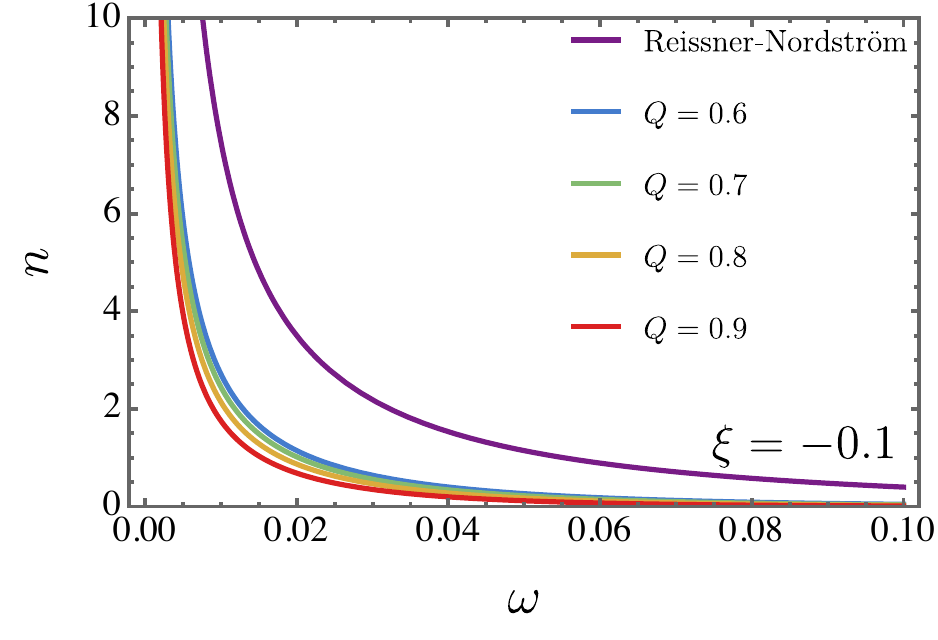}
    \caption{The magnitude of the particle density for different values of $\xi$ and $Q$.}
    \label{particledensity}
\end{figure}

Before proceeding, it is important to note that the calculations will be carried out without accounting for backreaction effects. The phenomenon of quantum tunneling allows particles inside the black hole to cross the event horizon. This tunneling probability can be derived, as shown in Refs. \cite{angheben2005hawking,kerner2006tunnelling,kerner2008fermions}. In this approach, our analysis centers specifically on radial paths. Consequently, near the horizon, the spacetime metric can be reduced to an effective two--dimensional form, simplifying the problem in this region
\ie
\mathrm{d}s^{2} = -f(r) \mathrm{d}t^{2} + \frac{\mathrm{d}r^{2}}{f(r)}.
\fe
As a result, the problem can be entirely addressed within the \(t-r\) plane. In this context, the dynamics of a scalar field \(\varphi\) with mass \(m_{\varphi}\) in a curved spacetime can be formulated by expressing the Klein--Gordon equation, which governs the field’s behavior, in the following manner:
\ie
 \hbar^{2} g^{\mu\nu} \nabla_\mu \nabla_\nu \varphi - m^2_{\varphi} \varphi = 0,
\fe
so that
\ie
-\partial_t^2 \varphi + f(r)^2 \partial_r^2 \varphi + \frac{1}{2} \partial_{r} f(r)^2 \partial_r \varphi - \frac{m^{2}_{\varphi}}{\hbar}f(r)\varphi = 0.
\fe
Using the Wentzel--Kramers--Brillouin (WKB) approximation, the solution to the previously mentioned equation can be obtained in the following form:
\ie
\varphi(t,r) = e^{-\frac{i}{\hbar} I(t,r)}.
\fe
Subsequently, the Hamilton--Jacobi equation can be expressed as
\ie
\left(\partial_t I\right)^2 - f(r)^2 \left(\partial_r I\right)^2 - m_{\varphi}^2 f(r) = 0,
\fe
Here, \( I(t,r) = -\omega t + W(r) \), where \(\omega\) represents the radiation frequency, and \(W(r)\) is the function describing the radial component of the action and \(W(r)\) is
\ie
W_{\pm}(r) = \pm \int \mathrm{d}r \frac{1}{f(r)} \sqrt{\omega^{2} - m_{\varphi}^{2} f(r)} \label{dablui}.
\fe
In this case, the symbols “$+$” and “$-$” correspond to the outgoing and ingoing solutions, respectively. Classically, the term \(W_{+}(r)\) is generally disallowed, as it describes trajectories that pass through the event horizon and move outward from \(r_{h}\). However, to examine Hawking radiation beyond the horizon, we focus on the outgoing solution, \(W_{+}(r)\). By applying an approximation for the function \(f(r)\) near the event horizon, around \(r_{h}\), we obtain the following expression:
\ie
f(r) = f(r_{h}) + f^{\prime}(r_{h})(r-r_{h}) + ..., 
\fe
so that Eq. (\ref{dablui}) simplifies to the following form:
\ie
W_{+}(r) = \frac{2i\pi \omega}{f^{\prime}(r_{h})}.
\fe
Thus, for a particle, the tunneling probability through the event horizon is directly related to the imaginary part of the action \(I(t,r)\). To put it differently:
\ie
\Gamma \simeq e^{-2\text{Im}I} = e^{-\frac{4\pi \omega}{f^{\prime}(r_{h})}},
\fe
As a result, the particle number density can be defined in terms of the tunneling rate as follows:
\ie
n = \frac{\Gamma}{1-\Gamma} = \frac{1}{e^{\frac{4 \pi  \omega }{f^{\prime}(r_{h})}}-1} = \frac{1}{e^{\frac{4 \pi  (r_{h})^3 \omega }{2 M r_{h}+Q^2 \left(\xi  (r_{h})^3-2\right)}}-1}.
\fe
In Fig. \ref{particledensity}, we illustrate the behavior of the particle density magnitude for various values of \(\xi\) and \(Q\). In the left panel, a decrease in \(\xi\) leads to a decrease in \(n\). Conversely, in the right panel, an increase in \(Q\) results in a lower magnitude of \(n\). Additionally, it is noteworthy that all results are compared against the Reisser--Nordström case for reference. It is worth mentioning that Hawking radiation was addressed for massive charged scalar field in the Kerr--Newman background as well \cite{Konoplya:2014sna}.

\section{Black hole evaporation}

\begin{figure}
    \centering
     \includegraphics[scale=0.6]{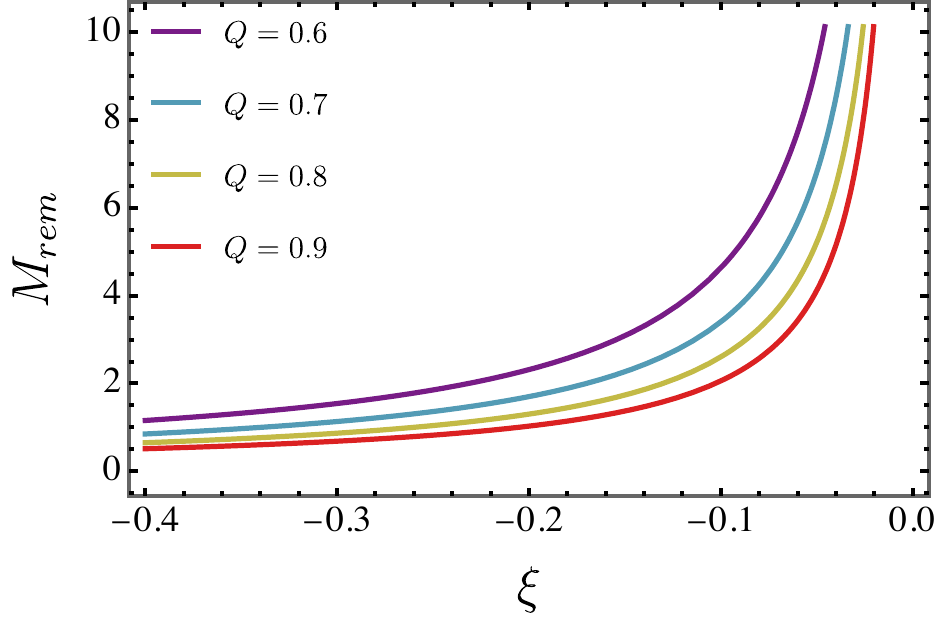}
    \caption{The remnant mass \( M_{rem} \) as a function of \( \xi \) for different values of \( Q \).}
    \label{remainingmass}
\end{figure}

This section is devoted to address the black hole evaporation when it reaches its final stage. Initially, we write the Hawking temperature a function of mass $M$
\ie
\begin{split}
\nonumber
T &  \, \approx \frac{1}{8 \pi  M}+\frac{3 \xi  Q^2}{4 \pi }.
\end{split}
\fe
In this context, we consider the extreme case \( T \to 0 \) to facilitate solving for \( M \) in the given expression. Additionally, to derive a simple analytical form for the remnant mass \( M_{rem} \), which is
\ie
M_{rem} = -\frac{1}{6 \xi  Q^2}.
\fe
To facilitate interpretation, we present Fig. \ref{remainingmass}. It is important to note that as \( \xi \) increases, \( M_{rem} \) increases. Additionally, an increase in \( Q \) leads to a lower magnitude of the remnant mass. To support the interpretation of the remnant mass, we present Table \ref{remmass}, which provides a quantitative analysis of the remnant mass.

\begin{table}[!h]
\begin{center}
\begin{tabular}{c c c || c c c} 
 \hline\hline\hline
 $\xi$ & $Q$ &  $M_{rem}$ & $\xi$ & $Q$ &  $M_{rem}$ \\ [0.2ex] 
 \hline
 -0.10  & {0.50} & { 6.66667   }   & -0.10  & {0.50} & { 6.66667 }    \\ 

 -0.11  & {0.50} &  {6.06061  }   & -0.10  & {0.60} & {4.62963  }   \\
 
 -0.12  & {0.50} & {5.55556  }   & -0.10  & {0.60} & {3.40136 }   \\
 
 -0.13  & {0.50} & { 5.12821  } & -0.10  & {0.80} & {  2.60417}   \\
 
 -0.14  & {0.50} & { 4.76190  }  & -0.10  & {0.90} & {  2.05761}   \\ 
 
  -0.15  & {0.50} & {4.44444  } & -0.10  & {0.99} & { 1.70051 }  \\ 
 [0.2ex] 
 \hline \hline \hline
\end{tabular}
\caption{\label{remmass} Quantitative analysis of the remanent mass $M_{rem}$ for different values of \( \xi \) and \( Q \).}
\end{center}
\end{table}

With these properties in hand, we can now address another important aspect that requires investigation: the black hole's lifetime. To analyze this, we express
\ie
\frac{\mathrm{d}M}{\mathrm{d}\tau} = - \alpha \sigma a T^{4}.
\fe

In this expression, \( a \) is the radiation constant, \( \sigma \) indicates the cross--sectional area, and \( \alpha \) stands for the greybody factor. It is worth noting that in an upcoming work, we shall conduct a detailed analysis of this aspect along with other features, such as gravitational lensing. In addition, within the geometric optics approximation, \( \sigma \) is equivalent to the photon capture cross--section, i.e., which is equivalent to $\pi \mathcal{R}^{2}$. Therefore, we obtain
\ie
\begin{split}
\frac{\mathrm{d}M}{\mathrm{d}\tau} = &    \frac{27 \, \Tilde{\Upsilon}}{4096 \left(\pi ^3 M^2\right)}-\frac{9 Q^2 \left[\Tilde{\Upsilon} \left(45 M^3 \xi -1\right)\right]}{4096 \left(\pi ^3 M^4\right)}  
\end{split}
\fe
with $\Tilde{\Upsilon} = a \alpha$. In this manner, it yields 
\ie
\begin{split}
\int_{0}^{t_{\text{evap}}} \Tilde{\Upsilon} \mathrm{d}\tau & = - \int_{M_{i}}^{M_{f}} 
\left[  \frac{27 \, \Tilde{\Upsilon}}{4096 \left(\pi ^3 M^2\right)}-\frac{9 Q^2 \left[\Tilde{\Upsilon} \left(45 M^3 \xi -1\right)\right]}{4096 \left(\pi ^3 M^4\right)}  \right]^{-1} \mathrm{d}M.
\end{split}
\fe
which leads to
\ie
\begin{split}
\label{evvapp}
t_{\text{evap}} = & \frac{1024 \pi ^3}{243 \Tilde{\Upsilon}} \left\{ 4 \sqrt{3} Q^3 \tanh ^{-1}\left(\frac{\sqrt{3} (M_{f} - M_{i})}{Q}\right) \right. \\
& \left. 3 \left[ 60 \xi  Q^4 (M_{f}-M_{i})^2+45 \xi  Q^2 (M_{f}-M_{i})^4-4 Q^2 (M_{f}-M_{i})-4 (M_{f}-M_{i})^3 \right. \right.\\
& \left. \left.   +\frac{10 \xi  Q^8}{Q^2-3 (M_{f}-M_{i})^2}+30 \xi  Q^6 \ln \left(3 (M_{f}-M_{i})^2-Q^2\right)     \right] \right\},
\end{split}
\fe
where $M_{i}$ and $M_{f}$ represent the initial and final masses, respectively, and $t_{\text{evap}}$ denotes the time corresponding to the final stage of the evaporation process. Notably, an analytical solution for black hole evaporation has been obtained. At this point, further analysis is needed. In the limit $\xi \to 0$, we recover
\ie
t_{\text{evap}}^{\xi \to 0} = -\frac{4096 \pi ^3 \left(3 (M_{f} - M_{i}) \left((M_{f} - M_{i})^2+Q^2\right)-\sqrt{3} Q^3 \tanh ^{-1}\left(\frac{\sqrt{3} (M_{f} - M_{i})}{Q}\right)\right)}{243 \Tilde{\Upsilon}} .
\fe
This expression corresponds to the evaporation lifetime of the Reissner--Nordström black hole. Moreover, in the limit $Q \to 0$, it reduces to the well-established result for the Schwarzschild spacetime \cite{heidari2024scattering,araujo2024particsle}
\ie
t_{\text{evap}}^{Q \to 0} = -\frac{4096 \pi ^3 (M_{f} - M_{i})^3}{81 \Tilde{\Upsilon}},
\fe
which defines the corresponding evaporation lifetime. Furthermore, as the black hole approaches its final stage, $M_{f} \to M_{rem}$, leading Eq. (\ref{evvapp}) to take the form
\ie
\begin{split}
\Tilde{t}_{\text{evap}} = & \frac{1024 \pi ^3}{243 \Tilde{\Upsilon}} \left\{ 135 M_{i}^4 \xi  Q^2 + 102 M_{i}^3 + \frac{57 M_{i}^2}{2 \xi  Q^2}+\frac{7 M_{i}}{2 \xi ^2 Q^4}+72 M_{i} Q^2+\frac{7}{\xi }+\frac{23}{144 \xi ^3 Q^6}  \right.\\
& \left. +180 M_{i}^2 \xi  Q^4+\frac{30 \xi  Q^8}{Q^2-3 \left(M_{i}+\frac{1}{6 \xi  Q^2}\right)^2}+4 \sqrt{3} Q^3 \tanh ^{-1}\left(\frac{\sqrt{3} \left(-M_{i}-\frac{1}{6 \xi  Q^2}\right)}{Q}\right)  \right.\\
& \left. +90 \xi  Q^6 \ln \left(3 \left(M_{i}+\frac{1}{6 \xi  Q^2}\right)^2-Q^2\right) \right\}.
\end{split}
\fe

\begin{figure}
    \centering
     \includegraphics[scale=0.48]{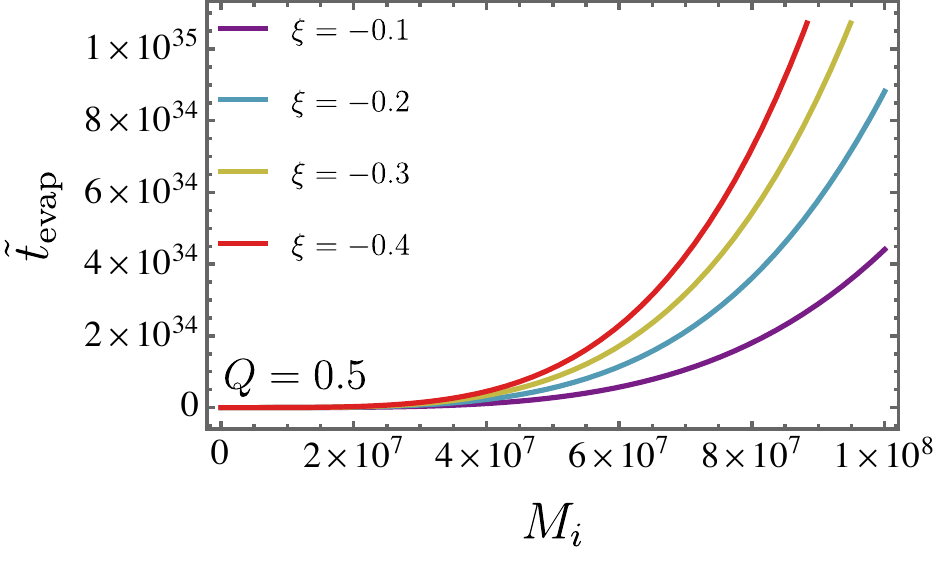} \includegraphics[scale=0.51]{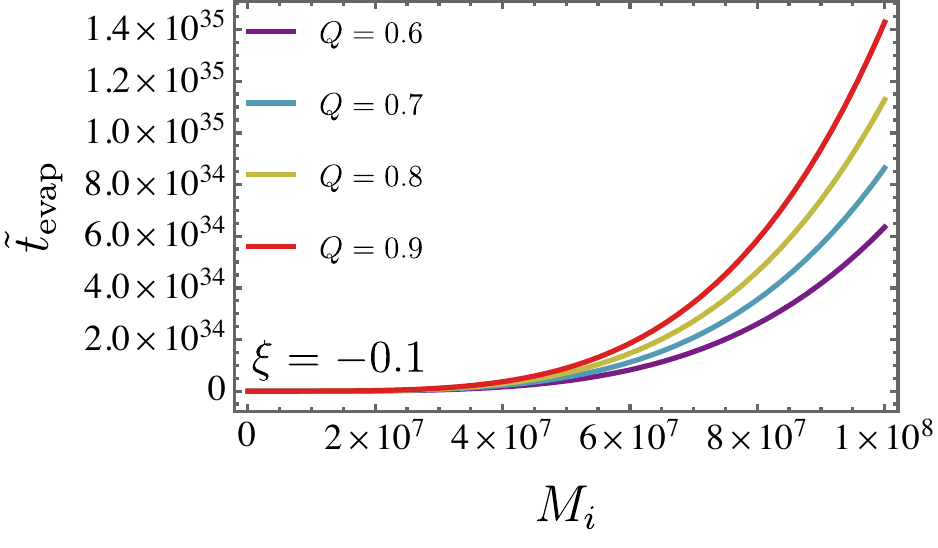}
    \caption{The evaporation lifetime $\Tilde{t}_{\text{evap}}$ as a function of $M_{i}$ for different values of $Q$ and $\xi$.}
    \label{evappp}
\end{figure}

To analyze the behavior of $\Tilde{t}_{\text{evap}}$, its magnitude is illustrated in Fig. \ref{evappp}. In general terms, a decrease in $\xi$ and an increase in $Q$ result in a longer evaporation timescale, particularly as $M_{i}$ increases.

Finally, an additional remark is considered. A recent study \cite{araujo2025remarks} suggested that, in a particular extension of the Reissner--Nordström scenario, black holes would absorb radiation than emit it, regardless of the presence of a cosmological constant. At least in the limit where $\Lambda$ vanishes, expanding the event horizon (assuming that $\xi$ is small) leads to a new expression for $r_h$ with constraints similar to those ones encountered in the Reissner--Nordström spacetime, in order to ensure real and positive defined values (e.g., $M > Q$).  

Under these conditions, the concern raised in Ref. \cite{araujo2025remarks}—and in the first arXiv version of this manuscript—about the possibility that the black hole might absorb radiation than emits it, depending on the parameter choices, is resolved. In other words, considering the Hawking temperature as a function of mass $M$, as shown in Fig. \ref{hawtempmass}, the black hole does emit radiation.


\section{Quasinormal modes}

In the ringdown phase of a black hole, a distinctive oscillatory behavior emerges, known as \textit{quasinormal} modes. These oscillations are not influenced by the initial disturbances but instead reflect the fundamental properties of the black hole and its surrounding spacetime. They represent the natural vibrations of the system, revealing its inherent characteristics and are independent of the specific initial conditions that triggered the perturbations \cite{Konoplya:2013rxa,Konoplya:2019hlu,Kokkotas:2010zd,Konoplya:2007zx,Konoplya:2011qq}.

Differently with what happens to \textit{normal} modes, which are associated with closed systems, quasinormal modes pertain to open systems instead, where energy is gradually lost through the emission of gravitational waves. In mathematical viewpoint, these modes correspond to the poles of the complex Green’s function and can be understood as solutions to the wave equation within the black hole's background spacetime.

Calculating the frequencies of quasinormal modes is often a challenging task due to the complexity of the underlying equations. These modes are governed by the wave equation in a spacetime characterized by a metric \( g_{\mu\nu} \) \cite{araujo2023analysis,heidari2023gravitational,araujo2024gravitational}. While exact analytical solutions are rare, various methods have been developed to approximate these frequencies.

\subsection{Scalar perturbations}

One of the most prominent techniques used for this purpose is the Wentzel--Kramers--Brillouin (WKB) method, which was initially developed by Will and Iyer \cite{iyer1987black,iyer1987black1} and later refined by Konoplya up to the sixth order \cite{konoplya2003quasinormal}. In our analysis, we focus on scalar field perturbations by solving the Klein--Gordon equation within a curved spacetime background to extract the relevant quasinormal frequencies
\ie
\frac{1}{\sqrt{-g}}\partial_{\mu}(g^{\mu\nu}\sqrt{-g}\partial_{\nu}\Phi) = 0.\label{Th4e3kleqin5Go55rdon}
\fe
It is important to mention that although the exploration of \textit{backreaction} effects in this context is an interesting feature of study, this manuscript does not cover that aspect. Instead, we focus on other key elements. Specifically, our primary objective is to analyze the scalar field as a small perturbation. Thereby, Eq. (\ref{Th4e3kleqin5Go55rdon}), turns out to be
\ie
\label{jekjsk}
\begin{split}
-& \frac{1}{f(r)} \frac{\partial^{2} \Phi}{\partial t^{2}} + \frac{1}{r^{2}} \left[  \frac{\partial}{\partial r} \left(  f(r) \, r^{2}  \frac{\partial \Phi}{\partial r}  \right)  \right] \\  + & \frac{1}{r^{2} \sin \theta}  \left[  \frac{\partial }{\partial \theta} \left( \sin \theta \frac{\partial}{\partial \theta} \Phi   \right)        \right] 
 +  \frac{1}{r^{2} \sin^{2}}  \frac{\partial^{2} \Phi}{\partial \phi^{2}} = 0,
\end{split}
\fe
where the determinant of the metric satisfies $\sqrt{-g} = r^{2}\sin\theta$. Owing to the system's spherical symmetry, the scalar field can be expanded in the form
\ie
\label{scalarfielddeco}
\Phi(t, r, \theta, \varphi) = \sum_{l=0}^{\infty} \sum_{m=-l}^{l}  Y_{lm}(\theta, \varphi) \frac{\Psi(t,r)}{r} ,
\fe
where \(Y_{lm}(\theta, \varphi)\) denotes the spherical harmonics. Under this decomposition, the radial component of Eq. (\ref{jekjsk}) can be reformulated as
\ie
\frac{\partial^{2}\Psi(t,r)}{\partial t^{2}}  + \frac{f(r)}{r} \left\{ \frac{\partial }{\partial r}  \left[ f(r) r^{2} \frac{\partial}{\partial r}  \left( \frac{\Psi(t,r)}{r} \right)   \right]     \right\} - f(r) \frac{l(l + 1)}{r^{2}}\Psi(t,r) = 0 .
\fe

\begin{figure}
    \centering
    \includegraphics[scale=0.6]{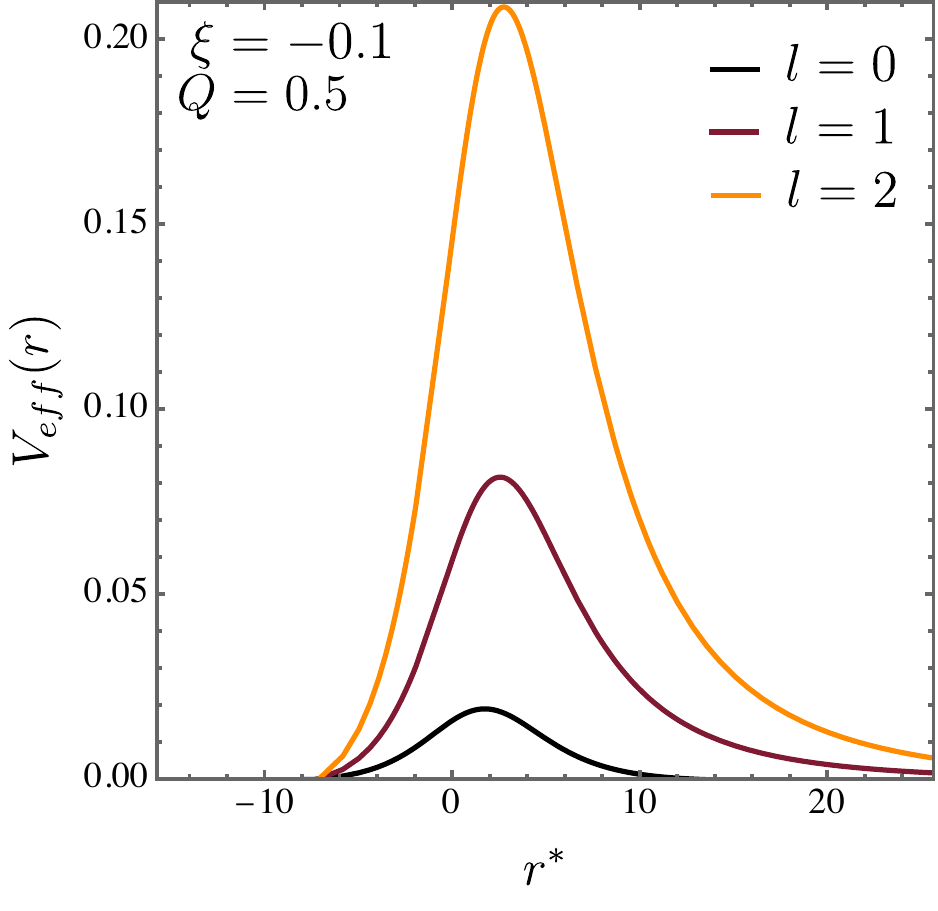}
    \caption{The effective potential $V_{eff}(r)$ for the scalar perturbations is depicted as a function of the tortoise coordinate $r^{*}$, specifically considering different values of $l$.}
    \label{veffscalarfiled}
\end{figure}

In this case, the spherical harmonics are represented by \( Y_{lm}(\theta, \varphi) \). By substituting the scalar field decomposition, as given in Eq. (\ref{scalarfielddeco}), into Eq. (\ref{Th4e3kleqin5Go55rdon}), the resulting equation takes on a Schrödinger--like form. This transformation introduces wave-like characteristics into the equation, rendering it particularly appropriate for the analysis we aim to perform. Thereby, we write
\ie
-\frac{\partial^{2} \Psi}{\partial t^{2}}+\frac{\partial^{2} \Psi}{\partial r^{*2}} + V_{eff}(r^{*})\Psi = 0.\label{schordingereq}
\fe
At this point, a remark is worth mentioning: the effective potential, \(V_{eff}\), commonly referred to as the \textit{Regge--Wheeler} potential, plays a significant role in encapsulating key details about the intrinsic remarks of the geometry of the black hole. To further simplify the analysis, we employ the tortoise coordinate \(r^*\), which smoothly extends across the entire spacetime, ranging from \(r^* \to \pm \infty\). This coordinate is defined through the relation \(\mathrm{d}r^* = \frac{1}{\sqrt{f(r)^2}} \mathrm{d}r\), leading to the following expression:
\ie
r^{*} = r + \frac{r_{2}^3 \ln (r-r_{2})}{(r_{2}-r_{3}) (r_{2}-r_{h})}-\frac{r_{3}^3 \ln (r-r_{3})}{(r_{2}-r_{3}) (r_{3}-r_{h})}+\frac{r_{h}^3 \ln (r-r_{h})}{(r_{2}-r_{h}) (r_{3}-r_{h})}.
\fe
Notice that, depending on the specific values of the parameters \(Q\), \(M\), and \(\xi\), the other horizons \(r_2\) and \(r_3\) might lose their physical significance. Following a series of algebraic manipulations, the effective potential takes the form:
\ie
\begin{split}
V_{eff}(r) = f(r) \left[ \frac{l (l+1)}{r^2}+\frac{\frac{2 M}{r^2}+3 \xi  Q^2-\frac{2 Q^2 \left(\xi  r^3+1\right)}{r^3}}{r} \right]
\end{split},
\fe
where $f(r) \equiv \left[  1 - \frac{2M}{r} + \frac{Q}{r^{2}}(1 + \xi r^{3})    \right]$. 
Fig. \ref{veffscalarfiled} illustrates the effective potential $V_{eff}$ as a function of the tortoise coordinate $r^{*}$ for different values of $l$.

\begin{table}[!h]
\begin{center}
\caption{\label{qnmtac0} The table displays the quasinormal modes, regarding scalar perturbations for $ l = 0$ as a function of the parameters \(\xi\) and \(Q\).}
\begin{tabular}{c| c | c | c} 
 \hline\hline\hline 
 \,\, $\xi$ \,\,\,\,\,\,  $Q$  & $\omega_{0}$ & $\omega_{1}$ & $\omega_{2}$  \\ [0.2ex] 
 \hline 
 -0.01, {0.99}  & {0.051981 - 0.203646$i$ }  &{0.029772 - 0.851015$i$} & {0.027769 - 2.117820$i$ }  \\
 
 -0.02,  {0.99}  & {0.319390 - 0.029622$i$} & {1.069930 - 0.020462$i$ }  &  {2.789960 - 0.016353$i$}  \\
 
 -0.03,  {0.99}  & {0.095925 - 0.089776$i$}  & {0.056693 - 0.370082$i$} &  {0.057294 - 0.795611$i$ } \\
 
 -0.04, {0.99}  & {0.033714 - 0.220596$i$} & {0.016423 - 0.962740$i$} & {0.007836 - 2.585400$i$} \\
 
 -0.05, {0.99}  &  {0.0149873 - 0.452733$i$ } & {0.009446 - 1.703900$i$} & {0.005592 - 4.410290$i$}  \\
   [0.2ex] 
 \hline\hline\hline 
 \,\, $\xi$ \,\,\,\,\,\,  $Q$  & $\omega_{0}$ & $\omega_{1}$ & $\omega_{2}$  \\ [0.2ex] 
 \hline 
 -0.01,  0.6  & 0.115325 - 0.101175$i$   & 0.0959357 - 0.339115$i$  & 0.175075 - 0.491897$i$  \\
 
 -0.01,  0.7  & 0.117284 - 0.101583$i$ & 0.0984806 - 0.339303$i$  &  0.164249 - 0.511014$i$  \\

 -0.01,  0.8  & 0.121881 - 0.100130$i$  & 0.105088 - 0.325194$i$ &  0.164139 - 0.480415$i$  \\
 
 -0.01, 0.9  & 0.120451 - 0.099259$i$ & 0.083942 - 0.360120$i$ & 0.074107 - 0.709462$i$ \\
 
 -0.01, {0.99}  & {0.051981 - 0.203646$i$ }  &{0.029772 - 0.851015$i$} & {0.027769 - 2.117820$i$ } \\
   [0.2ex] 
 \hline \hline \hline 
\end{tabular}
\end{center}
\end{table}

\begin{table}[!h]
\begin{center}
\caption{\label{qnmtac1} The table displays the quasinormal modes, regarding scalar perturbations for \( l = 1 \) as a function of the parameters \(\xi\) and \(Q\).}
\begin{tabular}{c| c | c | c} 
 \hline\hline\hline 
 \,\, $\xi$ \,\,\,\,\,\,  $Q$  & $\omega_{0}$ & $\omega_{1}$ & $\omega_{2}$  \\ [0.2ex] 
 \hline 
  -0.01, {0.99}  & {0.358583 - 0.087075$i$} & {0.332306 - 0.268461$i$} & {0.283522 - 0.475435$i$} \\
 
 -0.02,  {0.99}  & {0.342426 - 0.083211$i$ } & {0.320663 - 0.254602$i$}  &  {0.282074 - 0.439785$i$ }  \\
 
 -0.03,  {0.99}  & {0.326450 - 0.079076$i$}  & {0.313070 - 0.236969$i$} &  {0.297731 - 0.381071$i$}  \\
 
 -0.04, {0.99}  & {0.305713 - 0.075853$i$} & {0.266929 - 0.250455$i$} & {0.195061 - 0.527872$i$} \\
 
 -0.05, {0.99}  &  {0.290313 - 0.071041$i$} & {0.270115 - 0.220831$i$} & {0.230673 - 0.401278$i$}  \\
   [0.2ex] 
 \hline\hline\hline 
 \,\, $\xi$ \,\,\,\,\,\,  $Q$  & $\omega_{0}$ & $\omega_{1}$ & $\omega_{2}$  \\ [0.2ex] 
 \hline 
 -0.01,  0.6  & 0.308001 - 0.097311$i$ &  0.282811 - 0.303389$i$  & 0.252779 - 0.532660$i$   \\
 
 -0.01,  0.7  & 0.315174 - 0.096871$i$ & 0.291436 - 0.301245$i$  &  0.262809 - 0.527156$i$  \\

 -0.01,  0.8  & 0.325153 - 0.095905$i$  & 0.303291 - 0.297104$i$ &  0.276148 - 0.517348$i$  \\
 
 -0.01, 0.9  & 0.339620 - 0.093473$i$ & 0.319748 - 0.287540$i$ & 0.292397 - 0.496002$i$  \\
 
 -0.01, {0.99}  & {0.358583 - 0.087075$i$} & {0.332306 - 0.268461$i$} & {0.283522 - 0.475435$i$} \\
   [0.2ex] 
 \hline \hline \hline 
\end{tabular}
\end{center}
\end{table}

\begin{table}[!h]
\begin{center}
\caption{\label{qnmtac2} The table displays the quasinormal modes, regarding scalar perturbations for \( l = 2 \) as a function of the parameters \(\xi\) and \(Q\).}
\begin{tabular}{c| c | c | c} 
 \hline\hline\hline 
 \,\, $\xi$ \,\,\,\,\,\,  $Q$  & $\omega_{0}$ & $\omega_{1}$ & $\omega_{2}$  \\ [0.2ex] 
 \hline 
 -0.01, {0.99}  & {0.595716 - 0.086504$i$} & {0.580679 - 0.261580$i$} & {0.552021 - 0.442534$i$} \\
 
 -0.02,  {0.99}  & {0.569787 - 0.082622$i$} & {0.555847 - 0.249906$i$}  &  {0.528357 - 0.423752$i$} \\
 
 -0.03,  {0.99}  & {0.543321 - 0.078584$i$}  &{0.53118 - 0.237431$i$} &  {0.508297 - 0.400824$i$}  \\
 
 -0.04, {0.99}  & {0.516129 - 0.074397$i$} & {0.505359 - 0.224680$i$} & {0.485452 - 0.378710$i$} \\
 
 -0.05, {0.99}  &  {0.488109 - 0.070052$i$} & {0.478309 - 0.211606$i$} & {0.459838 - 0.357181$i$}  \\
   [0.2ex] 
 \hline\hline\hline 
 \,\, $\xi$ \,\,\,\,\,\,  $Q$  & $\omega_{0}$ & $\omega_{1}$ & $\omega_{2}$  \\ [0.2ex] 
 \hline 
 -0.01,  0.6  & 0.508924 - 0.0963663$i$   & 0.491378 - 0.293745$i$  & 0.461641 - 0.503458$i$   \\
 
 -0.01,  0.7  & 0.520895 - 0.0959584$i$ & 0.504371 - 0.292194$i$  &  0.476290 - 0.499878$i$  \\

 -0.01,  0.8  & 0.537543 - 0.0950556$i$  & 0.522358 - 0.289001$i$ &  0.496373 - 0.493093$i$  \\
 
 -0.01, 0.9  & 0.561793 - 0.0927849$i$  & 0.548120 - 0.281375$i$ & 0.524083 - 0.477982$i$ \\
 
 -0.01, {0.99}  & {0.595716 - 0.086504$i$} & {0.580679 - 0.261580$i$} & {0.552021 - 0.442534$i$} \\
   [0.2ex] 
 \hline \hline \hline 
\end{tabular}
\end{center}
\end{table}

The main goal at this stage is to obtain stationary solutions. To achieve this, we assume that the wave function \(\Psi(t,r)\) can be written in the form \(\Psi(t,r) = e^{-i\omega t} \psi(r)\), where \(\omega\) represents the associated frequency. This assumption facilitates the separation of the time-dependent part, enabling us to focus on the time--independent equation. In other words, we obtain
\ie
\frac{\partial^{2} \psi}{\partial r^{*2}} - \left[  \omega^{2} - V_{eff}(r^{*})\right]\psi = 0.\label{sserffs}
\fe
In order to effectively solve Eq. (\ref{sserffs}), it is essential to carefully account for the appropriate boundary conditions. For our particular setup, the solutions that meet these conditions are distinguished by their purely ingoing nature at the event horizon, ensuring physical consistency in the near--horizon region
\[
    \psi^{\text{in}}(r^{*}) \sim 
\begin{cases}
    \beta_{l}(\omega) e^{-i\omega r^{*}} & ( r^{*}\rightarrow - \infty)\\
    \alpha^{(-)}_{l}(\omega) e^{-i\omega r^{*}} + \alpha^{(+)}_{l}(\omega) e^{+i\omega r^{*}} & (r^{*}\rightarrow + \infty).
\end{cases}
\]
In our study, the complex constants \(\beta_l(\omega)\), \(\alpha^{(-)}_l(\omega)\), and \(\alpha^{(+)}_l(\omega)\) play a significant role. These constants are key to investigating the \textit{quasinormal} modes of a black hole, represented by frequencies \(\omega_{nl}\), which are determined by the condition \(\alpha^{(-)}_l(\omega_{nl}) = 0\). In this sense, they are characterized by their distinct behavior: they exhibit purely outgoing waves at spatial infinity and purely ingoing waves at the event horizon. The integers \(n\) and \(l\) denote the overtone and multipole numbers, respectively.

It is important to note that the spectrum of QNMs is derived from the eigenvalues of Eq. (\ref{sserffs}). To compute these frequencies, we utilize the WKB approximation, a semi--analytical method often used in quantum mechanics. This approach allows us to effectively study the behavior of quasinormal modes both near the event horizon and at large distances, providing crucial features about the black hole dynamics and the emission of gravitational waves.

The WKB method, originally introduced by Schutz and Will \cite{schutz1985black}, has become a widely used approach for calculating quasinormal modes associated with black hole perturbations. Over time, this method has been refined and extended, with significant contributions from Konoplya \cite{konoplya2003quasinormal, konoplya2004quasinormal}. This approach is particularly effective when the potential has a barrier-like structure, stabilizing to constant values as \(r^{*} \to \pm \infty\).

The calculation of quasinormal modes involves expanding the solution as a power series around the turning points where the potential reaches its maximum. By applying this method, the quasinormal mode frequencies can be obtained with a high degree of precision. The final expression for these frequencies, as derived by Konoplya, is given by:
\ie
\frac{i(\omega^{2}_{n}-V_{0})}{\sqrt{-2 V^{''}_{0}}} - \sum^{6}_{j=2} \Lambda_{j} = n + \frac{1}{2}.
\fe
Here, a critical term in this formulation is \(V''_0\), which represents the second derivative of the potential at its maximum point, \(r_0\). Additionally, constants \(\Lambda_j\) are functions of both the effective potential and its derivatives at this point, contributing to the precise determination of quasinormal mode frequencies.

If $Q \to 0$, the effective potential for scalar perturbations reduces to the Schwarzschild case. With these preliminary considerations in place, the quasinormal frequencies associated with scalar perturbations can now be analyzed. Fig. \ref{veffvector} illustrates the behavior of the effective potential $V_{eff}(r)$ as a function of the tortoise coordinate $r^{*}$. The potential exhibits a sine--shaped profile, which justifies the application of the WKB method for computing quasinormal modes.  

Tabs. \ref{qnmtac0}, \ref{qnmtac1}, and \ref{qnmtac2} present the damped frequencies for different values of $Q$ and $\xi$, corresponding to $l = 0$, $l = 1$, and $l = 2$, respectively. In general, for a fixed charge of $Q = 0.99$, decreasing $\xi$ results in less damped modes, as seen in $\omega_{0}$, $\omega_{1}$, and $\omega_{2}$. A similar trend is observed when increasing $Q$ for a fixed $\xi = -0.01$, leading to less damped oscillations in the same frequency modes.


\subsection{Vector perturbations}

Electromagnetic perturbations are analyzed using the tetrad formalism, following the approach detailed in \cite{chandrasekhar1998mathematical, Bouhmadi-Lopez:2020oia, Gogoi:2023kjt}. Within this framework, a tetrad basis $e_\mu^{a}$ is introduced in accordance with the black hole metric $g_{\mu\nu}$, satisfying the conditions:
\ie
\begin{split}
& e^{a}_\mu e^\mu_{b} = \delta^{a}_{b}, \, \, \, \,
e^{a}_\mu e^\nu_{a} = \delta^{\nu}_{\mu}, \, \, \, \,\\
& e^{a}_\mu = g_{\mu\nu} \eta^{a b} e^\nu_{b}, \, \, \, \,
g_{\mu\nu} = \eta_{a b}e^{a}_\mu e^{b}_\nu = e_{a\mu} e^{a}_\nu.
\end{split}
\fe

In the context of electromagnetic perturbations examined via the tetrad formalism, the field strength tensor satisfies the Bianchi identity, expressed as $\mathcal{F}_{[ab|c]} = 0$. This condition leads to the following relation:
\begin{align}
\left( r \sqrt{g_{tt}(r)}\, \mathcal{F}_{t \phi}\right)_{,r} + r \sqrt{g_{rr}(r)}\,
\mathcal{F}_{\phi r, t} &=0,  \label{edem1} \\
\left( r \sqrt{g_{tt}(r)}\, \mathcal{F}_{ t \phi}\sin\theta\right)_{,\theta} + r^2
\sin\theta\, \mathcal{F}_{\phi r, t} &=0.  \label{edem2}
\end{align}

Consequently, the conservation equation can be expressed as
\ie
\eta^{b c}\! \left( \mathcal{F}_{a b} \right)_{|c} =0.
\fe

Notice that, within spherical polar coordinates, this equation can be rewritten as:
\ie  \label{edem3}
\left( r \sqrt{g_{tt}(r)}\, \mathcal{F}_{\phi r}\right)_{,r} + \sqrt{g_{tt}(r) g_{rr}(r)}%
\, \mathcal{F}_{\phi \theta,\theta} + r \sqrt{g_{rr}(r)}\, \mathcal{F}_{t \phi, t} = 0.
\fe

In this formulation, the vertical bar and comma indicate intrinsic and directional derivatives corresponding to the tetrad indices. Utilizing Eqs. \eqref{edem1} and \eqref{edem2}, together with the time differentiation of Eq. \eqref{edem3}, the following expression emerges:
\ie  \label{edem4}
\begin{split}
& \left[ \sqrt{g_{tt}(r) g_{rr}(r)^{-1}} \left( r \sqrt{g_{tt}(r)}\, \mathcal{F}
\right)_{,r} \right]_{,r} \\
& + \dfrac{g_{tt}(r) \sqrt{g_{rr}(r)}}{r} \left( \dfrac{%
\mathcal{F}_{,\theta}}{\sin\theta} \right)_{,\theta}\!\! \sin\theta - r 
\sqrt{g_{rr}(r)}\, \mathcal{F}_{,tt} = 0.
\end{split}
\fe
Define $F = \mathcal{F}_{t \phi } \sin\theta$. Applying Fourier decomposition $(\partial_t \rightarrow -i \omega)$ and introducing a redefinition of the field as $F(r,\theta) = F(r) Y_{,\theta} / \sin\theta$, where $Y(\theta)$ represents the Gegenbauer function \cite{g1,g2,g3,g5,g6}, Eq. \eqref{edem4} can be rewritten in the following form:
\ie  
\begin{split}
\label{edem5}
& \left[ \sqrt{g_{tt}(r) g_{rr}(r)^{-1}} \left( r \sqrt{g_{tt}(r)}\, F
\right)_{,r} \right]_{,r} \\
& + \omega^2 r \sqrt{g_{rr}(r)}\, F -
g_{tt}(r) \sqrt{g_{rr}(r)} r^{-1} l (l + 1)\, F = 0.
\end{split}
\fe

Introducing the redefinition \(\psi^{\, \text{v}} \equiv r \sqrt{g_{tt}(r)} \, F\), Eq. \eqref{edem5} is reformulated into a Schrödinger--like equation, expressed as
\ie
\partial^2_{r_*} \psi^{\, \text{v}} + \omega^2 \psi^{\, \text{v}} = V^{\, \text{v}}_{eff}(r) \psi^{\, \text{v}},
\fe
such that the effective potential corresponding to the vectorial perturbation is given by
\ie  
V_{eff}^{\, \text{v}}(r) = g_{tt}(r) \, \dfrac{l ( l + 1 )}{r^2}.
\fe

\begin{figure}
    \centering
    \includegraphics[scale=0.6]{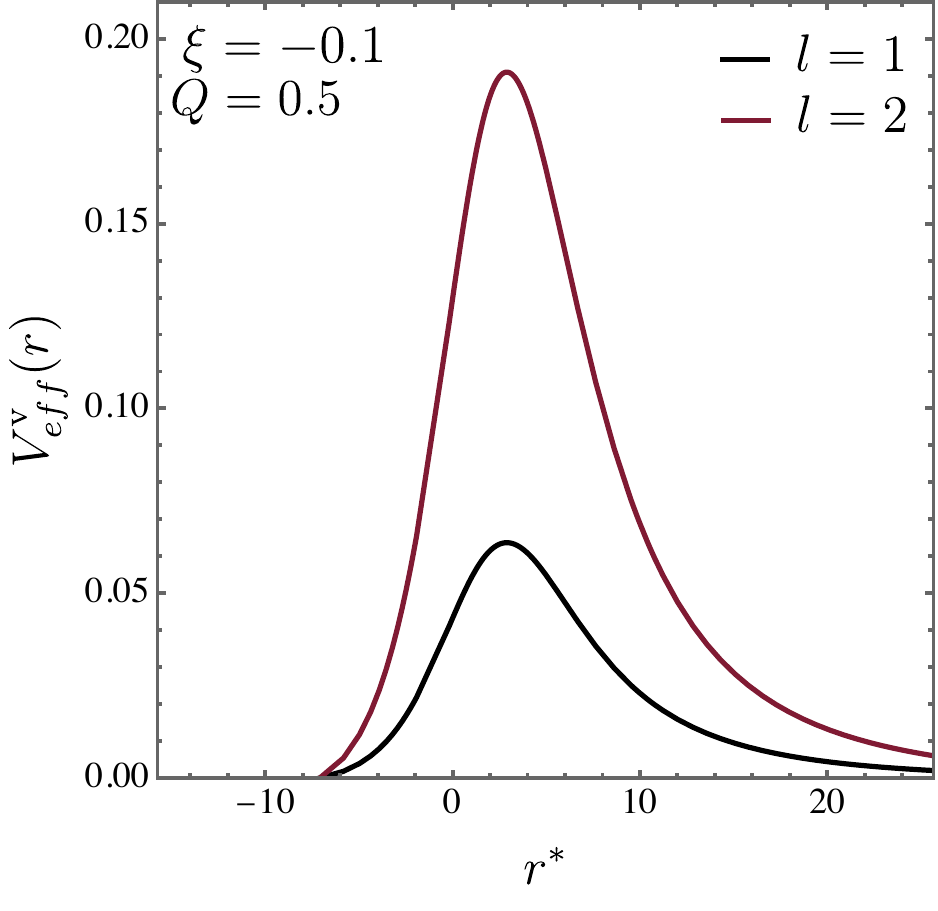}
    \caption{The effective potential for vector perturbations $V_{eff}^{\text{v}}(r)$ is depicted as a function of the tortoise coordinate $r^{*}$, specifically considering different values of $l$.}
    \label{veffvector}
\end{figure}

When $Q \to 0$, the effective potential governing vector perturbations reduces to the Schwarzschild limit. Likewise, setting $\xi \to 0$ retrieves the Reissner--Nordström case, as expected. The behavior of the effective potential $V_{eff}^{\text{v}}(r)$ as a function of the tortoise coordinate $r^{*}$ is depicted in Fig. \ref{veffvector}.  

With these preliminary considerations, the analysis of quasinormal frequencies for vector perturbations can proceed. As previously observed for other type of perturbation discussed previously in this work (the scalar ones), the effective potential follows a sine--like profile, making it suitable for applying the WKB method to determine quasinormal modes.  

The damped frequencies for various values of $Q$ and $\xi$ are listed in Tables \ref{qnmtac2vectorial1} and \ref{2qnmtac1vectorial2}, corresponding to $l=1$ and $l=2$, respectively. For a fixed charge of $Q = 0.99$, decreasing $\xi$ leads to weaker damping in the modes, as indicated by $\omega_{0}$. Similarly, when $Q$ increases while keeping $\xi = -0.01$ constant, the damping behavior follows the same trend observed for scalar perturbations.

\begin{table}[!h]
\begin{center}
\caption{\label{qnmtac2vectorial1} The table displays the quasinormal modes, regarding vector perturbations for $l = 1$ as a function of the parameters \(\xi\) and \(Q\).}
\begin{tabular}{c| c | c | c} 
 \hline\hline\hline 
 \,\, $\xi$ \,\,\,\,\,\,  $Q$  & $\omega_{0}$ & $\omega_{1}$ & $\omega_{2}$  \\ [0.2ex] 
 \hline 
  -0.01, 0.99  & 0.315276 - 0.083936$i$ & 0.259646 - 0.287513$i$ & 0.171670 - 0.635994$i$  \\
 
 -0.02,  0.99  & 0.302001 - 0.080398$i$ & 0.247676 - 0.277702$i$  &  0.163377 - 0.621900$i$  \\
 
 -0.03,  0.99  & 0.295008 - 0.074954$i$  & 0.290633 - 0.216352$i$ & 0.312683 - 0.299824$i$  \\
 
 -0.04, 0.99  & 0.270640 - 0.073817$i$ & 0.197753 - 0.288720$i$  & 0.119394 - 0.723054$i$  \\
 
 -0.05, 0.99  &  0.262538 - 0.068046$i$ & 0.230067 - 0.222709$i$ & 0.171212 - 0.456595$i$  \\
   [0.2ex] 
 \hline\hline\hline 
 \,\, $\xi$ \,\,\,\,\,\,  $Q$  & $\omega_{0}$ & $\omega_{1}$ & $\omega_{2}$  \\ [0.2ex] 
 \hline 
 -0.01,  0.6  & 0.264162 - 0.092755$i$ & 0.234108 - 0.292490$i$   & 0.198571 - 0.521580$i$  \\
 
 -0.01,  0.7  & 0.271746 - 0.092555$i$ & 0.243544 - 0.290865$i$  &  0.210186 - 0.515998$i$  \\

 -0.01,  0.8  & 0.282297 - 0.091902$i$  & 0.256332 - 0.287694$i$ &  0.225031 - 0.508012$i$  \\
 
 -0.01, 0.9  & 0.298010 - 0.089736$i$ & 0.275986 - 0.277551$i$ & 0.248117 - 0.479898$i$  \\
 
 -0.01, 0.99  & 0.315276 - 0.083936$i$ & 0.259646 - 0.287513$i$ & 0.171670 - 0.635994$i$ \\
   [0.2ex] 
 \hline \hline \hline 
\end{tabular}
\end{center}
\end{table}

\begin{table}[!h]
\begin{center}
\caption{\label{2qnmtac1vectorial2} The table displays the quasinormal modes, regarding vector perturbations for $l = 2$ as a function of the parameters \(\xi\) and \(Q\).}
\begin{tabular}{c| c | c | c} 
 \hline\hline\hline 
 \,\, $\xi$ \,\,\,\,\,\,  $Q$  & $\omega_{0}$ & $\omega_{1}$ & $\omega_{2}$  \\ [0.2ex] 
 \hline 
 -0.01, 0.99  & 0.572665 - 0.084905$i$ & 0.560286 - 0.255427$i$ & 0.544423 - 0.421465$i$   \\
 
 -0.02,  0.99  & 0.548513 - 0.081141$i$ &0.535028 - 0.245128$i$  &  0.510547 - 0.412853$i$  \\
 
 -0.03, 0.99  & 0.524024 - 0.077181$i$  & 0.512332 - 0.232894$i$ & 0.492486 - 0.390356$i$   \\
 
 -0.04, 0.99  & 0.498440 - 0.073121$i$ & 0.484533 - 0.222167$i$  & 0.452025 - 0.384656$i$ \\
 
 -0.05, 0.99  &  0.472908 - 0.068772$i$  & 0.465601 - 0.206545$i$ & 0.459469 - 0.338932$i$  \\
   [0.2ex] 
 \hline\hline\hline 
 \,\, $\xi$ \,\,\,\,\,\,  $Q$  & $\omega_{0}$ & $\omega_{1}$ & $\omega_{2}$  \\ [0.2ex] 
 \hline 
 -0.01,  0.6  & 0.483370 - 0.094753$i$ & 0.464681 - 0.289247$i$  & 0.433021 - 0.497037$i$  \\
 
 -0.01,  0.7  & 0.495554 - 0.094408$i$ & 0.477965 - 0.287866$i$  &  0.448120 - 0.493658$i$  \\

 -0.01,  0.8  & 0.512510 - 0.093584$i$  & 0.496381 - 0.284884$i$ &  0.468836 - 0.487156$i$  \\
 
 -0.01, 0.9  & 0.537300 - 0.091386$i$ & 0.522871 - 0.277447$i$  & 0.497483 - 0.472367$i$  \\
 
 -0.01, 0.99  & 0.572665 - 0.084905$i$ & 0.560286 - 0.255427$i$ & 0.544423 - 0.421465$i$ \\
   [0.2ex] 
 \hline \hline \hline 
\end{tabular}
\end{center}
\end{table}


\subsection{Tensor perturbations}

Additionally, without relying on a specific fundamental theory, the master equations were obtained under the assumption that both the Klein-Gordon and Maxwell equations hold. However, it is important to emphasize that the conservation laws governing these test fields may not necessarily be preserved in certain gravitational models unless the matter sector remains minimally coupled to the metric $g_{\mu\nu}$. 

To analyze axial gravitational perturbations (odd--perturbations), both the gravitational field equations and the stress--energy tensor must be perturbed. However, in the absence of a well--defined underlying theory, an alternative approach is required. In this work, the solution is considered within the framework of the Einstein equation, incorporating an effective stress-energy tensor. Notably, a similar methodology has been applied in different contexts throughout the literature \cite{nicolini2006noncommutative, ashtekar2018quantum, ashtekar2018quantum2, baruah2025quasinormal}. From a phenomenological perspective, the effective stress--energy tensor associated with the black hole solution can be modeled as an anisotropic fluid:
\ie
T_{\mu\nu}=\left(\rho+p_2\right)u_\mu u_\nu+\left(p_1-p_2\right)x_\mu x_\nu+p_2 g_{\mu\nu}.\label{anisopf}
\fe

In this formulation, $\rho$ represents the energy density as observed in the comoving frame of the fluid. The quantities $u^\mu$ and $x^\mu$ correspond to the timelike four--velocity and a spacelike unit vector perpendicular to both $u^\mu$ and the angular directions, respectively. Within Eq. \eqref{anisopf}, the terms $p_1$ and $p_2$ denote the radial and tangential pressures. Additionally, the vectors $u^\mu$ and $x^\mu$ satisfy the following conditions
\ie
u_\mu u^\mu=-1\,,\qquad x_\mu x^\mu=1\,.\label{fourvelocity}
\fe
In this framework, the metric \( g_{\mu\nu} \) is responsible for raising and lowering indices. Within the comoving frame, the four--velocity and the spacelike unit vector can be expressed as \( u^\mu = (u^t, 0, 0, 0) \) and \( x^\mu = (0, x^r, 0, 0) \), respectively. By considering Eq. (\ref{fourvelocity}), the following relation is obtained:
\ie
u_t^2 = g_{tt}(r )u_tu^t = -g_{tt}(r)\,,\qquad x_r^2 = g_{rr}(r)x_rx^r = g_{rr}(r)\,.
\fe

Notice that, taking into account the background level, the components of the stress--energy tensor are given by:
\begin{align}
T_{tt}&= - g_{tt}(r)\rho\,,\qquad T_t^t=-\rho\,,\\
T_{rr}& = g_{rr}(r)p_1\,,\qquad T_r^r=p_1\,,\\
&T_{\theta}^{\theta}=T_\varphi^\varphi=p_2\,.
\end{align}
The quantities $\rho$, $p_1$, and $p_2$ depend on the radial coordinate \( r \) and can be explicitly determined by computing the components of the Einstein tensor associated with the spacetime.

To investigate the quasinormal modes of a static and spherically symmetric black hole, we introduce a perturbation in the spacetime, modifying it into a non--stationary, axisymmetric configuration. The perturbed metric takes the following form \cite{chen2019gravitational}:
\begin{align}
\mathrm{d}s^2=&-e^{2\nu}\left(\mathrm{d}x^0\right)^2+e^{2\psi}\left(\mathrm{d}x^1-\sigma \mathrm{d}x^0-q_2\mathrm{d}x^2-q_3\mathrm{d}x^3\right)^2\nonumber\\&+e^{2\mu_2}\left(\mathrm{d}x^2\right)^2+e^{2\mu_3}\left(\mathrm{d}x^3\right)^2\,.\label{metricg}
\end{align}
The functions \(\nu\), \(\psi\), \(\mu_2\), \(\mu_3\), \(\sigma\), \(q_2\), and \(q_3\) depend on the time coordinate \(t\) (\(t = x^0\)), the radial coordinate \(r\) (\(r = x^2\)), and the polar angle \(\theta\) (\(\theta = x^3\)). Due to the axisymmetric nature of the system, these metric functions are independent of the azimuthal coordinate \(\varphi\) (\(\varphi = x^1\)). The notation adopted here follows that of Ref.~\cite{chen2019gravitational}. It is important to note that in a static, spherically symmetric background, the functions \(q_2\), \(q_3\), and \(\sigma\) vanish. Consequently, when linearizing the field equations, these quantities must be treated as first--order perturbations.

To advance the analysis, the tetrad formalism is employed, introducing a tetrad basis associated with the metric \eqref{metricg}. For a detailed overview of this approach, Ref.~\cite{chen2019gravitational} provides a comprehensive introduction:
\begin{align}
e^{\mu}_{0}&=\left(e^{-\nu},\sigma e^{-\nu},0,0\right)\,,\nonumber\\
e^{\mu}_{1}&=\left(0, e^{-\psi}, 0,0\right)\,,\nonumber\\
e^{\mu}_{2}&=\left(0, q_2e^{-\mu_2}, e^{-\mu_2},0\right)\,,\nonumber\\
e^{\mu}_{3}&=\left(0, q_3e^{-\mu_3}, 0, e^{-\mu_3}\right)\,.\label{tetradbasis111}
\end{align}

It is important to notice that, in this formalism, tetrad indices are enclosed in parentheses to differentiate them from standard tensor indices. The tetrad approach projects all relevant quantities, originally defined in the coordinate basis of \( g_{\mu\nu} \), onto a chosen frame associated with \( \eta_{ab} \) using the corresponding tetrad basis. Typically, \( \eta_{ab} \) is taken as the Minkowski metric for convenience. With this framework, any vector or tensor field can be expressed within the tetrad frame, where it is represented in terms of its tetrad components
\begin{align}
A_{\mu}&=e_{\mu}^{a}A_{a}\,,\quad A_{a}=e_{a}^{\mu}A_{\mu}\,,\nonumber\\
B_{\mu\nu}&=e_{\mu}^{a}e_{\nu}^{b}B_{ab}\,,\quad B_{ab}=e_{a}^{\mu}e_{b}^{\nu}B_{\mu\nu}\,.
\end{align}

Within the tetrad framework, the expression for the perturbed stress--energy tensor of an anisotropic fluid takes the form
\begin{align}
\delta T_{ab}=&\,(\rho+p_2)\delta(u_{a}u_{b})+(\delta\rho+\delta p_2)u_{a}u_{b}\nonumber\\
&+(p_1-p_2)\delta(x_{a}x_{b})+(\delta p_1-\delta p_2)x_{a}x_{b}\nonumber\\&+\delta p_2\eta_{ab}.
\end{align}
Applying the conditions on \( u^\mu \) and \( x^\mu \), namely Eq.~\eqref{fourvelocity} along with the orthogonality relation \( u^\mu x_\mu = 0 \), it is found that the axial components of the perturbed stress--energy tensor in the tetrad frame are identically zero:
\begin{align}
\delta T_{10}&=\delta T_{12}=\delta T_{13}=0\,.
\end{align}

Within the tetrad framework, the Einstein equation can be reformulated as:
\ie
R_{ab}-\frac{1}{2}\eta_{ab}R=8\pi T_{ab}\,.\label{eineq}
\fe
Given that the axial components of the perturbed stress-energy tensor are zero, the master equation governing axial perturbations follows directly from the condition \( R_{ab}|_{\text{axial}} = 0 \). By performing the necessary algebraic transformations to obtain this equation (refer to the Appendix of Ref. \cite{chen2019gravitational} for details), the resulting gravitational effective potential is expressed as \cite{baruah2025quasinormal}
\ie
V_{eff}^{\text{t}}(r) = g_{tt}(r) \left[ \dfrac{2}{r^2} \left( \dfrac{1}{g_{rr}(r)} - 1 \right) + \dfrac{l(l+1)}{r^2} - \dfrac{1}{r \sqrt{g_{tt}(r) g_{rr}(r)}} \left( \dfrac{\mathrm{d}}{\mathrm{d}r} \sqrt{g_{tt}(r) g_{rr}^{-1}(r)} \right) \right],
\fe
or, more explicitly
\ie
V_{eff}^{\text{t}}(r) = g_{tt}(r) \left( \frac{l (l+1)}{r^2}+\frac{Q^2 \left(\xi  r^3+4\right)-6 M r}{r^4} \right).
\fe

If \( Q \to 0 \), the effective potential for odd-parity tensor perturbations reduces to the Schwarzschild case. Likewise, setting \( \xi \to 0 \) restores the axial perturbation behavior corresponding to the Reissner–Nordström solution. With these preliminaries established, the analysis of quasinormal frequencies for tensorial perturbations can proceed.  

\begin{figure}
    \centering
    \includegraphics[scale=0.6]{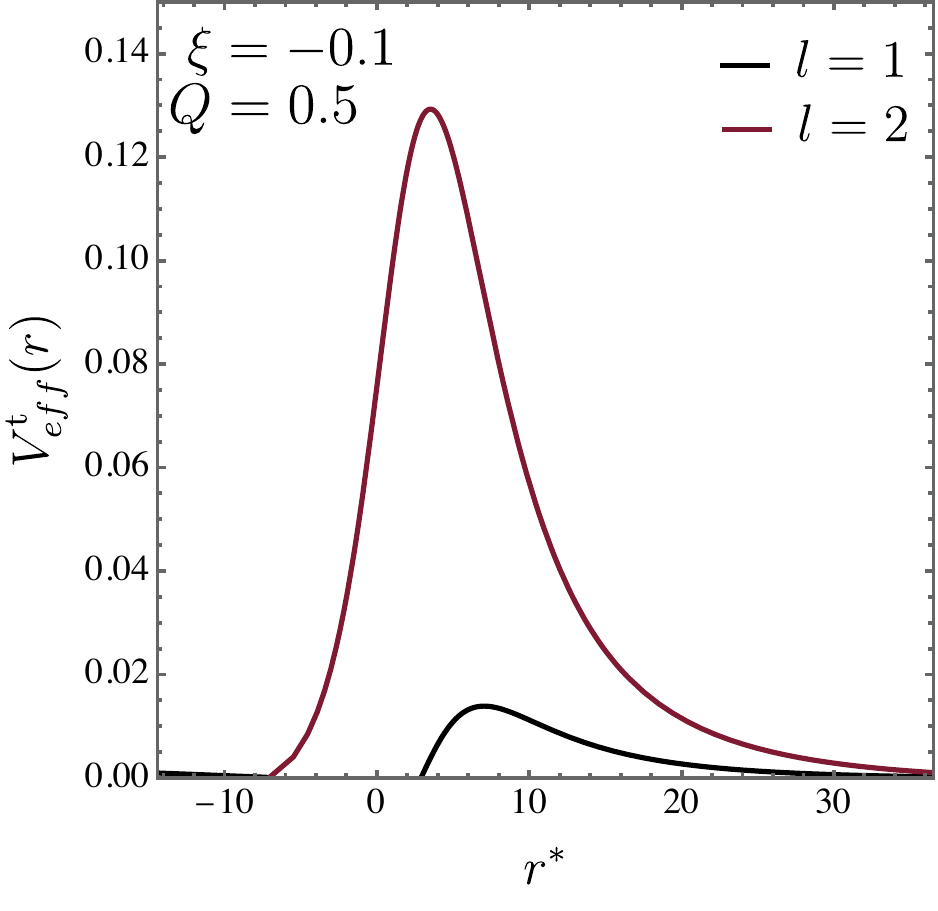}
    \caption{The effective potential for tensor perturbations $V_{eff}^{\text{t}}(r)$ is depicted as a function of the tortoise coordinate $r^{*}$, specifically considering different values of $l$.}
    \label{vefftensor}
\end{figure}

Fig. \ref{vefftensor} illustrates the effective potential \( V_{eff}^{\text{t}}(r) \) as a function of the tortoise coordinate \( r^{*} \). As observed in previous perturbation analyses, this potential exhibits a sine-like profile, making the WKB method a suitable approach for computing quasinormal modes.  

Tabs. \ref{qnmtac1tensor} and \ref{qnmtac2tensor} present the damped frequencies for various values of \( Q \) and \( \xi \) for \( l=1 \) and \( l=2 \), respectively. In general, for a fixed charge of \( Q = 0.99 \), decreasing \( \xi \) results in weaker damping of the modes, as reflected in \( \omega_{0} \) and \( \omega_{1} \). Similarly, increasing \( Q \) while keeping \( \xi = -0.01 \) fixed leads to a comparable reduction in damping for these oscillations.

\begin{table}[!h]
\begin{center}
\caption{\label{qnmtac1tensor} The table displays the quasinormal modes, regarding tensor perturbations for $l = 1$ as a function of the parameters \(\xi\) and \(Q\).}
\begin{tabular}{c| c | c | c} 
 \hline\hline\hline 
 \,\, $\xi$ \,\,\,\,\,\,  $Q$  & $\omega_{0}$ & $\omega_{1}$ & $\omega_{2}$  \\ [0.2ex] 
 \hline 
 -0.01,  0.99  & 0.465716 - 0.080185$i$  & 0.440404 - 0.247612$i$ & 0.380003 - 0.452700$i$  \\
 
 -0.02,  0.99  & 0.448975 - 0.076333$i$ & 0.452668 - 0.221361$i$  &  0.510702 - 0.310252$i$  \\
 
 -0.03,  0.99  & 0.427207 - 0.073113$i$  & 0.415918 - 0.219942$i$ & 0.402720 - 0.360627$i$  \\
 
 -0.04, 0.99  & 0.406695 - 0.069415$i$ & 0.398220 - 0.207956$i$ & 0.393776 - 0.335087$i$ \\
 
 -0.05, 0.99  &  0.383423 - 0.065898$i$  & 0.357090 - 0.207796$i$ & 0.285816 - 0.414563$i$  \\
   [0.2ex] 
 \hline\hline\hline 
 \,\, $\xi$ \,\,\,\,\,\,  $Q$  & $\omega_{0}$ & $\omega_{1}$ & $\omega_{2}$  \\ [0.2ex] 
 \hline 
 -0.01,  0.6  & 0.394552 - 0.088827$i$ & 0.370803 - 0.272583$i$   &  0.329887 - 0.473206$i$  \\
 
 -0.01,  0.7  & 0.404439 - 0.088584$i$ & 0.382252 - 0.271544$i$  &  0.344285 - 0.470296$i$  \\

 -0.01,  0.8  & 0.418189 - 0.087914$i$  & 0.398032 - 0.269081$i$ &  0.363758 - 0.464591$i$  \\
 
 -0.01, 0.9  & 0.43826 - 0.085981$i$ & 0.420501 - 0.262457$i$ & 0.389961 - 0.451040$i$  \\
 
-0.01,  0.99  & 0.465716 - 0.080185$i$  & 0.440404 - 0.247612$i$ & 0.380003 - 0.452700$i$ \\
   [0.2ex] 
 \hline \hline \hline 
\end{tabular}
\end{center}
\end{table}

\begin{table}[!h]
\begin{center}
\caption{\label{qnmtac2tensor} The table displays the quasinormal modes, regarding tensor perturbations for $l = 2$ as a function of the parameters \(\xi\) and \(Q\).}
\begin{tabular}{c| c | c | c} 
 \hline\hline\hline 
 \,\, $\xi$ \,\,\,\,\,\,  $Q$  & $\omega_{0}$ & $\omega_{1}$ & $\omega_{2}$  \\ [0.2ex] 
 \hline 
 -0.01,  0.99  & 0.465716 - 0.080185$i$  & 0.440404 - 0.247612$i$ & 0.380003 - 0.452700$i$   \\
 
 -0.02,  0.99  & 0.448975 - 0.076333$i$  & 0.452668 - 0.221361$i$  &  0.510702 - 0.310252$i$  \\
 
 -0.03,  0.99  & 0.427207 - 0.073113$i$  & 0.415918 - 0.219942$i$ & 0.402720 - 0.360627$i$  \\
 
 -0.04, 0.99  & 0.406695 - 0.069415$i$ & 0.39822 - 0.207956$i$ & 0.393776 - 0.335087$i$ \\
 
 -0.05, 0.99  &  0.383423 - 0.065898$i$ & 0.35709 - 0.207796$i$ & 0.285816 - 0.414560$i$  \\
   [0.2ex] 
 \hline\hline\hline 
 \,\, $\xi$ \,\,\,\,\,\,  $Q$  & $\omega_{0}$ & $\omega_{1}$ & $\omega_{2}$  \\ [0.2ex] 
 \hline 
 -0.01,  0.6  & 0.394552 - 0.088827$i$ & 0.370803 - 0.272583$i$  & 0.329887 - 0.473206$i$  \\
 
 -0.01,  0.7  & 0.404439 - 0.088584$i$ & 0.382252 - 0.271544$i$  &  0.344285 - 0.470296$i$  \\

 -0.01,  0.8  & 0.418189 - 0.087914$i$  & 0.398032 - 0.269081$i$ &  0.363758 - 0.464591$i$  \\
 
 -0.01, 0.9  & 0.438260 - 0.085981$i$ & 0.420501 - 0.262457$i$ & 0.389961 - 0.451040$i$  \\
 
 -0.01,  0.99  & 0.465716 - 0.080185$i$  & 0.440404 - 0.247612$i$ & 0.380003 - 0.452700$i$ \\
   [0.2ex] 
 \hline \hline \hline 
\end{tabular}
\end{center}
\end{table}


\section{Time--Domain solution}

Investigating scalar, vector, and tensor perturbations in the time domain is essential for assessing the influence of the quasinormal spectrum on time-dependent scattering phenomena. Due to the intricate nature of the effective potential, an accurate numerical technique is required to analyze its behavior. To achieve this, the characteristic integration scheme introduced by Gundlach et al. \cite{Gundlach:1993tp} is employed, offering a robust framework for exploring quasinormal modes in dynamical scattering and their relevance to black hole physics.

The method described in Refs. \cite{Baruah:2023rhd, Bolokhov:2024ixe, Guo:2023nkd, Yang:2024rms, Gundlach:1993tp, Skvortsova:2024wly, Shao:2023qlt} relies on the introduction of light-cone coordinates, given by \( \Tilde{u} = t - r^{*} \) and \( \Tilde{v} = t + r^{*} \). Expressing the system in terms of these coordinates simplifies the wave equation, making the analysis more efficient. Under this framework, the wave equation can be rewritten as
\ie
\left(4 \frac{\partial^{2}}{\partial \Tilde{u} \partial \Tilde{v}} + V(\Tilde{u},\Tilde{v})\right) \Tilde{\psi} (\Tilde{u},\Tilde{v}) = 0.
\fe

A reliable method for numerically solving the equation involves discretizing the system through a combination of the finite-difference approach and supplementary computational techniques, ensuring greater accuracy and stability
\ie
\Tilde{\psi}(\Tilde{N}) = -\Tilde{\psi}(\Tilde{S}) + \Tilde{\psi}(\Tilde{W}) + \Tilde{\psi}(\Tilde{E}) - \frac{\Tilde{h}^{2}}{8}\Tilde{V}(\Tilde{S})[\Tilde{\psi}(\Tilde{W}) + \Tilde{\psi}(\Tilde{E})] + \mathcal{O}(\Tilde{h}^{4}).
\fe

To define the coordinate points, we introduce the following notation: \( \Tilde{S} = (\Tilde{u}, \Tilde{v}) \), \( \Tilde{W} = (\Tilde{u} + \Tilde{h}, \Tilde{v}) \), \( \Tilde{E} = (\Tilde{u}, \Tilde{v} + \Tilde{h}) \), and \( \Tilde{N} = (\Tilde{u} + \Tilde{h}, \Tilde{v} + \Tilde{h}) \), where \( \Tilde{h} \) denotes the grid spacing parameter. The null surfaces \( \Tilde{u} = \Tilde{u}_{0} \) and \( \Tilde{v} = \Tilde{v}_{0} \) serve as key reference points for initializing the computational domain. In this analysis, the initial conditions along the null surface \( \Tilde{u} = \Tilde{u}_{0} \) are characterized by a Gaussian distribution centered at \( \Tilde{v} = \Tilde{v}_{c} \) with a specified width parameter \( \sigma \)
\ie
\Tilde{\psi}(\Tilde{u} = \Tilde{u}_{0},\Tilde{v}) = A e^{-(\Tilde{v}-\Tilde{v}_{0})^{2}}/2\sigma^{2}, \,\,\,\,\,\, \Tilde{\psi}(\Tilde{u},\Tilde{v}_{0}) = \Tilde{\psi}_{0}.
\fe
To accomplish the calculations, the initial setup is employed by \( \Tilde{\psi}(\Tilde{u}, \Tilde{v}_{0}) = \Tilde{\psi}_{0} \) at \( \Tilde{v} = \Tilde{v}_{0} \), where \( \Tilde{\psi}_{0} \) is taken as zero for convenience, without any impact on generality. The integration follows a stepwise progression along surfaces of constant \( \Tilde{u} \), with \( \Tilde{v} \) increasing according to the prescribed null data. For computational efficiency, a scalar test field is introduced with \( M = 1 \). The initial profile adopts a Gaussian shape centered at \( \Tilde{v} = 0 \), featuring a width of \( \sigma = 1 \) and an initial amplitude set to zero. The numerical grid spans the range \( \Tilde{u} \in [0, 1000] \) and \( \Tilde{v} \in [0, 1000] \), with a spacing parameter \( \Tilde{h} = 0.1 \).


\subsection{Scalar perturbations}

\begin{figure}
    \centering
    \includegraphics[scale=0.5]{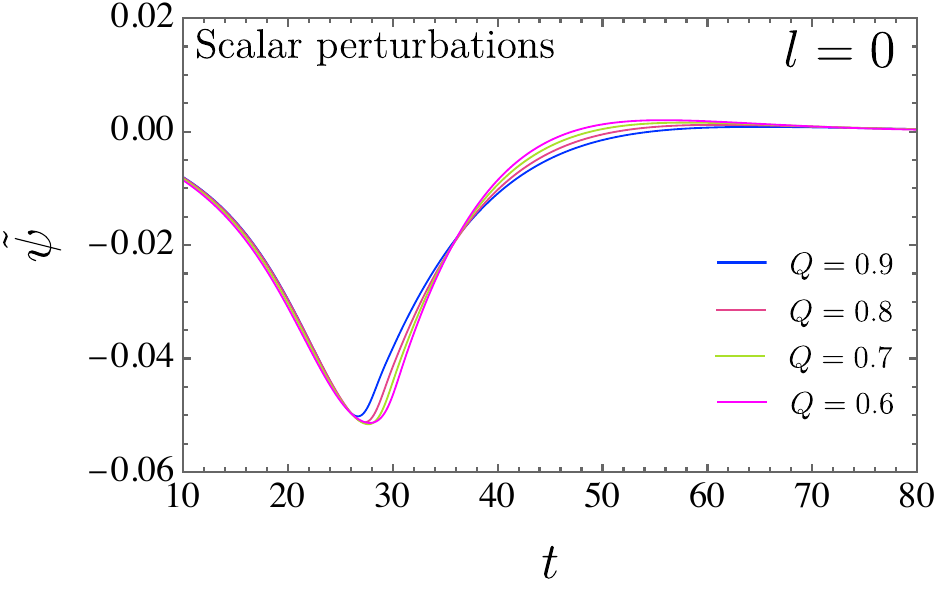}
    \includegraphics[scale=0.5]{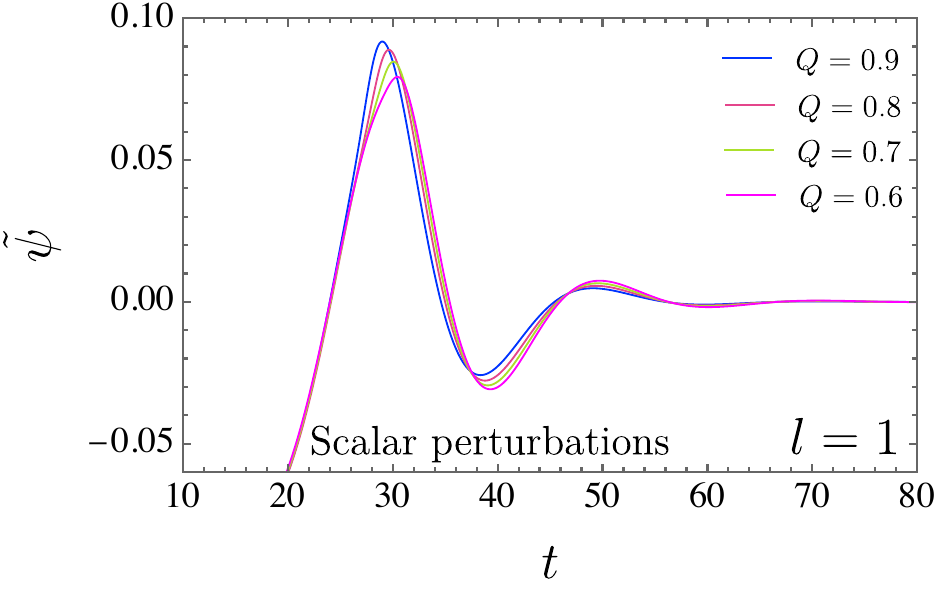}
    \includegraphics[scale=0.5]{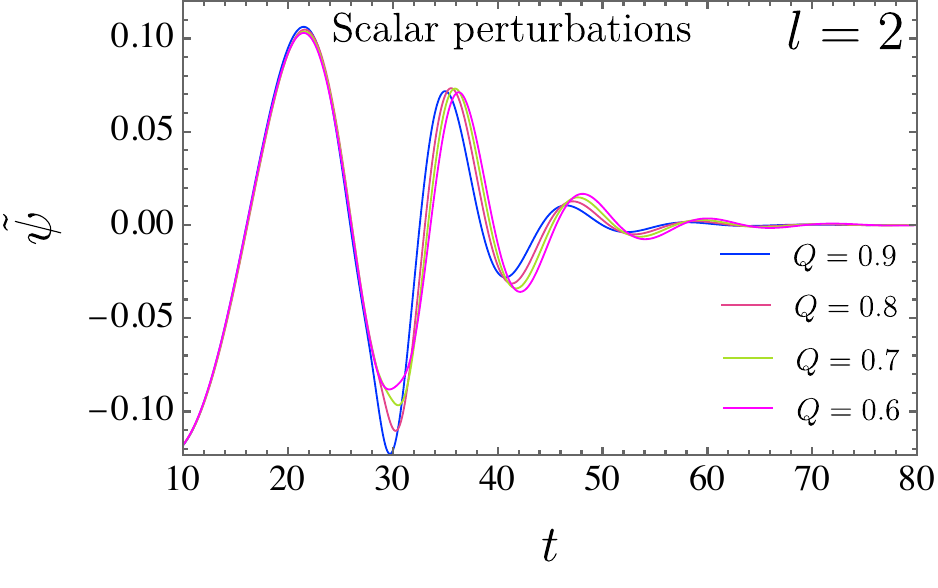}
    \caption{The waveform $\Tilde{\psi}$ is plotted for the scalar perturbations as a function of time $t$ for different values of $Q$, with $\xi$ fixed at $-0.001$. The analysis covers $Q = 0.6, 0.7, 0.8, 0.9$, with results presented for $l = 0$ (top left panel), $l = 1$ (top right panel) and $l = 2$ (bottom panel).}
    \label{scalarpsi}
\end{figure}

This section explores the time--domain evolution of scalar perturbations. In Fig. \ref{scalarpsi}, the waveform \( \Tilde{\psi} \) is plotted as a function of time \( t \) for different values of \( Q \), keeping \( \xi \) fixed at \(-0.001\). The selected values of \( Q \) are 0.6, 0.7, 0.8, and 0.9, with results shown for \( l = 0 \) (top--left panel), \( l = 1 \) (top--right panel), and \( l = 2 \) (bottom panel). The waveforms exhibit a damped oscillatory behavior over time.  

\begin{figure}
    \centering
    \includegraphics[scale=0.5]{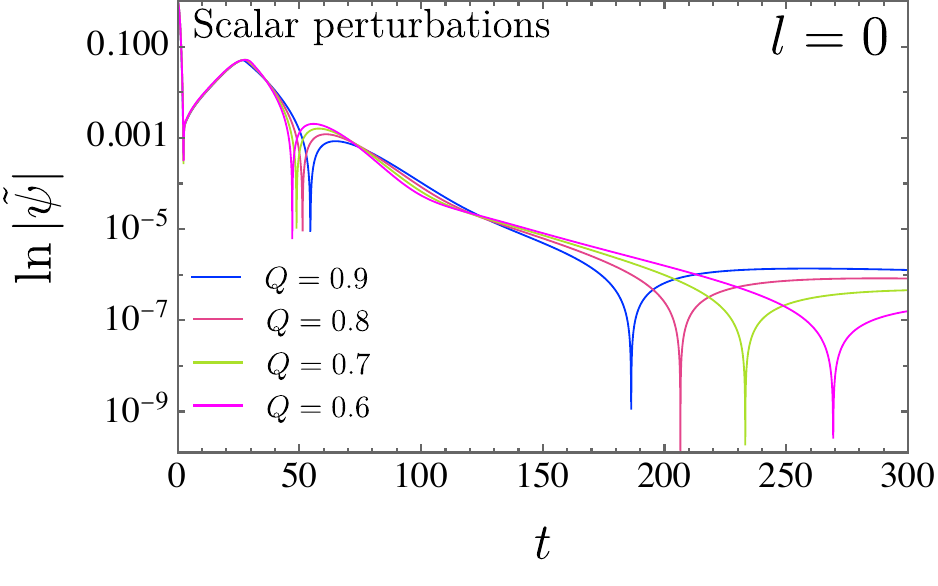}
    \includegraphics[scale=0.5]{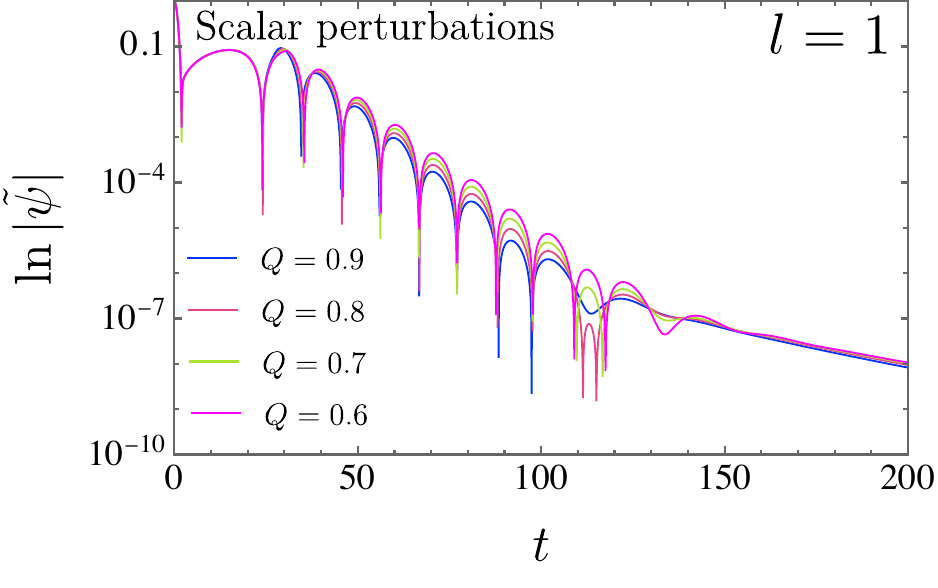}
    \includegraphics[scale=0.5]{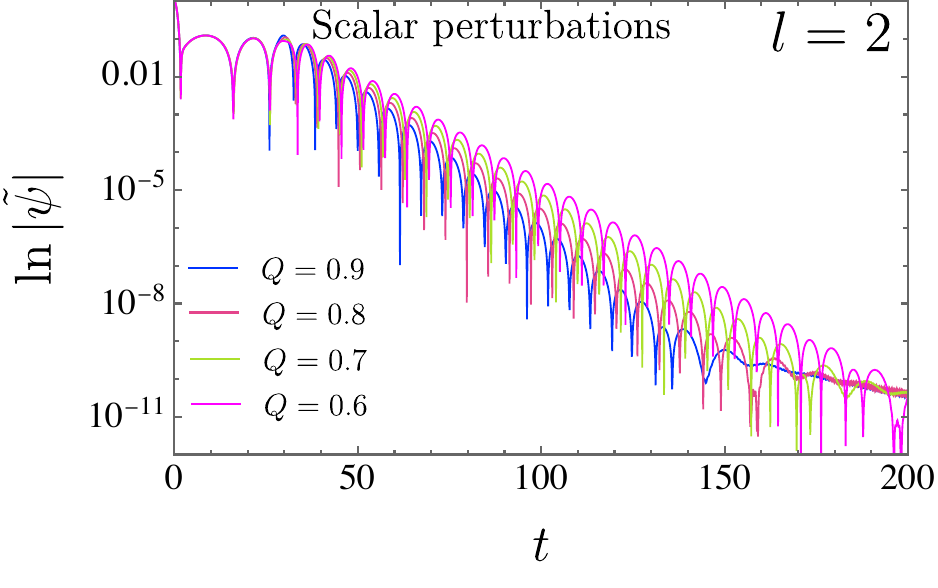}
    \caption{The $\ln|\Tilde{\psi}|$ is plotted for the scalar perturbations as a function of time $t$ for different values of $Q$, with $\xi$ fixed at $-0.001$. The analysis covers $Q = 0.6, 0.7, 0.8, 0.9$, with results presented for $l = 0$ (top left panel), $l = 1$ (top right panel) and $l = 2$ (bottom panel).}
    \label{scalarlnpsi}
\end{figure}

Fig. \ref{scalarlnpsi} shows the evolution of \( \ln|\Tilde{\psi}| \) for the same values of \( Q \) and angular momentum modes. The damping trend remains visible, and at late times, power-law tails emerge, a well--known characteristic following the quasinormal phase.  

\begin{figure}
    \centering
    \includegraphics[scale=0.5]{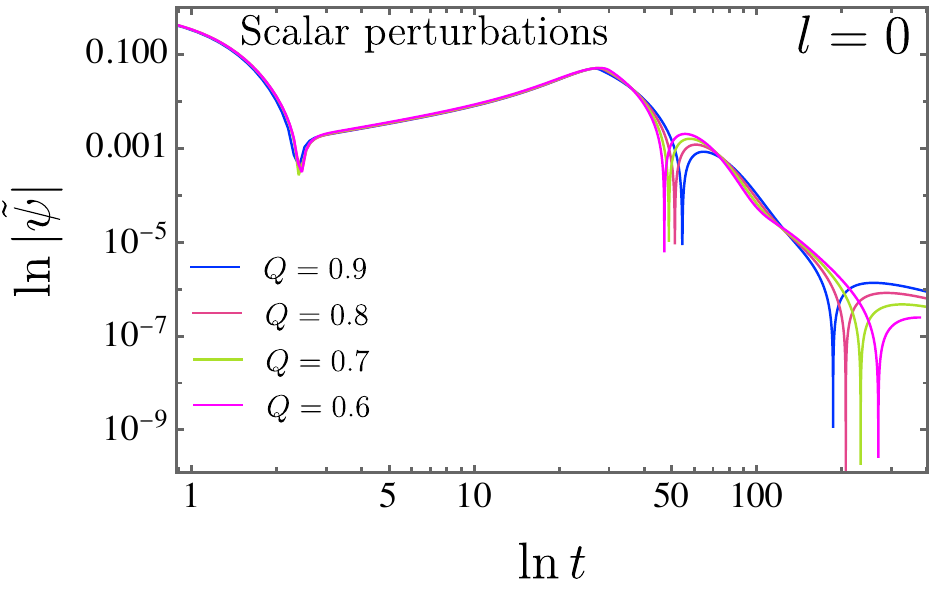}
    \includegraphics[scale=0.5]{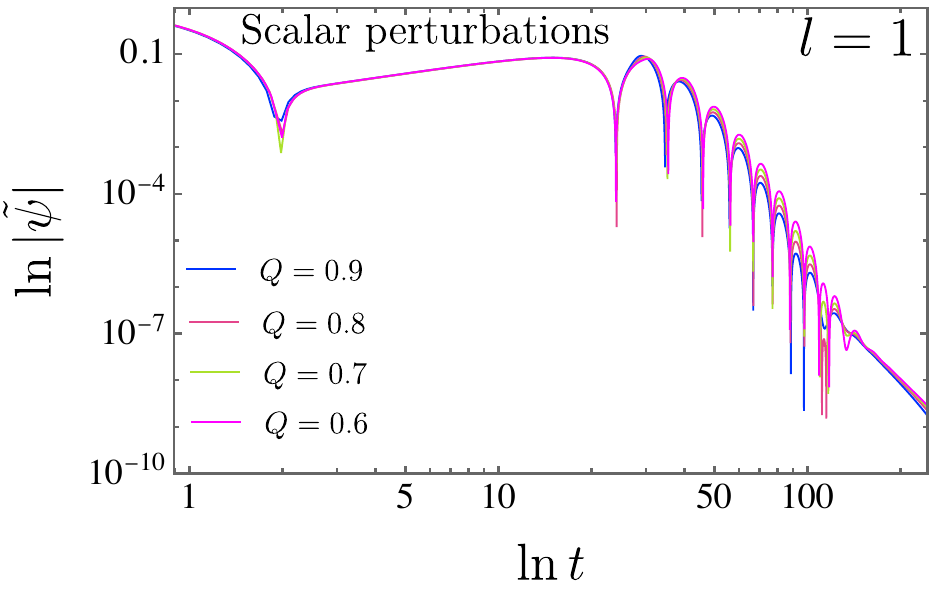}
    \includegraphics[scale=0.5]{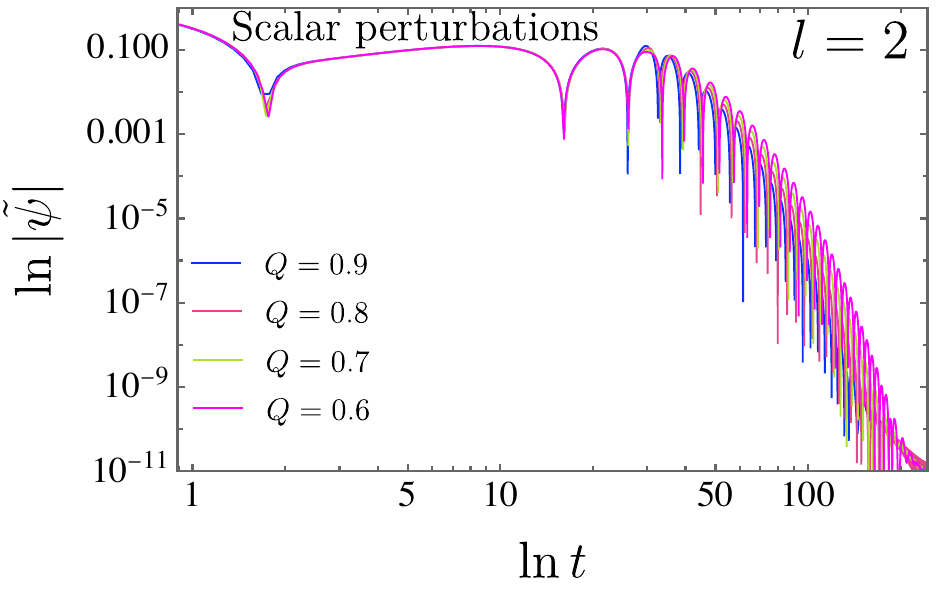}
    \caption{The $\ln|\Tilde{\psi}|$ is plotted for the scalar perturbations as a function $\ln|t|$ for different values of $Q$, with $\xi$ fixed at $-0.001$. The analysis covers $Q = 0.6, 0.7, 0.8, 0.9$, with results presented for $l = 0$ (top left panel), $l = 1$ (top right panel) and $l = 2$ (bottom panel).}
    \label{scalarloglogpsi}
\end{figure}

To further extend the analysis, Fig. \ref{scalarloglogpsi} presents a log--log plot of \( \Tilde{\psi} \) against \( t \), maintaining the same range of \( Q \) values and angular modes, with panels corresponding to \( l = 0 \) (top-left), \( l = 1 \) (top-right), and \( l = 2 \) (bottom).

\subsection{Vector perturbations}

\begin{figure}
    \centering
    \includegraphics[scale=0.5]{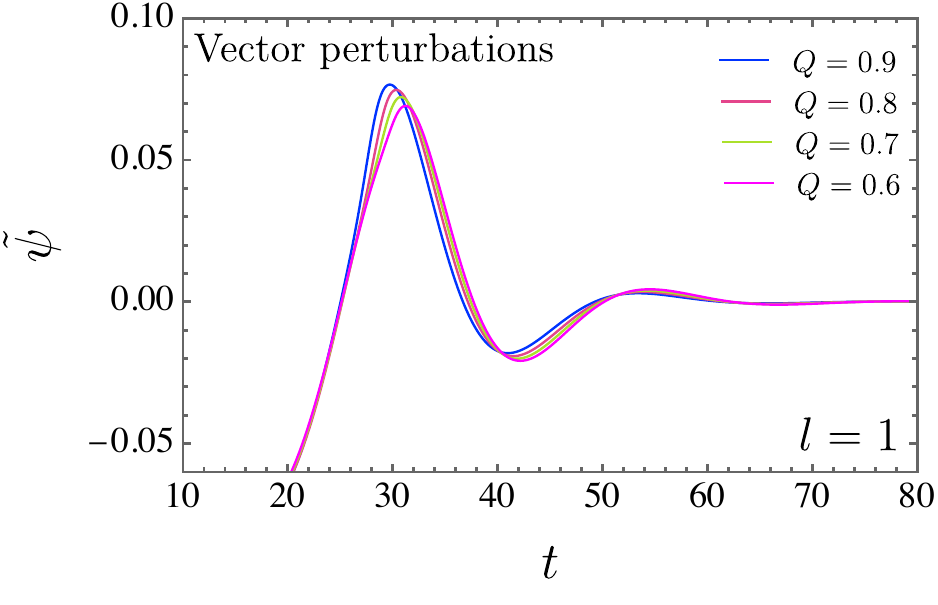}
    \includegraphics[scale=0.5]{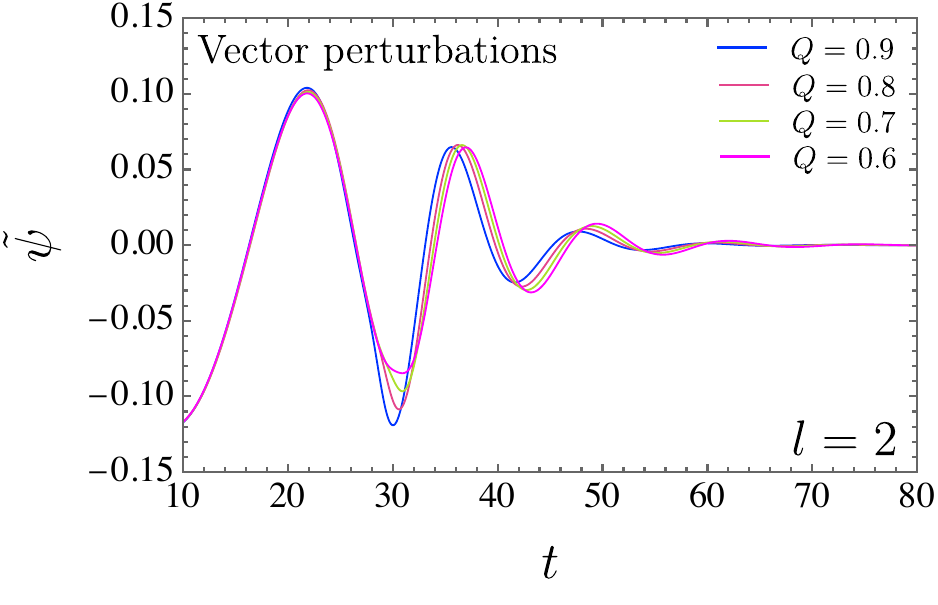}
    \caption{The waveform $\Tilde{\psi}$ is plotted for the vector perturbations as a function of time $t$ for different values of $Q$, with $\xi$ fixed at $-0.001$. The analysis covers $Q = 0.6, 0.7, 0.8, 0.9$, with results presented for $l = 1$ (left panel) and $l = 2$ (right panel).}
    \label{vectorpsi}
\end{figure}

This section focuses on the time--domain behavior of vector perturbations. Fig. \ref{vectorpsi} presents the evolution of the waveform \( \Tilde{\psi} \) over time \( t \) for different values of \( Q \), while keeping \( \xi \) fixed at \(-0.001\). The cases examined include \( Q = 0.6, 0.7, 0.8, 0.9 \), with results shown for various angular indices: \( l = 0 \) (left panel), \( l = 1 \) (middle panel), and \( l = 2 \) (right panel). The oscillations diminish over time, displaying a characteristic damping pattern.  

\begin{figure}
    \centering
    \includegraphics[scale=0.5]{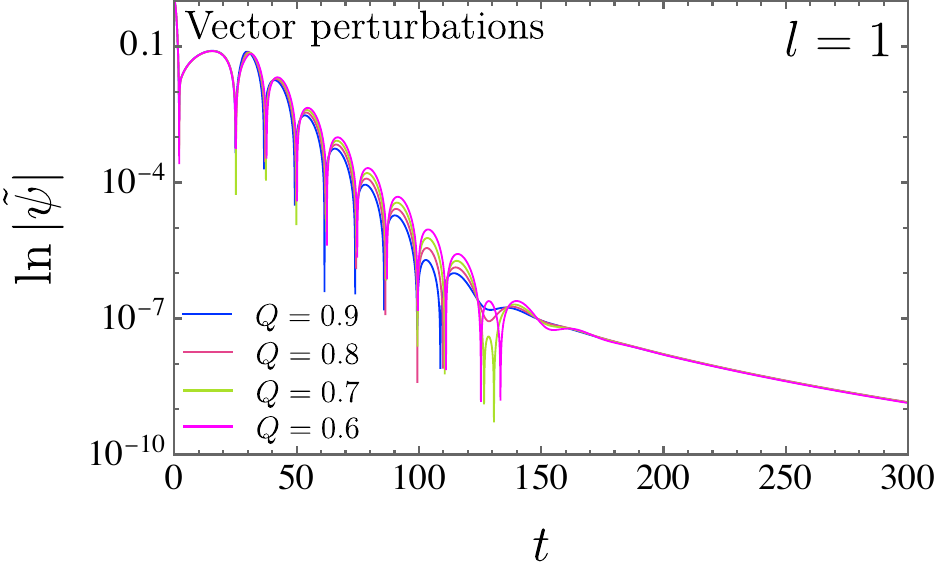}
    \includegraphics[scale=0.5]{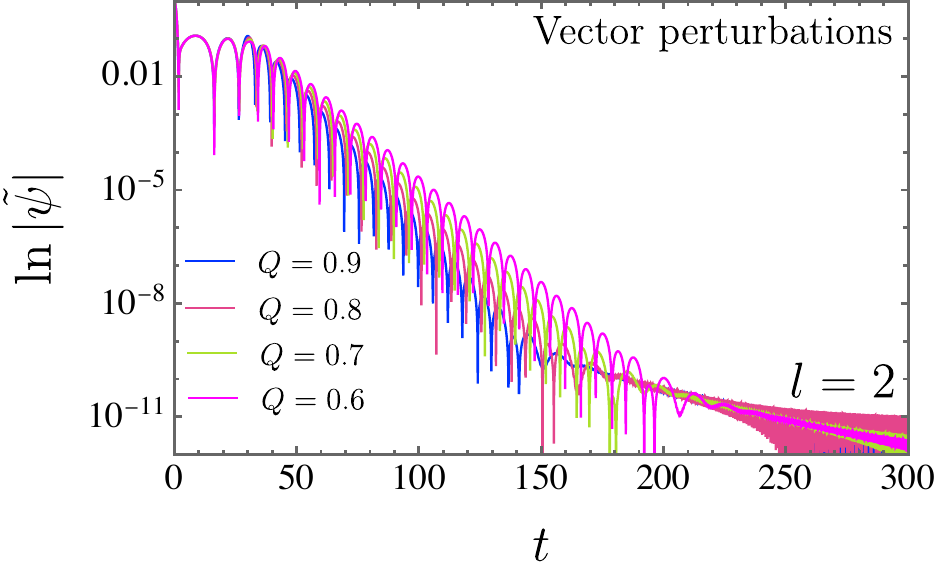}
    \caption{The $\ln|\Tilde{\psi}|$ is plotted for the vector perturbations as a function of time $t$ for different values of $Q$, with $\xi$ fixed at $-0.001$. The analysis covers $Q = 0.6, 0.7, 0.8, 0.9$, with results presented for $l = 1$ (left panel) and $l = 2$ (right panel).}
    \label{vectorlnpsi}
\end{figure}

To further investigate the decay behavior, Fig. \ref{vectorlnpsi} plots \( \ln|\Tilde{\psi}| \) as a function of \( t \) for the same parameter set. The damping trend remains consistent across all cases, and at later times, a transition to a power-law tail emerges, indicating the expected behavior following the quasinormal phase.  

\begin{figure}
    \centering
    \includegraphics[scale=0.5]{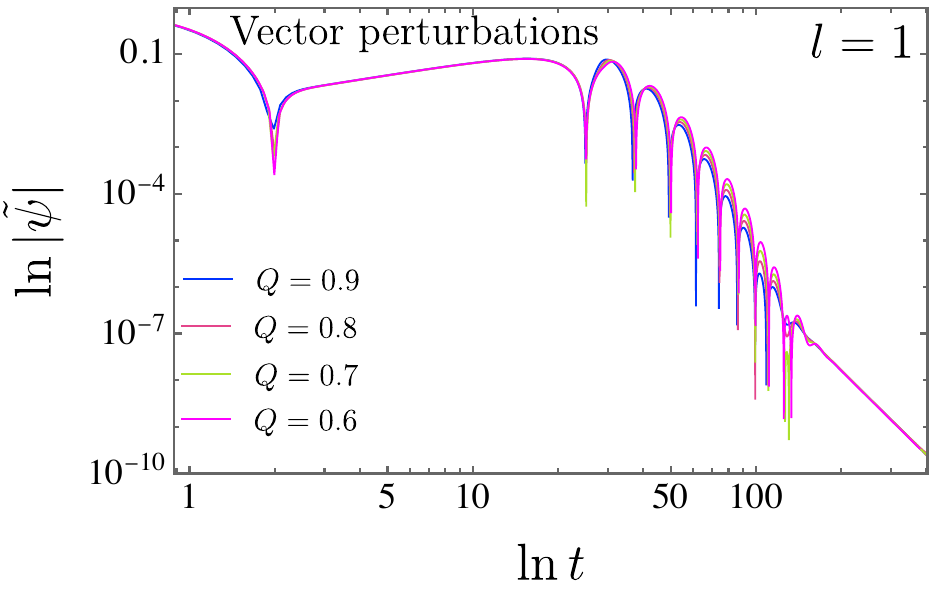}
    \includegraphics[scale=0.5]{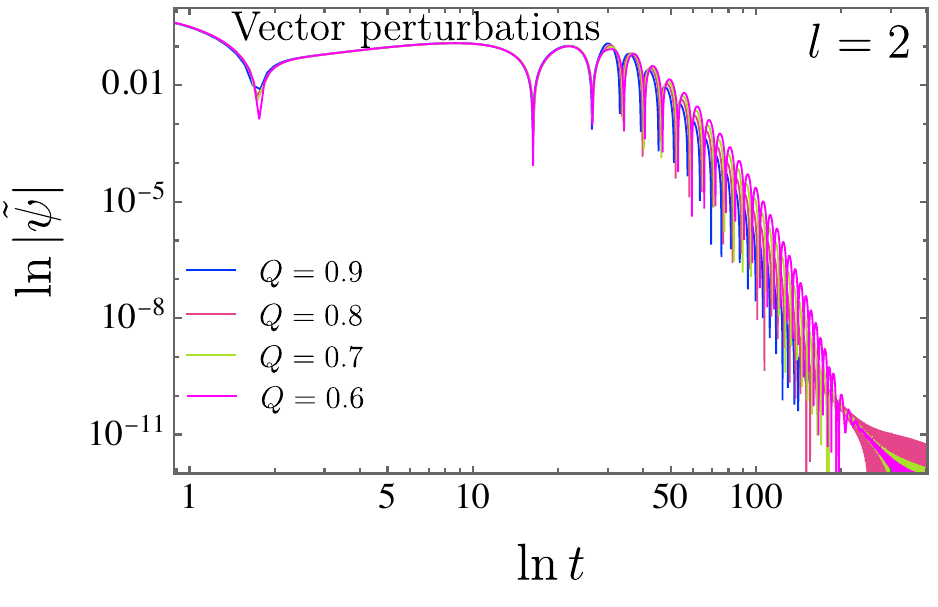}    \caption{The $\ln|\Tilde{\psi}|$ is plotted for the vector perturbations as a function $\ln|t|$ for different values of $Q$, with $\xi$ fixed at $-0.001$. The analysis covers $Q = 0.6, 0.7, 0.8, 0.9$, with results presented for $l = 1$ (left panel) and $l = 2$ (right panel).}
    \label{vectorloglogpsi}
\end{figure}

For an extended perspective, Fig. \ref{vectorloglogpsi} provides a logarithmic--scale analysis through a \( \ln \)–\( \ln \) representation of \( \Tilde{\psi} \) versus \( t \). This approach, considering the same values of \( Q \) and angular modes, emphasizes the asymptotic decay characteristics and reinforces the presence of power--law tails in the late--time signal.

\subsection{Tensor perturbations}

\begin{figure}
    \centering
    \includegraphics[scale=0.5]{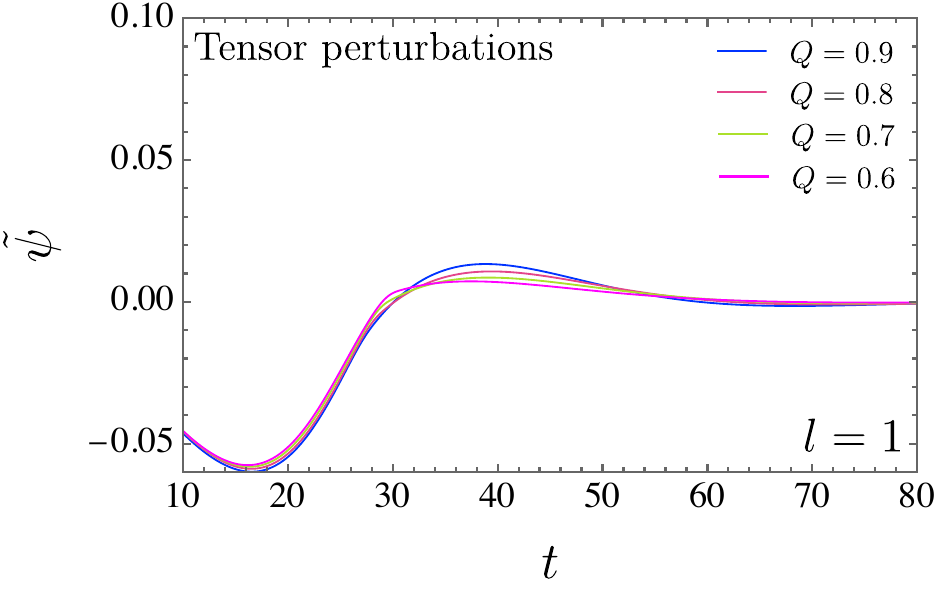}
    \includegraphics[scale=0.5]{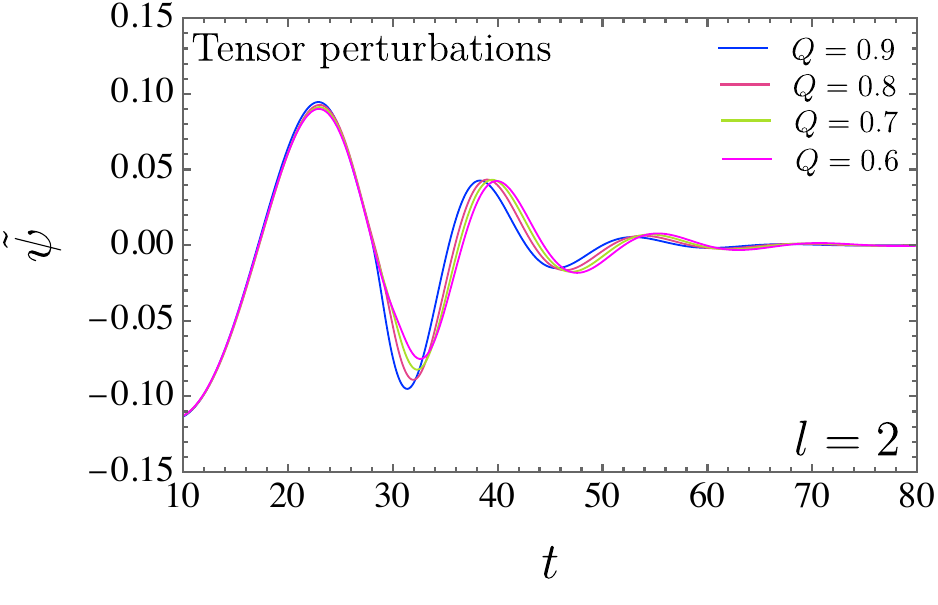}
    \caption{The waveform $\Tilde{\psi}$ is plotted for the tensor perturbations as a function of time $t$ for different values of $Q$, with $\xi$ fixed at $-0.001$. The analysis covers $Q = 0.6, 0.7, 0.8, 0.9$, with results presented for $l = 1$ (left panel) and $l = 2$ (right panel). }
    \label{tensorpsi}
\end{figure}

This section examines the time-domain evolution of tensor perturbations. Fig. \ref{tensorpsi} depicts the waveform \( \Tilde{\psi} \) as a function of time \( t \) for different values of \( Q \), keeping \( \xi \) fixed at \(-0.001\). The analysis covers the cases \( Q = 0.6, 0.7, 0.8, 0.9 \), with results shown for various angular modes: \( l = 0 \) (left panel), \( l = 1 \) (middle panel), and \( l = 2 \) (right panel). The perturbations exhibit a damped oscillatory pattern, with amplitudes gradually decreasing over time.  

\begin{figure}
    \centering
    \includegraphics[scale=0.5]{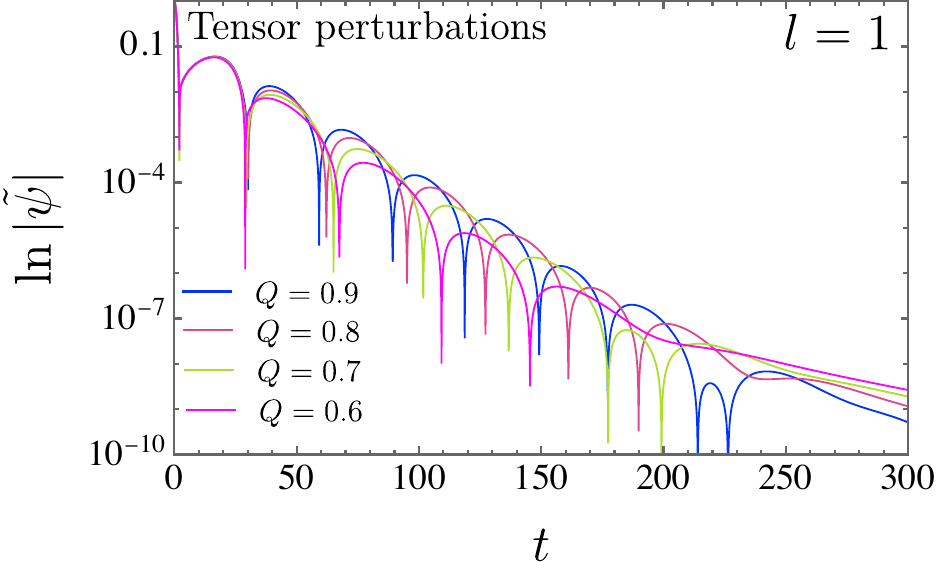}
    \includegraphics[scale=0.5]{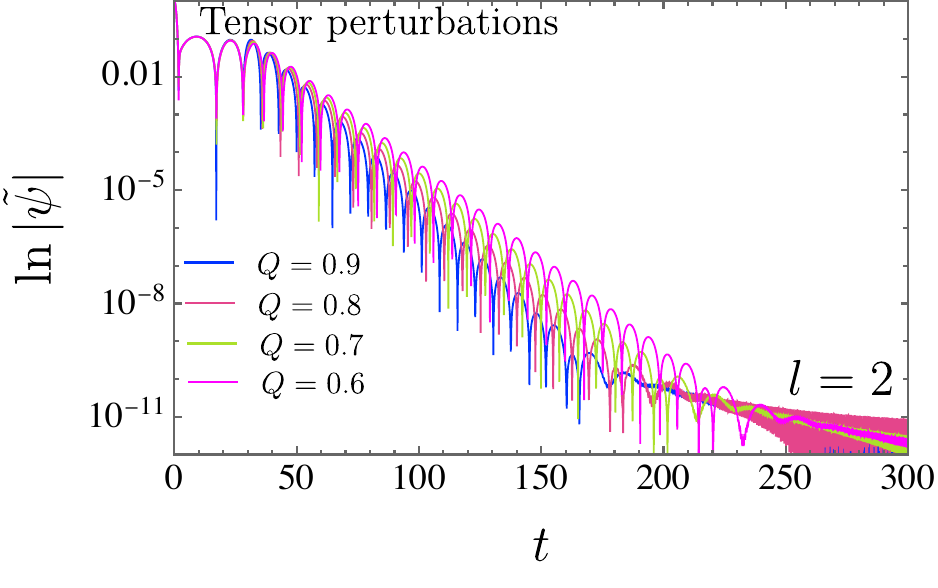}
    \caption{The $\ln|\Tilde{\psi}|$ is plotted for the tensor perturbations as a function of time $t$ for different values of $Q$, with $\xi$ fixed at $-0.001$. The analysis covers $Q = 0.6, 0.7, 0.8, 0.9$, with results presented for $l = 1$ (left panel) and $l = 2$ (right panel).}
    \label{tensorlnpsi}
\end{figure}

To further investigate the decay behavior, Fig. \ref{tensorlnpsi} illustrates the logarithmic amplitude \( \ln|\Tilde{\psi}| \) as a function of \( t \) for the same set of parameters. The attenuation remains consistent across all cases, and at later times, a transition to a power--law tail becomes apparent, signaling the expected post--quasinormal phase.  

\begin{figure}
    \centering
    \includegraphics[scale=0.5]{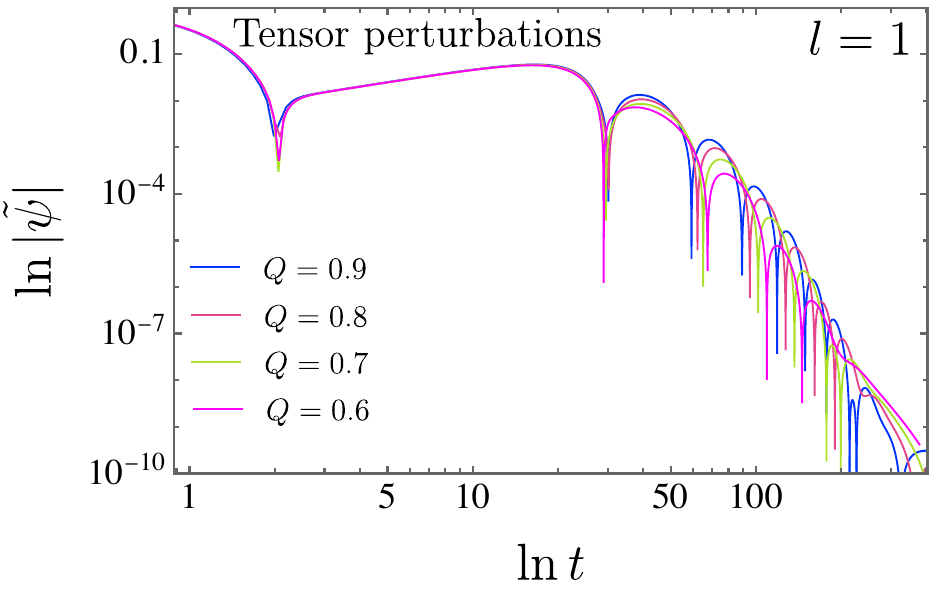}
    \includegraphics[scale=0.5]{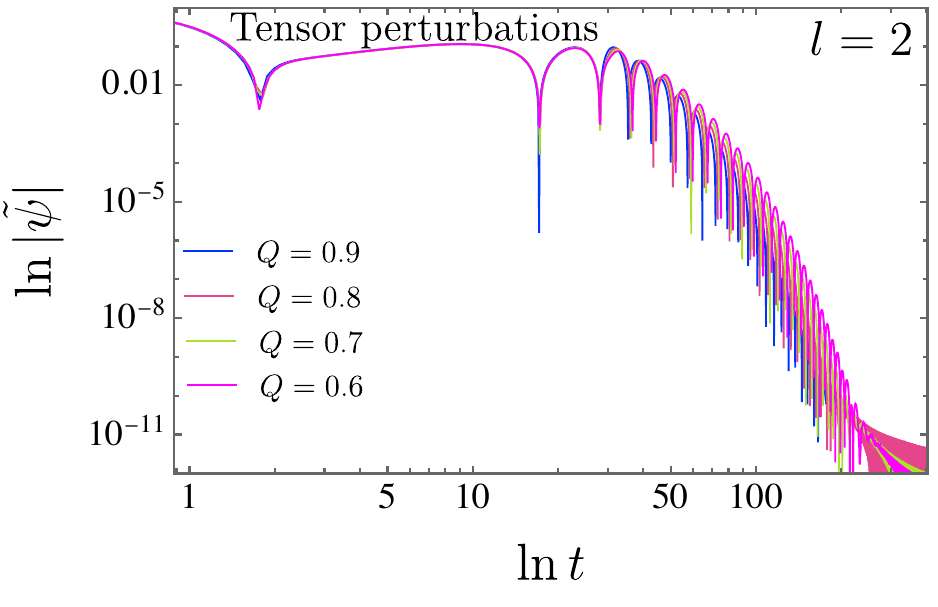}    \caption{The $\ln|\Tilde{\psi}|$ is plotted for the tensor perturbations as a function $\ln|t|$ for different values of $Q$, with $\xi$ fixed at $-0.001$. The analysis covers $Q = 0.6, 0.7, 0.8, 0.9$, with results presented for $l = 1$ (left panel) and $l = 2$ (right panel).}
    \label{tensorloglogpsi}
\end{figure}

For an extended perspective, Fig. \ref{tensorloglogpsi} presents a \( \ln \)–\( \ln \) plot of \( \Tilde{\psi} \) against \( t \), offering a detailed view of the asymptotic decay regime. By exploring the same range of \( Q \) values and angular indices, this representation confirms the emergence of power--law tails at late times, further highlighting their role in the evolution of tensor perturbations.

\section{Conclusion}

This paper was devoted to analyze a nonlinear electromagnetic generalization of the Reissner--Nordström black hole recently introduced in the literature \cite{galindo2024nonlinear}. We began by presenting the corresponding metric, alongside the event horizon. To identify any physical singularities, we computed the Kretschmann scalar, which confirmed the existence of a singularity as \( r \to 0 \). The Ricci scalar exhibited similar behavior, showing the same divergence in the same limit.

Next, we examined the trajectory of light by evaluating the photon spheres and the resulting shadows. Observational data from the Event Horizon Telescope (EHT) were used to establish limits on the parameter $\xi$, which governed the nonlinear modification of the Reissner--Nordström black hole. Additionally, we assessed the stability of critical orbits. The analysis using the Gaussian curvature criterion confirmed that the orbit outside the event horizon ($r_{1c}$) was unstable. For completeness, we numerically computed the geodesics. Additionally, we explored the time delay. Regarding time delay, as \( \xi \) decreased, the delay \( \Delta T \) also decreased.

To further investigate the static solution, we examined the thermodynamic behavior of the theory, focusing on the Hawking temperature and heat capacity. Broadly speaking, the magnitude of the temperature \( T \) increased with increasing \( \xi \), whereas an increase in \( Q \) caused a reduction in temperature. A similar trend was observed for the heat capacity \( C_V \), where changes in \( \xi \) and \( Q \) resulted in shifts in its values. Furthermore, Hawking radiation was derived as a tunneling process, with the corresponding particle density, $n$, also computed. When $Q$ increased, there existed a reduction of $n$.

The quasinormal modes were determined using the WKB approximation in order to obtain the system's characteristic oscillations under scalar, vector, and tensor perturbations. For scalar modes, a decrease in $\xi$ and an increase in $Q$ resulted in weaker damping for $\omega_{0}$, $\omega_{1}$, and $\omega_{2}$. A similar pattern was observed for vector modes, where $\omega_{0}$ exhibited reduced damping. Likewise, for tensor modes, the same trend was identified, with $\omega_{0}$ and $\omega_{1}$ displaying less damped oscillations.  

Additionally, the time--domain analysis was carried out to examine the evolution of these perturbations. The results confirmed the quasinormal mode behavior, showing progressively damped frequencies and a characteristic tail at late times. The expected power--law decay was also evident in the late--time regime.

As a future direction, studies on gravitational lensing via the extended \textit{Gauss--Bonnet} theorem \cite{Li:2020wvn}, along with scattering effects and \textit{greybody} factors for both fermions and bosons, are also promising areas for development. These and other ideas were under investigation.


\section*{Acknowledgments}
\hspace{0.5cm}

A. A. Araújo Filho acknowledges support from the Conselho Nacional de Desenvolvimento Científico e Tecnológico (CNPq) and the Fundação de Apoio à Pesquisa do Estado da Paraíba (FAPESQ) under grant [150891/2023-7]. The author also expresses gratitude to C. Molina and N. Heidari for assistance with computational calculations and for the fruitful discussions during the revision of this paper.

\section{Data Availability Statement}

Data Availability Statement: No Data associated in the manuscript


\bibliographystyle{ieeetr}
\bibliography{main}

\end{document}